\documentclass[reprint, aps, prd, letterpaper, noshowpacs, amsmath, %
amssymb, amsfonts, nofootinbib, floatfix, twoside]{revtex4-1}
\pdfoutput=1

\usepackage[T3,T1]{fontenc}
\usepackage{mathrsfs} % For \mathscr fonts
\usepackage{bm} % For \bm fonts
\makeatletter
\@ifclassloaded{beamer}
  { % if this is a beamer document, use sans-serif fonts
    \typeout{UsePackages: Detected beamer}
    \usepackage{tgheros}
    
  }
  { % otherwise:
    \typeout{UsePackages: Did not detect beamer}
    \ifx\asybeamer\undefined % use times for articles
    \typeout{UsePackages: Detected article}
    \usepackage[varg]{txfonts} % For math
    \usepackage{tgtermes} % For text
    \else % use beamer fonts for asy documents with beamer
    \typeout{Fonts: Detected asy for beamer}
    \usepackage{tgheros}
    
    \fi
  }
\usepackage{microtype} % For nicer alignment of text
\def\MT@register@subst@font{
  \MT@exp@one@n\MT@in@clist\font@name\MT@font@list
  \ifMT@inlist@\else\xdef\MT@font@list{\MT@font@list\font@name,}\fi}
\makeatother

\usepackage{graphicx} %
\usepackage[x11names,svgnames,rgb]{xcolor} %
\usepackage{xspace}
\usepackage{braket}
\usepackage{accents}
\usepackage{siunitx}
\usepackage{mathtools}
\DeclareSymbolFontAlphabet{\mathrm}{operators}

\definecolor{CiteColor}{rgb}{0.18039, 0.18824, 0.57255}
\definecolor{UrlColor} {rgb}{0.741, 0.173, 0.000}
\definecolor{DarkUrlColor} {rgb}{0.500, 0.110, 0.000}
\definecolor{LinkColor}{rgb}{0.25098, 0.47843, 0.04706}
\makeatletter %
\newcommand{\ShowFont}{%
  \typeout{The main font is \f@encoding \space \f@family \space %
    \f@series \space \f@shape \space at \f@size pt.}%
  \typeout{The math font sizes are \tf@size pt (main), \sf@size pt %
    (script), and \ssf@size pt (scriptscript).}%
  \typeout{The linewidth is \the\linewidth}} %
\makeatother %
\usepackage{xargs}

\usepackage[farskip=0pt,caption=false]{subfig}
\usepackage{enumitem}

\makeatletter

\def\@seccntformat#1{\csname the#1\endcsname.~}%

\def\section{%
  \@startsection {section}
  {1} {\z@} {0.55cm \@plus1ex \@minus .02ex}%
    {0.225cm} { \normalfont\bfseries \centering}%
}%
\def\subsection{%
  \@startsection {subsection}
  {2} {\z@ } {0.45cm \@plus 0.8ex \@minus 0.2ex}%
  {0.1125cm}{\normalfont \bfseries \centering }}
\def\subsubsection{%
  \@startsection {subsubsection}
  {3} {\z@ } {0.4cm \@plus 0.6ex \@minus 0.1ex}%
  {0.075cm}{\normalfont \it \centering }}

\newcommand{\surnames}[1]{\def\@surnamelist{#1}\relax}
\surnames{\ }
\ifx \@shorttitle \@empty
\else
  \def\@oddhead{\small \MakeUppercase{\@shorttitle} \hfill}
\fi
\ifx \@surnamelist \@empty
\else
  \def\@evenhead{\small \MakeUppercase{\@surnamelist} \hfill}
\fi
\def\@oddfoot{\reset@font\hfil\thepage\hfil}%
\def\@evenfoot{\reset@font\hfil\thepage\hfil}%
\g@addto@macro\maketitle{\global\@specialpagetrue\gdef\@specialstyle{plain}}

\usepackage[colorlinks, plainpages=false,
hyperfigures=true]{hyperref}
\hypersetup{linkcolor=LinkColor}
\hypersetup{citecolor=CiteColor}
\hypersetup{urlcolor=UrlColor}
\hypersetup{setpagesize=false}
\hypersetup{pdfborder=0 0 0}
\makeatother

\let\Originalddefinition\d
\renewcommand{\d}{\ensuremath{\mathrm{d}}}

\newcommand{\e}{\ensuremath{\mathrm{e}}}
\let\Originalidefinition\i
\renewcommand{\i}{\ensuremath{\mathrm{i}}}
\DeclareSymbolFont{wasy}{U}{wasy}{m}{n}
\DeclareMathSymbol{\thorn}{\mathord}{wasy}{105}
\DeclareMathSymbol{\Thorn}{\mathord}{wasy}{106}
\newcommand\Eth{\text{\DH}}
\newcommand\Alpha{\mathrm{A}}
\newcommand\Beta{\mathrm{B}}
\newcommand\Epsilon{\mathrm{E}}
\newcommand\Varepsilon{\mathit{E}}
\newcommand\Zeta{\mathrm{Z}}
\newcommand\Eta{\mathrm{H}}
\newcommand\Vartheta{\varTheta}
\newcommand\Iota{\mathrm{I}}
\newcommand\Kappa{\mathrm{K}}
\newcommand\Mu{\mathrm{M}}
\newcommand\Nu{\mathrm{N}}
\chardef\omicron=111 % For insurance against upcasing
\chardef\Omicron=79  % For insurance against downcasing
\newcommand{\Varpi}{\varPi}
\newcommand\Rho{\mathrm{P}}
\newcommand\Varrho{\mathit{P}}
\newcommand\Varsigma{\varSigma}
\newcommand\Varphi{\varPhi}
\newcommand\Tau{\mathrm{T}}
\newcommand\Chi{\mathrm{X}}
\newcommand\MakeAllUppercase[1]{%
  \begingroup
  \let\eth\Eth
  \let\thorn\Thorn
  \let\alpha\Alpha
  \let\beta\Beta
  \let\gamma\Gamma
  \let\delta\Delta
  \let\epsilon\Epsilon
  \let\varepsilon\Varepsilon
  \let\zeta\Zeta
  \let\eta\Eta
  \let\theta\Theta
  \let\vartheta\Vartheta
  \let\iota\Iota
  \let\kappa\Kappa
  \let\lambda\Lambda
  \let\mu\Mu
  \let\nu\Nu
  \let\xi\Xi
  \let\omicron\Omicron
  \let\pi\Pi
  \let\varpi\Varpi
  \let\rho\Rho
  \let\varrho\Varrho
  \let\sigma\Sigma
  \let\varsigma\Varsigma
  \let\tau\Tau
  \let\upsilon\Upsilon
  \let\phi\Phi
  \let\varphi\Varphi
  \let\chi\Chi
  \let\psi\Psi
  \let\omega\Omega
  \MakeUppercase{#1}%
  \endgroup
}

\let\Originalcdefinition\c
\renewcommand{\c}{\mathrm{c}}

\newcommand{\MSun}{\ensuremath{M_\odot}\xspace}

\DeclareSIUnit{\strain}{strain}
\DeclareSIPrePower{\root}{1/2}
\DeclareSIUnit{\parsec}{pc}
\DeclareSIUnit{\yr}{yr}
\DeclareSIUnit{\year}{yr}
\DeclareSIUnit{\lightyear}{ly}
\DeclareSIUnit{\SolarMass}{\ensuremath{\MSun}}
\DeclareSIUnit{\Mass}{\ensuremath{M}}
\newcommand{\abs} [1]{\left\lvert{#1}\right\rvert}

\DeclareMathOperator{\artanh}{artanh}

\newcommand{\defined}{\coloneqq}
\newcommand{\identically}{\equiv}
\newcommand{\roughly}{\mathord{\sim}} % Different from \sim in spacing
\newcommand{\asymptoticallyequal}{\simeq}

\newcommand{\scriplus}{\ensuremath{\mathscr{I}^{+}}}
\newcommand{\scriminus}{\ensuremath{\mathscr{I}^{-}}}

\newcommand{\D}{\ensuremath{\mathfrak{D}} }

\newcommand{\sYlm}[1]{\ensuremath{\scripts{_{s}}{Y}{_{#1}}}}

\newcommand{\mTwoYlm}[1]{\ensuremath{\scripts{_{-2}}{Y}{_{#1}}}}

\DeclareSymbolFont{tipa}{T3}{tipa}{m}{n}
\DeclareMathAccent{\ibreve}{\mathalpha}{tipa}{'020}
\newcommand{\Rotated}[1]{\ensuremath{\ibreve{#1}}}

\newcommandx{\structure}[3][2={}, 3={}]{f^{#1 #2}_{#3}}

\newcommand{\SOp}[1]{\ensuremath{\mathrm{SO}^{+}(#1)}}

\newcommand{\PSL}[1]{\ensuremath{\mathrm{PSL}(#1)}}

\newcommand{\bms}{\ensuremath{\mathfrak{bms}}}

\newcommand{\h}{\ensuremath{h} }

\newcommand{\FourVector}[1]{\bm{#1}}
\newcommand{\Vector}[1]{\FourVector{#1}}

\newcommand{\fourvec}[1]{\FourVector{#1}}
\newcommand{\threevec}{\Vector}

\newcommand{\rapidity}{\varphi}

\makeatletter
\newcommand{\foreign}[1]{\textit{#1}} % {\textrm{#1}}
\newcommand{\etal}{\foreign{et~al}\@ifnextchar{\relax}{.\relax}{\ifx\@let@token.\else\ifx\@let@token~.\else.\@\xspace\fi\fi}}
\newcommand{\etc}{\foreign{etc}\@ifnextchar{\relax}{.\relax}{\ifx\@let@token.\else\ifx\@let@token~.\else.\@\xspace\fi\fi}}
\newcommand{\eg}{\foreign{e.g}\@ifnextchar{\relax}{.\relax}{\ifx\@let@token.\else\ifx\@let@token~.\else.\@\xspace\fi\fi}}
\newcommand{\ie}{\foreign{i.e}\@ifnextchar{\relax}{.\relax}{\ifx\@let@token.\else\ifx\@let@token~.\else.\@\xspace\fi\fi}}
\newcommand{\perse}{\foreign{per~se}\xspace}
\newcommand{\cf}{\foreign{cf}\@ifnextchar{\relax}{.\relax}{\ifx\@let@token.\else\ifx\@let@token~.\else.\@\xspace\fi\fi}}
\makeatother

\newcommand{\pN}{\text{PN}\xspace}

\newcommand{\Poincare}{Poincar{\'{e}}\xspace}

\definecolor{NoteColor}{rgb}{0.900, 0.218, 0.000}

\definecolor{NewColor}{rgb}{0,.55,0}

\newcommand{\software}[1]{\texttt{#1}}

\makeatletter
\newcommand{\CapName}[1]{\textbf{#1}.}
\newcommand{\ShowDimensions}{%
  \typeout{The font encoding is \f@encoding}        %
  \typeout{The font family is \f@family}            %
  \typeout{The font series is \f@series}            %
  \typeout{The font shape is \f@shape}              %
  \typeout{The font size is \f@size}                %
  \typeout{The baselineskip is \f@baselineskip}     %
  \typeout{The math font size is \tf@size}          %
  \typeout{The math script size is \sf@size}        %
  \typeout{The math scriptscript size is \ssf@size} %
  \typeout{The linewidth is \the\linewidth}         %
  \typeout{The textwidth is \the\textwidth}         %
}
\newcommand{\prefixscripts}[2]{%
  \@mathmeasure\z@\displaystyle{#2}%
  \global\setbox\@ne\vbox to\ht\z@{}\dp\@ne\dp\z@
  \setbox\tw@\box\@ne
  \@mathmeasure4\displaystyle{\copy\tw@#1}%
  \@mathmeasure6\displaystyle{#2}%
  \dimen@-\wd6 \advance\dimen@\wd4 \advance\dimen@\wd\z@
  \hbox to\dimen@{}{\kern-\dimen@\box4\box6}%
}
\newcommand{\scripts}[3]{%
  \@mathmeasure\z@\displaystyle{#2}%
  \global\setbox\@ne\vbox to\ht\z@{}\dp\@ne\dp\z@
  \setbox\tw@\box\@ne
  \@mathmeasure4\displaystyle{\copy\tw@#1}%
  \@mathmeasure6\displaystyle{#2#3}%
  \dimen@-\wd6 \advance\dimen@\wd4 \advance\dimen@\wd\z@
  \hbox to\dimen@{}{\kern-\dimen@\box4\box6}%
}
\makeatother

\makeatletter
\let\protect\relax
{\catcode`\|=\active
  \xdef\InnerProduct{\protect\expandafter\noexpand\csname InnerProduct \endcsname}
  \expandafter\gdef\csname InnerProduct \endcsname#1{%
    \begingroup
    \ifx\SavedDoubleVert\relax
    \let\SavedDoubleVert\|\let\|\IpDoubleVert
    \fi
    \mathcode`\|32768\let|\IPVert
    \left({#1}\right)
    \endgroup
  }
}
\def\IPVert{\@ifnextchar|{\|\@gobble}% turn || into \|
     {\egroup\,\mid@vertical\,\bgroup}}
\def\IPDoubleVert{\egroup\,\mid@dblvertical\,\bgroup}
\let\SavedDoubleVert\relax
\def\midvert{\egroup\mid\bgroup}
\def\SetVert{\@ifnextchar|{\|\@gobble}% turn || into \|
    {\egroup\;\mid@vertical\;\bgroup}}
\def\SetDoubleVert{\egroup\;\mid@dblvertical\;\bgroup}
\def\mid@vertical{\mskip1mu\vrule\mskip1mu}
\def\mid@dblvertical{\mskip1mu\vrule\mskip2.5mu\vrule\mskip1mu}
\makeatother

\input{Macros_GeometricAlgebra}
\newcommand{\Cornell}{\affiliation{Cornell Center for Astrophysics and
    Planetary Science, Cornell University, Ithaca, New York 14853,
    USA}} %

\DeclareMathAlphabet{\mathbfsf}{\encodingdefault}{\sfdefault}{bx}{sl}
\newcommand{\A}{\ensuremath{\mathscr{A}}}
\newcommand{\B}{\ensuremath{\mathscr{B}}}
\renewcommand{\O}{\ensuremath{\mathscr{O}}}
\newcommand{\N}{\ensuremath{\mathscr{N}}}

\newcommand{\supertranslations}{\mathbfsf{T}}
\newcommand{\scritime}{\ensuremath{u}}
\newcommand{\observertime}{\ensuremath{\tau}}
\newcommand{\distance}{\ensuremath{r}}

\newcommand{\observer}{\mathscr{O}}
\newcommand{\point}{p}

\newcommand{\anticelestialsphere}{\mathscr{S}^+}

\newcommand{\waveform}{\ensuremath{f}}
\newcommand{\directionvec}{\fourvec{r}}
\newcommand{\ellnonzero}{\ell_\text{nz}}
\newcommand{\mnonzero}{m_\text{nz}}

\newcommand{\BoostRotor}{\rotor{B}'}
\newcommand{\FrameRotor}{\rotor{R}_{\text{f}}}

\newcommand{\finite}{\MakeAllUppercase}
\newcommand{\asymptotic}{\MakeLowercase}
\newcommand{\metric}{\gamma}
\newcommand{\st}{\alpha}
\newcommand{\sfunc}{w}
\newcommand{\SpinPhase}{\lambda}
\newcommand{\ellmax}{\ell_{\text{max}}}

\newcommand{\SXSwaveforms}{http://www.black-holes.org/waveforms}
\newcommand{\xCoM}{\fourvec{x}_{\text{CoM}}}

\newcommand{\arxivAbs}{http://arxiv.org/abs/1509.00862}

\allowdisplaybreaks[1]

\begin{document}

%%%%%%%%%%%%%%%%

\graphicspath{%
  {Plots/}%
  % More directories are added in braces, without commas between
}

\title[Transformations of asymptotic gravitational-wave data]
{Transformations of asymptotic gravitational-wave data}

% \makeatletter \@booleantrue\frontmatterverbose@sw \makeatother

\surnames{Boyle}%, \protect\etal}
\author{Michael Boyle} \Cornell

\date{\today}

\begin{abstract}
  Gravitational-wave data is gauge dependent.  While we can restrict
  the class of gauges in which such data may be expressed, there will
  still be an infinite-dimensional group of transformations allowed
  while remaining in this class, and almost as many different---though
  physically equivalent---waveforms as there are transformations.
  This paper presents a method for calculating the effects of the most
  important transformation group, the Bondi-Metzner-Sachs (BMS) group,
  consisting of rotations, boosts, and supertranslations (which
  include time and space translations as special cases).  To a
  reasonable approximation, these transformations result in simple
  coupling between the modes in a spin-weighted spherical-harmonic
  decomposition of the waveform.  It is shown that waveforms from
  simulated compact binaries in the publicly available SXS waveform
  catalog contain unmodeled effects due to displacement and drift of
  the center of mass, accounting for mode-mixing at typical levels of
  \SI{1}{\percent}.  However, these effects can be mitigated by
  measuring the average motion of the system's center of mass for a
  portion of the inspiral, and applying the opposite transformation to
  the waveform data.  More generally, controlling the BMS
  transformations will be necessary to eliminate the gauge ambiguity
  inherent in gravitational-wave data for both numerical and
  analytical waveforms.  Open-source code implementing BMS
  transformations of waveforms is included along with this paper in
  the supplemental materials.
\end{abstract}


% 04.25.-g, % Approximation methods; equations of motion
% 04.25.D-, % Numerical relativity
% 04.25.dc, % NR studies of crit. behavior, sing.'s, cosmic censorsh.
% 04.25.dg, % NR studies of black holes and black-hole binaries
% 04.25.dk, % NR studies of other relativistic binaries
% 04.25.Nx, % PN approximation; perturbation theory; etc.
% 04.30.-w, % Gravitational waves
% 04.30.Db, % Wave generation and sources
% 04.30.Nk, % Wave propagation and interactions
% 04.30.Tv, % Gravitational-wave astrophysics
% 04.80.Nn, % Gravitational wave detectors and experiments

\maketitle

%%%%%%%%%%%%%%%%%%%%%%%%%%%%%%%%%%%%%%%%%%%%%%%%%%%%%%%%%%%%%%%%%%%%%%
%%%%%%%%%%%%%%%%%%%%%%%%%%%%%%%%%%%%%%%%%%%%%%%%%%%%%%%%%%%%%%%%%%%%%%
\section{Introduction}
\label{sec:Introduction}

As the era of gravitational-wave astronomy approaches, models of
gravitational waveforms from physical systems become crucial to the
extraction of scientific results from the data.  The basic goal of
this effort is to make the claim that a waveform measured in a
detector corresponds to some particular physical model.  But a
treacherous gulf lies between any waveform and its corresponding
physical model, abounding in subtle and delicate challenges---not
least of which is the gauge flexibility of general relativity.  This
paper describes the gauge transformations most relevant to studies of
gravitational waves and shows how to calculate their effects on
waveforms.  We will see that, in order to obtain accurate waveform
models, we must account for gauge effects.

The literature on gravitational-wave analysis almost universally
allows for two standard gauge ambiguities: time translations and phase
rotations.  For example, the standard technique of matched filtering
involves optimizing the match over the time and phase of the
signal~\cite{Finn1992, Cutler1994}.  Similarly, comparisons between
numerical evolutions, between numerical and analytical waveforms, and
between different approximate analytical waveforms have generally
allowed for time and phase offsets~\cite{Buonanno2007, Boyle2007}.
These transformations alter the waveforms, but in well behaved ways
which can be expressed fairly simply as functions of the
transformations.  More recently, the harder problem of analyzing
precessing systems has required generalizing phase rotations to
include the full three-dimensional rotation group, which induces
slightly more complicated---though still well
understood---transformations of the waveforms~\cite{Schmidt2011,
  OShaughnessy2011a, Boyle2011, Boyle2014}.

Because of the essential diffeomorphism invariance of general
relativity, it might seem that the natural endpoint of this
progression would include all possible gauge transformations.  This
would be problematic, to say the least, because accounting for the
effects of arbitrary diffeomorphisms on a waveform would be
intractable.  Fortunately, by making certain standard approximations,
we can avoid accounting for the \emph{complete} diffeomorphism
freedom, and restrict to a smaller gauge group.  The end result will
be somewhat larger than the familiar \Poincare group---in fact
infinitely so, at least in principle---yet entirely tractable and far
smaller than the diffeomorphism group.

To see how this is possible, we must first note that near-field
effects in the waveforms (effects appearing at second order in the
distance between the emitter and observer) should be quite small in
data collected in the vicinity of Earth, because even the
leading-order waveform will be hard to detect.  Thus, the model
waveforms only need to capture asymptotic features of the radiation
far from the source.  In particular, we assume that the model
spacetime is asymptotically flat, and calculate the asymptotic
waveform in the limit of future null infinity, $\scriplus$, which is
described below.  Though it is not believed that our universe is
asymptotically flat, this is a useful construction approximating an
isolated source when the intervening curvature is typically small.
The gravitational-wave signal observed by a detector in the vicinity
of Earth will then be very well approximated by the waves along some
geodesic of $\scriplus$---up to a scaling related to distance from the
source.  The benefit of assuming such an asymptotic structure is that
it allows us to impose certain conditions on the gauge, the most
common of which is called Bondi gauge~\cite{Bondi1962, Sachs1962a,
  Newman1962a, Moreschi1987, Deadman2009, Helfer2010}.

Essentially, Bondi gauge consists of a special class of coordinates
that manifest the asymptotic behavior of the spacetime such that the
metric and its derivatives, when expressed in these coordinates,
approach those of Minkowski spacetime at large radii.  The allowed
gauge transformations are symmetry transformations of this metric,
which form a group known as the Bondi-Metzner-Sachs (BMS)
group~\cite{Bondi1962, Sachs1962, Sachs1962a, Newman1966,
  Moreschi1987, Frauendiener2004, Adamo2009}.  This group simply
extends the \Poincare group with generalized translations.  Bondi
coordinates exist in a neighborhood of $\scriplus$ for any
asymptotically flat system~\cite{Moreschi1987}, and any two Bondi
coordinate systems are related by some BMS
transformation~\cite{Sachs1962a}.  This means that the BMS group
encompasses all possible gauge transformations we need to be concerned
with when discussing the limits of an asymptotically flat spacetime.

Bondi gauge also has a particularly nice feature related to the
inertial observers in a neighborhood of $\scriplus$.  At very large
radii, curves of constant spatial coordinates parametrized by the
retarded-time coordinate are nearly timelike geodesics---becoming more
exactly geodesic at larger radii.  Thus, if we extract a quantity on
$\scriplus$ along a simple curve of constant spatial coordinate, we
approximate the signal an inertial observer measures as a function of
proper time (up to the usual amplitude scaling with radius).
Moreover, in the approximately flat asymptotic region, any two
inertial observers are related by an element of the \Poincare group.
But that is a subgroup of the BMS group, so we can use the BMS group
to easily generate all possible signals that might be measured by any
inertial observer.

Taken together, these facts mean that Bondi gauge is not only
sufficiently general to describe any signal observed at great distance
from a source in the asymptotically flat approximation, but is also a
convenient choice that allows us to construct waveforms using simple
curves and BMS transformations.  We will therefore assume that any
waveform is expressed in Bondi gauge, and narrow our focus to the BMS
group.  These concepts are reviewed pedagogically in
Sec.~\ref{sec:BMSGroup}; the eager reader may prefer to skim that
section, and simply refer to Eqs.~\eqref{eq:BMS-transformation} for
the key expressions describing the BMS transformations.

Having understood the BMS group itself, we will then need to
understand its effects on waveforms, which can be separated into two
parts.  First is the effect at a single spacetime event.  The
transformation changes the differential structure in a neighborhood of
that event, and since the gravitational field is fundamentally a
measure of that differential structure, it should come as no surprise
that the waveform will change under a transformation.  To make these
ideas more precise, however, we will need a careful treatment of
asymptotic flatness.  Section~\ref{sec:AsymptoticFlatness} will review
a convenient formalism for developing asymptotic flatness, then use
that formalism to calculate the transformation properties of a
waveform at a point.

Of course, a waveform is not simply measured at a single spacetime
event: a gravitational-wave detector will measure it along (or in the
vicinity of) some worldline, whereas model waveforms are typically
expressed over an extended portion of the (future) celestial sphere of
the source, as a function of time.  In practical terms, this means
expressing the waveform as a function of some coordinate system, and
that coordinate system also changes under a BMS transformation.  So
the second part of a BMS transformation involves rewriting the
waveform as a function of these new coordinates.  This is a fairly
simple bookkeeping exercise in principle, but involves numerous
delicate manipulations and various minor subtleties for practical
implementation, as discussed in Sec.~\ref{sec:impl-bms-transf}.

Portions of the BMS group have been discussed previously in the
context of transforming gravitational waveforms produced by numerical
simulations.  Gualtieri \etal~\cite{Gualtieri2008} considered
rotations and boosts, neglecting quantities of order $v^{2}/c^{2}$.
Kelly and Baker~\cite{Kelly2013} looked at the effect of
supertranslations on ringdown modes, to first order in the
time-derivative of the waveform.  This paper, however, presents an
exact algorithm for the full BMS group.  The practical implementation
of the algorithm is only limited by numerical precision and the
accuracy of interpolation of the input waveform as a function of time.

Section~\ref{sec:underst-effects-bms} will give a brief overview of
the size of these effects, for various types of BMS transformations
and simple waveforms.  Basic analytical arguments will show that the
leading-order coupling due to a supertranslation will be proportional
to the size of the translation and the dominant frequency of the
coupled mode (or generally, the mode's logarithmic derivative); for
boosts the leading-order coupling will be proportional to the speed of
the boost.  In both cases, the constants of proportionality are
typically of order unity, though there are various geometric factors
involved.

In Sec.~\ref{sec:RemovingDriftFromNumericalWaveforms}, mode coupling
will be demonstrated for a full waveform from a numerical simulation
of a binary black-hole system in the public waveform catalog
maintained by the SXS collaboration~\cite{Mroue2013, sxs_catalog}.
This example system is chosen for its seeming symmetry, being
equal-mass and nonprecessing, though with a spin on one black hole
aligned with the orbital angular velocity.  Indeed, simple
coordinate-based measures suggest that the center of mass only strays
from the coordinate origin by about $0.1M$ over the course of the
simulation (where $M$ is the total mass of the system, and is
implicitly multiplied by the appropriate factor of
$G_{\text{N}} / c^{2}$ to provide a unit of distance).  Nonetheless,
we will find that near merger more than \SI{1}{\percent} of each
mode---most notably the dominant $(2,2)$ mode---will mix into other
modes.  In fact, in the raw data the $(3,3)$ and $(3,1)$ modes are
completely dominated by power leaking in from the $(2,2)$ mode.
Overall, up to \SI{30}{\percent} of the amplitude of the third-largest
mode in the data, $(2,1)$, is composed of leakage from the $(2,2)$
mode.  These couplings give rise to curious features in the smaller
modes that are not present in the post-Newtonian model of this system,
for example.  The mode couplings, and resulting curious features, can
be dramatically decreased by measuring the motion of the center of
mass (in simulation coordinates), and applying the opposite
transformation to the waveform.

The example system was chosen so that we will be able to see clearly
that the unexpected features are removed.  More complicated
systems---in particular, precessing systems---will have more
complicated waveforms, but also larger anomalous motion of the center
of mass.  A survey of the entire SXS catalog suggests that the center
of mass in more complicated simulations will drift from the origin by
larger amounts, up to $8M$ for the most extreme system.  This implies
correspondingly larger mode couplings for these systems.  Ossokine
\etal~\cite{Ossokine2015} showed that it is possible to greatly reduce
the size of these displacements by adjusting the initial data,
improving the outlook for future simulations.  Nonetheless, the
current SXS waveform catalog must still be adjusted, and any recoil
that develops during future evolutions will need to be accounted for.
Moreover, boosts and translations only account for six of the
infinitely many degrees of freedom in the BMS group; the general
supertranslations in particular are still uncontrolled, even after
eliminating the drift of the center of mass.

Given the amount of work that needs to be done to account for gauge
effects in numerical waveforms, it is reasonable to wonder how this
affects searches for gravitational waves in detector data.  To
understand this issue we will need to know more about BMS
transformations of waveforms, and so we delay the full discussion
until the end of the paper, Sec.~\ref{sec:effects-data-analys}.  The
upshot is that, while we must take BMS transformations into account
when constructing waveforms, we do not need to search over all
waveforms generated by the BMS group.  For an isolated observer, the
gauge ambiguities reduce to time translation and Lorentz rotations,
which are already known effects.

Open-source code, in the form of a Python module \software{scri}, is
provided in the supplemental materials along with this
paper~\cite{boyle2016supplement}.  It implements the BMS
transformations of the most common gravitational waveforms, including
the Newman-Penrose quantity $\asymptotic{\psi}_{4}$, the Bondi news
function, the shear spin coefficient $\asymptotic{\sigma}$, and the
transverse-traceless metric perturbation $\asymptotic{h}$---as well as
the remaining Newman-Penrose quantities $\asymptotic{\psi}_{0}$
through $\asymptotic{\psi}_{3}$.  Several appendices describe details
about various constructions and calculations from Geometric
Algebra~\cite{Hestenes2002, Doran2010, Hestenes2015} used in this
paper and in the \software{scri} module, including the method of
implementing a boost of the sphere described in
Sec.~\ref{sec:RotationsAndBoosts}.  A final appendix details the crude
method of measuring and removing the center-of-mass drift found in the
numerical data, which is used in
Sec.~\ref{sec:RemovingDriftFromNumericalWaveforms}.

\subsection*{A note on conventions}

More extensive description of the conventions used in this paper are
given in the appendices, but a few basic comments are appropriate
here.  We will assume that all transformations are proper and
orthochronous; both spatial orientation and the direction of time are
preserved.  In general, a strictly improper transformation can be
written as the product of a proper transformation and a parity
operator.  The effects of parity operations on modes of a
spin-weighted spherical-harmonic decomposition are described in
Appendix~B of Ref.~\cite{Boyle2014}.  Similarly, anachronous
transformations can be written as the product of an orthochronous
transformation and the time-reversal operator.  The effect of the
time-reversal operator is to simply negate the time coordinate, and in
some cases to change the sign and labeling of waveform quantities.
Since these can be dealt with separately in ways that are already
understood, we dispense with them entirely, and will not bother to
repeat below that all transformations in this paper are proper and
orthochronous.

Points on the sphere will be labeled interchangeably by the usual
spherical coordinates $(\theta, \phi)$, by the standard stereographic
coordinate $\zeta$, or by the unit spatial vector $\directionvec$
pointing in that direction from the origin when the sphere is
considered as being embedded in Euclidean 3-space and centered on the
origin.  While stereographic coordinates are the preferred
representation throughout much of the literature, and do occasionally
simplify theoretical calculations, they are unsuited for practical
computations because of their infinite range and the nature of the
point at infinity.  Perhaps surprisingly, $(\theta, \phi)$ presents a
more useful parametrization for practical applications.  Despite the
coordinate singularities in its representation of the sphere $S^{2}$,
it actually provides a \emph{non}-singular parametrization of a
portion of the rotation group that covers $S^{2}$, which is more
relevant for dealing with spin-weighted functions.  As a result,
software packages such as \software{spinsfast}~\cite{Huffenberger2010,
  Huffenberger} that implement numerical routines involving
spin-weighted spherical harmonics use the $(\theta, \phi)$
representation.  Consequently, the same representation is used by the
\software{scri} package accompanying this paper.

%%%%%%%%%%%%%%%%%%%%%%%%%%%%%%%%%%%%%%%%%%%%%%%%%%%%%%%%%%%%%%%%%%%%%%
%%%%%%%%%%%%%%%%%%%%%%%%%%%%%%%%%%%%%%%%%%%%%%%%%%%%%%%%%%%%%%%%%%%%%%
\section{The BMS group}
\label{sec:BMSGroup}

We introduce the Bondi-Metzner-Sachs (BMS) group in a simple and
familiar setting.  This will provide a common basis and useful
motivation for the coming sections.  Though more of the formalism of
asymptotically flat spacetime will be needed below, for our purposes
in this section, it will be sufficient to consider the standard
compactification of Minkowski space.  In particular, this
compactification will provide all the understanding needed for more
general asymptotically flat spacetimes.  We will use null rays to
relate coordinates of timelike geodesics at finite radius to
coordinates at future null infinity.  By relating timelike geodesics
to each other, we will then be able to describe the effect of a BMS
transformation on the coordinates of future null infinity.

%%%%%%%%%%%%%%%%%%%%%%%%%%%%%%%%%%%%%%%%%%%%%%%%%%%%%%%%%%%%%%%%%%%%%%
\begin{figure}
  \includegraphics{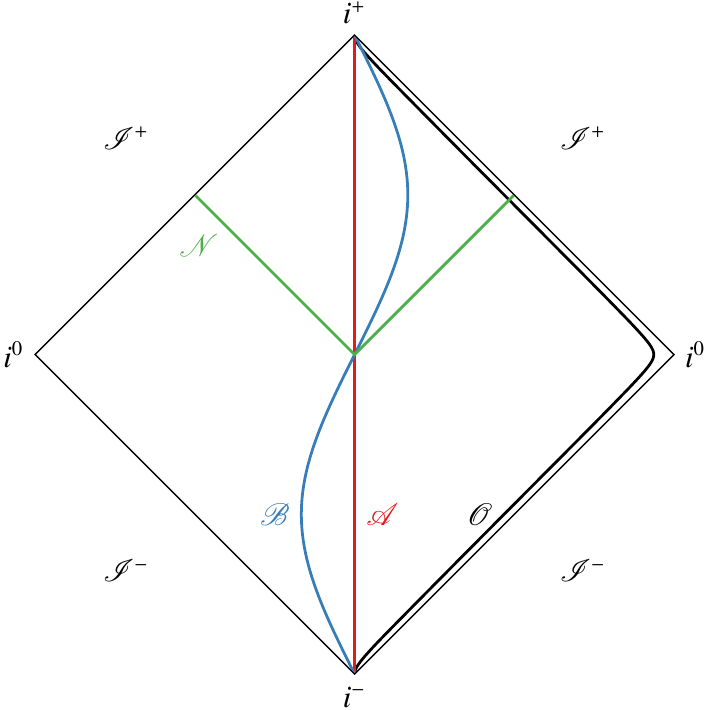}
  \caption{ \label{fig:BoostDiagram} %
    \CapName{Extending local coordinates to $\scriplus$} This
    conformal diagram shows the worldlines of a pair of inertial
    emitters, with emitter $\B$ moving at speed $0.5c$ relative to
    emitter $\A$, and a distant observer $\O$ stationary with respect
    to $\A$.  The origins of the emitters' coordinate systems coincide
    at $\observertime_{\A}=\observertime_{\B}=0$.  We construct the
    null cone $\N$ emanating from that event, which allows us to
    extend coordinates to $\scriplus$.  The intersection of $\N$ with
    $\scriplus$ is a sphere; all points on that sphere are assigned
    time coordinate $\scritime_{\A}=0$ by emitter $\A$ and
    $\scritime_{\B}=0$ by emitter $\B$.  Each point is also labeled by
    the direction of the null generator extending from the given
    emitter to that point.  Note that a rotation obviously does not
    affect the set of points comprising $\N$, though the labeling of
    points will change---except for the points along the axis of
    rotation.  Similarly, a boost leaves the null cone invariant, but
    will change the labeling of any point not along the boost velocity
    vector, as discussed in Sec.~\ref{sec:RotationsAndBoosts}.  The
    observer $\O$ can also be assigned coordinates based on the null
    rays emitted by $\A$ and, in the limit of very large separation,
    any field it observes will approach the field observed on
    $\scriplus$, up to a scaling based on radius.  This is the basic
    motivation for using asymptotically flat spacetimes to model
    radiation.  %
  }
\end{figure}
%%%%%%%%%%%%%%%%%%%%%%%%%%%%%%%%%%%%%%%%%%%%%%%%%%%%%%%%%%%%%%%%%%%%%%

The conformal diagram displayed in Fig.~\ref{fig:BoostDiagram}
provides the standard picture~\cite{Penrose1965a, Hawking1976,
  Frauendiener2004}.  Here, $i^{-}$ and $i^{+}$ are past and future
timelike infinity; $i^{0}$ is spacelike infinity; $\scriminus$ is past
null infinity; and $\scriplus$ is future null infinity.  We will be
concerned almost exclusively with $\scriplus$, as that is the
asymptotic limit of outgoing gravitational radiation.  Also shown in
the diagram is a pair of emitters, $\A$ and $\B$, traveling along
timelike geodesics.  We can extend coordinates defined in a
neighborhood of an inertial emitter to coordinates throughout the
spacetime and to $\scriplus$ using null generators.  For example,
suppose $\A$ emits a null ray at proper time $\observertime_{\A}=0$ in
a direction given in local coordinates by the angular coordinates
$(\theta, \phi)$, or equivalently the stereographic coordinate
$\zeta$.  Any point at finite distance along that ray can be assigned
coordinates $(\scritime, \distance, \zeta)$, where
$\scritime=\observertime_{\A}$ is the retarded time and $\distance$ is
an affine parameter along the geodesic---in Minkowski spacetime, we
can think of this as the distance between the emitter and that point
as measured in the frame of the emitter.  The future limit of the null
ray will represent a unique point on $\scriplus$; we typically assign
that point the coordinates $(\scritime, \zeta)$, dropping
$\distance$ because it will, of course, be infinite.  Continuing in
this way for all directions, emitter $\A$ can provide
coordinates for the entire null cone $\N$ and, in particular, the
sphere $\anticelestialsphere$ given by the intersection of $\N$ with
$\scriplus$.

Of course, coordinates can equivalently be constructed in the same way
by emitter $\B$.  The set of points $\anticelestialsphere$ will
naturally be the same in both cases; any relative rotation of the two
emitters will simply take one null ray into another, and a boost
leaves null rays invariant.  But the coordinates labeling each null
ray---hence the coordinates labeling each point on
$\anticelestialsphere$---will be different for the two systems
whenever $\A$ and $\B$ are related by any Lorentz transformation.  We
discuss the effect of these transformations in
Sec.~\ref{sec:RotationsAndBoosts}.

But the Lorentz transformations only relate a subset of possible
emitters, and hence a subset of possible coordinate systems on
$\scriplus$.  In Minkowski spacetime, we are familiar with
translations as the remaining freedom relating coordinate systems.
However, a translation at finite radius has somewhat surprising
effects on the coordinates at $\scriplus$.  This is explained more
fully in Sec.~\ref{sec:Translations}.  The conclusion will be that a
translation is equivalent to an offset of the retarded time
$\scritime$ that depends on direction, though in a simple way.

The surprising result of early studies~\cite{Bondi1962, Sachs1962a}
was that the familiar Lorentz and translation groups are not
sufficient for describing all of the asymptotic symmetries of
asymptotically flat spacetimes.  It turns out that a focus on null
cones in Minkowski spacetime is too restrictive.  Since we are only
prescribing the asymptotic behavior, we can disregard all but a
neighborhood of $\scriplus$---which means that our null ``cones'' need
no longer look like cones, in the sense that the generators need not
meet at a point.  In general, then, the simple angular dependence of
the offset of the retarded time $\scritime$ induced by translations
must be generalized to an arbitrary (smooth) function of the angles.
These transformations are referred to as \textit{supertranslations}.
Together with the Lorentz group,\footnote{To be precise, the
  supertranslations form a normal subgroup $\supertranslations$ of the
  BMS group, and the factor group of the BMS group by
  $\supertranslations$ is precisely the Lorentz group $\SOp{3,1}$.
  However, the Lorentz group is not a \emph{normal} subgroup.  Thus,
  the BMS group is the semidirect product
  $\supertranslations \rtimes \SOp{3,1}$.} these form the complete BMS
group, as discussed further in Sec.~\ref{sec:complete-bms-group}.

\subsection{Rotations and boosts}
\label{sec:RotationsAndBoosts}

%%%%%%%%%%%%%%%%%%%%%%%%%%%%%%%%%%%%%%%%%%%%%%%%%%%%%%%%%%%%%%%%%%%%%%
\begin{figure*}
  \includegraphics{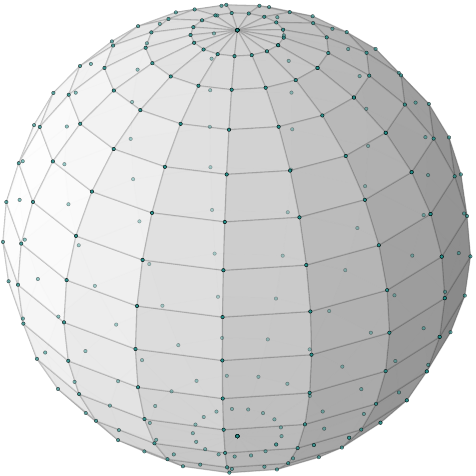} \hfil
  \includegraphics{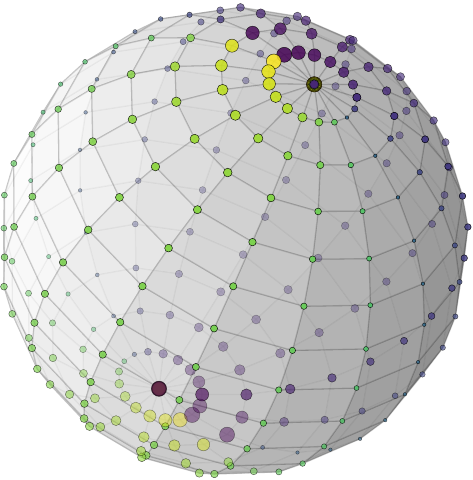} \hfil
  \includegraphics{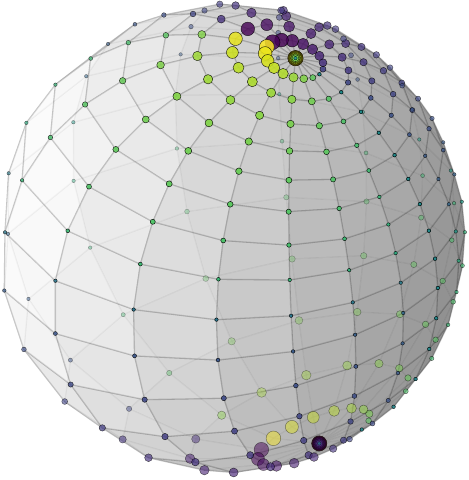} \hfil
  \hspace{20pt}
  % \raisebox{2pt}{\includegraphics[height=135pt]{SphereGrid_Scaler}}
  \raisebox{2pt}{\includegraphics[height=135pt]{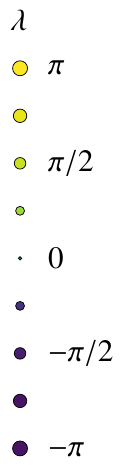}}
  \caption{ \label{fig:BoostedGrids} %
    \CapName{Transformation of a grid under Lorentz rotations} To
    decompose a spin-weighted field into spin-weighted spherical
    harmonics, the values of the field are needed on the
    colatitude-longitude grid of frame $\B$ seen on the left-hand
    side.  If the field is known in frame $\A$, and $\B$ is simply
    rotated relative to $\A$, then the appropriate values can be found
    by evaluating the field in $\A$ on the grid shown in the center.
    The points on the two grids represent the same physical points.
    Similarly if $\B$ is boosted relative to $\A$ (in this example
    with velocity $0.5c$ to the right and out of the page), the values
    can be found by evaluating the field in $\A$ on the grid shown on
    the right-hand side.  However, for spin-weighted fields, the
    locations of the points alone are not sufficient; we also need to
    know the relative alignment angle between the tangent basis
    constructed by $\A$ and the tangent basis constructed by $\B$ at
    each point.  This angle is the spin phase $\SpinPhase$ described
    in the text, represented here by the size and color of the marker
    at each grid point.  The transformed grid positions and
    $\SpinPhase$ values are calculated using
    Eq.~\eqref{eq:boost-rotor}.  %
  }
\end{figure*}
%%%%%%%%%%%%%%%%%%%%%%%%%%%%%%%%%%%%%%%%%%%%%%%%%%%%%%%%%%%%%%%%%%%%%%

We first confine ourselves to a single null cone, and the
corresponding sphere $\anticelestialsphere$ on $\scriplus$, by
considering emitters with coordinate systems having identical origins
but which are related by elements of the Lorentz subgroup: rotations
and boosts.  The situation is depicted in Fig.~\ref{fig:BoostDiagram},
where two emitters give off null rays at the same spacetime event.  It
is, of course, one of the fundamental conclusions of basic special
relativity that boosts take null rays to null rays.  Thus, the
collection of all null rays (the null cone) originating at a
particular spacetime event will be invariant under boosts.  However,
the coordinates assigned to the direction of a given null ray within
that collection will change under a boost.  In the same way, a
rotation maps the null cone onto itself, while simply changing the
coordinates of individual rays.  Our objective in this section, then,
is to find how directions in one coordinate system map to directions
in another system under rotations and boosts.

We begin with the simpler case, in which the coordinate systems of
$\A$ and $\B$ are simply related by some known rotation, with no
relative boost.  For a scalar field, if we suppose that the field is
known in frame $\A$, we can find the value of the field at any point
$\directionvec_{\B}$ in frame $\B$ by simply rotating that point back
to $\directionvec_{\A}$ in $\A$ and evaluating the field there.
However, for spin-weighted fields, there is an additional
complication.  A spin-weighted field at a point is defined with
respect to the basis of the tangent space to the sphere at that
point---usually represented by a complex tangent vector $\fourvec{m}$.
But if $\A$ and $\B$ are rotated relative to one another, the tangent
vector $\fourvec{m}_{\A}$ at $\directionvec_{\A}$ will also be rotated
relative to the tangent vector $\fourvec{m}_{\B}$ at
$\directionvec_{\B}$ by some angle $\SpinPhase$, referred to as the
``spin phase''.  [This factor $\SpinPhase$ is defined more precisely
in Appendix~\ref{sec:rotor-boost}.]  The situation is depicted in
Fig.~\ref{fig:BoostedGrids}, where a standard grid is shown in frame
$\B$ on the left-hand side, and a grid representing the same physical
points in the coordinates of $\A$ is shown in the center, along with
the spin phase.

A simple way of dealing with the complication of the spin phase is
described in Appendix~B of Ref.~\cite{Boyle2014}.  The essential idea
is to evaluate spin-weighted fields directly in terms of a rotation
operator.  Thus, if the field value is needed at $(\theta, \phi)$ in
$\B$, this is represented by a rotation operator
$\rotor{R}_{\theta, \phi}$, described in
Appendix~\ref{sec:spherical-coordinates} [of this paper].  Now, if
frame $\B$ is obtained from frame $\A$ by a frame-rotation
$\FrameRotor$, then the value of the field can be found by evaluating
the field in $\A$ at $\FrameRotor\, \rotor{R}_{\theta, \phi}$.  The
spin phase is automatically accounted for.

Similarly, we can find the value of the field in $\B$ if it is related
to $\A$ by a pure boost.  We assume that $\B$ moves with respect to
$\A$ with three-velocity $\fourvec{v}$, and use the conventions
established in the appendices to directly compare components in the
two frames.  Suppose that $\A$ measures an angle $\Theta_{\A}$ between
$\fourvec{v}$ and the spatial component, $\directionvec$ of some null
direction.  That is, we have
$\cos \Theta_{\A} = \fourvec{v} \cdot \directionvec /
\abs{\fourvec{v}} \abs{\directionvec}$.
Similarly $\B$ measures an angle $\Theta_{\B}$ between $\fourvec{v}$
and the spatial component of that same null direction.  Note that the
spatial subspaces will, of course, generally be different for the two
frames, except along the axis containing $\fourvec{v}$.  Nonetheless,
we can relate the angles measured in the two frames, as shown in
Appendix~\ref{sec:rotor-boost}, by the formula
\begin{equation}
  \label{eq:boost-angle-relationship}
  \tan \frac{\Theta_{\B}}{2} = \e^{\rapidity} \tan \frac{\Theta_{\A}}{2},
\end{equation}
where $\rapidity = \artanh \abs{\fourvec{v}}$ is the usual rapidity
parameter.

We can use this equation to transform a physical scalar field measured
in one frame into the other frame.  Suppose that this physical field
is known on $\anticelestialsphere$ as a function of the null direction
measured by $\A$, and we wish to know the value of the field in some
null direction $\directionvec_{\B}$ as measured by $\B$, noting that
the angle between $\fourvec{v}$ and $\directionvec_{\B}$ is
$\Theta_{\B}$.  We first take the direction $\directionvec_{\B}'$ in
the frame of $\A$ having the same components with respect to the basis
of $\A$ as $\directionvec_{\B}$ has with respect to the basis of $\B$,
even though this is a different frame.  We then rotate this vector in
the $\fourvec{v}$-$\directionvec_{\B}'$ plane until we arrive at a new
vector $\directionvec_{\A}$ that makes an angle with $\fourvec{v}$ of
$\Theta_{\A}$, satisfying Eq.~\eqref{eq:boost-angle-relationship}.
The physical field measured at $\scriplus$ by $\A$ in this direction
is the same as the physical field as measured by $\B$, and is thus the
result we sought.

Again, there are complications involved with spin-weighted fields.
However, as shown in Appendix~\ref{sec:rotor-boost}, these
complications are automatically dealt with when using the
rotation-operator approach described above.  The scenario is
illustrated in Fig.~\ref{fig:BoostedGrids}, where the grid in $\B$ is
shown on the left-hand side, and the same grid of physical points is
shown in the coordinate system of $\A$ on the right-hand side.  The
basic idea is the same: the grid points are simply moved around the
sphere and associated with some spin phase $\SpinPhase$.  Of course,
in this case, the points are moved in different ways, and $\SpinPhase$
is a different function of position.  Nonetheless, it is still
beneficial to evaluate the field in $\A$ directly in terms of the
rotation operator.  Here, however, rather than the constant
frame-rotation operator $\FrameRotor$ accounting for the difference
between frames $\A$ and $\B$, we need to use a position-dependent
rotation operator $\BoostRotor$.  Using the notation of quaternions,
we can write this operator as
\begin{equation}
  \label{eq:boost-rotor}
  \BoostRotor(\theta, \phi) = \exp \left[ \frac{\Theta_{\B} -
      \Theta_{\A}} {2} \frac{ \directionvec_{\B}' \times \fourvec{v} }
    {\abs{ \directionvec_{\B}' \times \fourvec{v} }} \right],
\end{equation}
where $\directionvec_{\B}'$ is the unit spatial vector in the
$(\theta, \phi)$ direction of frame $\A$, $\Theta_{\B}$ is the angle
between that vector and $\fourvec{v}$, and $\Theta_{\A}$ is related to
it by Eq.~\eqref{eq:boost-angle-relationship}.  This operator
represents a rotation through $\Theta_{\B} - \Theta_{\A}$ about the
$\directionvec_{\B}' \times \fourvec{v}$ axis.  In this case, the
field can be evaluated in $\A$ from
$\BoostRotor\, \rotor{R}_{\theta, \phi}$.  More generally, an
arbitrary element of the Lorentz group can be written as the product
of a frame rotation and a boost, $\BoostRotor\, \FrameRotor$, then the
field can be evaluated from
$\BoostRotor\, \FrameRotor\, \rotor{R}_{\theta, \phi}$.

It is worth exhibiting the effect of a Lorentz transformation in terms
of stereographic coordinates.  Though ill suited to numerical
computations, the stereographic formalism is common throughout the
literature, and there is a certain effectiveness that comes with
familiarity.  In particular, if the sphere $\anticelestialsphere$ is
parametrized by the stereographic coordinate $\zeta_{\A}$ in frame
$\A$ and by $\zeta_{\B}$ in frame $\B$, then under a general Lorentz
transformation the two are related by\footnote{The stereographic
  coordinates are usually thought of as elements of the complex plane
  augmented by adjoining the point at infinity, also known as the
  Riemann sphere.  In this form, the transformation shown here is
  usually known as a M\"obius transformation---an element of the
  M\"obius group, which is isomorphic to the group of conformal
  transformations of the sphere, the projective special linear group
  $\PSL{2, \mathbb{C}}$, and the proper orthochronous Lorentz group
  $\SOp{3,1} \cong \SOp{1,3}$.}
\begin{equation}
  \label{eq:stereographic_lorentz}
  \zeta_{\B} = \frac{a\, \zeta_{\A} + b} {c\, \zeta_{\A} + d},
\end{equation}
where $(a, b, c, d)$ is a collection of complex coefficients
satisfying $a\,d - b\,c = 1$.  Because of its compactness and
familiarity, the representation in stereographic coordinates is useful
for descriptions, and occasionally for deriving results.  We will
encounter this formalism again in Sec.~\ref{sec:AsymptoticFlatness},
though the stereographic coordinates themselves will not appear in the
final results.  For all other purposes, quaternions and related
formalism (discussed in more detail in the appendices) will be used
because of their computational and formal superiority.

Finally, we also note that this transformation of a spin-weighted
field under Lorentz transformations is only part of the story.  More
generally, different fields will mix with each other because
$\fourvec{m}$ remains neither tangent to the sphere nor even purely
spatial under a Lorentz transformation.  For example, to calculate the
transformation of the Newman-Penrose quantity $\asymptotic{\psi}_{3}$
on $\scriplus$ we will also need a contribution from
$\asymptotic{\psi}_{4}$.  And that simple behavior is only a result of
the peeling theorem; at finite radii all Newman-Penrose quantities
could mix with each other under a Lorentz transformation.  This will
be discussed further in Sec.~\ref{sec:transformations}.  Throughout
the remainder of this section, however, we will be able to focus
solely on the movement of points at which a field is evaluated.

\subsection{Translations and supertranslations}
\label{sec:Translations}

%%%%%%%%%%%%%%%%%%%%%%%%%%%%%%%%%%%%%%%%%%%%%%%%%%%%%%%%%%%%%%%%%%%%%%
\begin{figure}
  \includegraphics{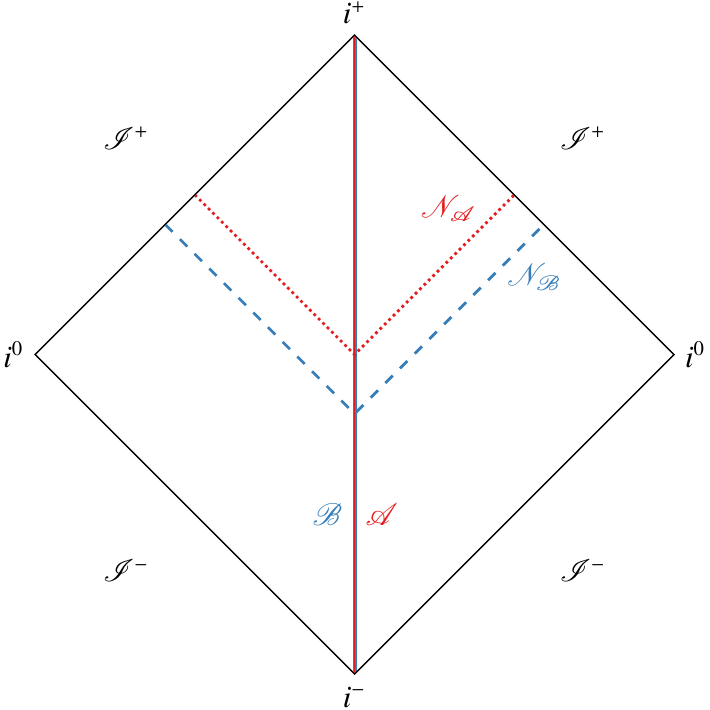}
  \caption{ \label{fig:TimeTranslation} %
    \CapName{Effect of time translation on coordinates of $\scriplus$}
    Here, we see two different local coordinate systems extended to
    $\scriplus$: $\A$ and $\B$ represent the same emitter with the
    same spatial coordinates, but different origins for the time
    coordinate.  The two null cones correspond to the two origins of
    the time coordinate.  We see that the time translation
    $\observertime_{\B} = \observertime_{\A} - \delta \observertime$
    affects the time coordinates on $\scriplus$ isotropically---in
    fact, the transformation of the retarded-time coordinate is simply
    $\scritime_{\B} = \scritime_{\A} - \delta \observertime$.  %
  }
\end{figure}
%%%%%%%%%%%%%%%%%%%%%%%%%%%%%%%%%%%%%%%%%%%%%%%%%%%%%%%%%%%%%%%%%%%%%%

Now, having understood the transformations that preserve the light
cone, we can move on to more general transformations---though still
considering only inertial emitters in Minkowski space.  In particular,
we have translations of both time and space.  Generalizing these, we
will be led to the encompassing notion of supertranslations.

It is instructive to begin with the simple case of time translations.
As noted in the introduction to this section, every point on a null
cone originating at emitter $\A$, at a proper time of
$\observertime_{\A}$, is assigned the same retarded time
$\scritime_{\A}=\observertime_{\A}$; similarly
$\scritime_{\B}=\observertime_{\B}$.  Now, if the emitters' time
scales are related by a simple time translation such that
$\observertime_{\B} = \observertime_{\A} - \delta t$, we clearly have
the simple relation between retarded-time coordinates
$\scritime_{\B} = \scritime_{\A} - \delta t$.  This is depicted in
Fig.~\ref{fig:TimeTranslation}.  The notable feature of this
transformation is that it is isotropic; the change in the
retarded-time coordinate does not depend on the direction.  This
seemingly trivial observation is important because it is not true of
space translations, and generalizing this notion will be key to
understanding the broader class of supertranslations.

\begin{figure}
  \includegraphics{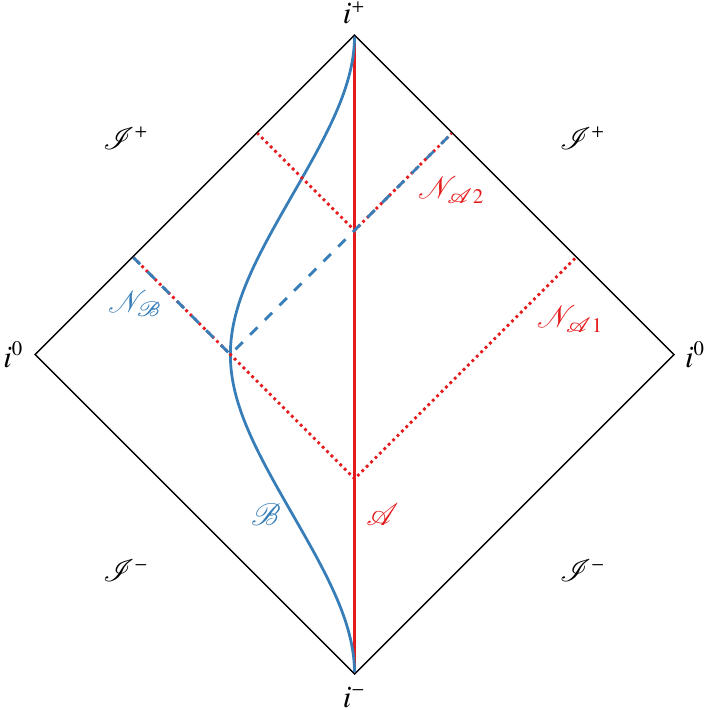}
  \caption{ \label{fig:SpaceTranslation} %
    \CapName{Effect of space translation on coordinates of
      $\scriplus$} Here, $\A$ and $\B$ represent emitters displaced
    relative to each other.  A single null cone emanates from $\B$ and
    intersects $\scriplus$ in two points.  The same points of
    $\scriplus$ are found on two separate null cones emitted by $\A$.
    Thus, a space translation has a non-isotropic effect on the
    retarded time coordinates of $\scriplus$.  More generally,
    allowing for all three spatial dimensions, the effect of a
    translation $\delta \fourvec{x}$ will transform the retarded-time
    coordinate in a direction $\directionvec$ as
    $\scritime_{\B} = \scritime_{\A} + \delta \fourvec{x} \cdot
    \directionvec$.  %
  }
\end{figure}

We can now consider space translations as depicted in
Fig.~\ref{fig:SpaceTranslation}.  Emitter $\B$ is simply displaced
from $\A$ by a spatial vector $\delta \fourvec{x}$ but the two are
stationary with respect to each other.  The null cone $\N_{\B}$
emanates from the origin of $\B$, $\scritime_{\B} = 0$, and intersects
$\scriplus$ at two points on this diagram.  Those same points of
$\scriplus$ are on null rays from two separate null cones of
$\A$---one in the $-\fourvec{x}$ direction with retarded time
$\scritime_{\A1} = - \abs{\delta \fourvec{x}}$, the other in the
$\fourvec{x}$ direction emitted at
$\scritime_{\A2} = \abs{\delta \fourvec{x}}$.  Of course, these two
points correspond to the two points of the sphere $S^{0}$ because the
three spatial dimensions have been collapsed to one in this simple
diagram.  More generally, for any point on the sphere $S^{2}$ the
relationship between the retarded time coordinates is
$\scritime_{\B} = \scritime_{\A} + \delta \fourvec{x} \cdot
\directionvec$.

%%%%%%%%%%%%%%%%%%%%%%%%%%%%%%%%%%%%%%%%%%%%%%%%%%%%%%%%%%%%%%%%%%%%%%
%%%%%%%%%%%%%%%%%%%%%%%%%%%%%%%%%%%%%%%%%%%%%%%%%%%%%%%%%%%%%%%%%%%%%%

We can combine these two transformation laws into a single law for
general spacetime translations:
\begin{equation}
  \label{eq:spacetime_translation}
  \scritime_{\B} = \scritime_{\A} - \sum_{\ell\in\{0,1\}}
  \sum_{m=-\ell}^{\ell} \st^{\ell,m}\, Y_{\ell,m}(\theta, \phi),
\end{equation}
where
\begin{subequations}
  \begin{align}
    \st^{0, 0} &= \sqrt{4\pi}\, \delta t, \\
    \st^{1, -1} &= -\sqrt{ \frac{2\pi} {3} }\, (\delta x + \i\,
                     \delta y), \\
    \st^{1, 0} &= -\sqrt{ \frac{4\pi} {3} }\, \delta z, \\
    \st^{1, 1} &= -\sqrt{ \frac{2\pi} {3} }\, (-\delta x + \i\,
                     \delta y),
  \end{align}
\end{subequations}
using $\delta \fourvec{x} = (\delta x, \delta y, \delta z)$.  Note
that the sum over $\ell$ is restricted to $\{0, 1\}$ here.  This
suggests the final generalization we need to arrive at the BMS group:
expanding the range of the sum over $\ell$ to all positive integers,
while retaining the condition that
$\st^{\ell,m} = (-1)^{m} \bar{\st}^{\ell, -m}$ to ensure that the
retarded-time coordinate remains real.  More precisely, we construct a
transformation of the coordinates such that
\begin{equation}
  \label{eq:supertranslation}
  \scritime' = \scritime - \st,
\end{equation}
where $\st$ is any real-valued function on the sphere.  To simplify
later analyses, we can also add the conditions that $\st$ be
square-integrable and twice-differentiable.  This
transformation---which encompasses spacetime translations---is
referred to as a \textit{supertranslation}.  It can be shown that
supertranslations are asymptotic symmetries of asymptotically flat
spacetimes~\cite{Bondi1962, Sachs1962a}, and thus are indeed members
of the BMS group.

One way of thinking about supertranslations is to imagine a network of
observers located on a sphere surrounding the source.  Ideally, we
could combine the signals detected by these observers, but to do so we
would need some idea of how their time coordinates compared to each
other; we would need to have some synchronization between their
clocks.  But if we now move the network to $\scriplus$, such a
synchronization becomes impossible.  We could supply a separate time
offset to each observer without changing the physics.  Roughly
speaking, a supertranslation is just the limit of this
direction-dependent time translation where there is a different
observer in every possible direction.

Supertranslations present an interesting departure from the other,
more basic, types of transformations constituting the familiar
\Poincare group.  If $\scritime$ is constructed as given above by
light cones emitting from an inertial world line $\A$, then we know
(by construction) that the null rays generating a surface of constant
$\scritime$ meet in a common point---the vertex of the null cone.  On
the other hand, if the function $\st$ has any $\ell>1$ components, the
null rays generating a surface of constant $\scritime'$, as given by
Eq.~\eqref{eq:supertranslation}, \emph{do not} meet in a common point.
This is why the notation changed in Eq.~\eqref{eq:supertranslation},
dropping the subscripts denoting the emitter, because in general we do
not require the retarded time to be constructed by an emitter.

\begin{figure}
  % \tikzset{external/remake next=true}
  \includegraphics{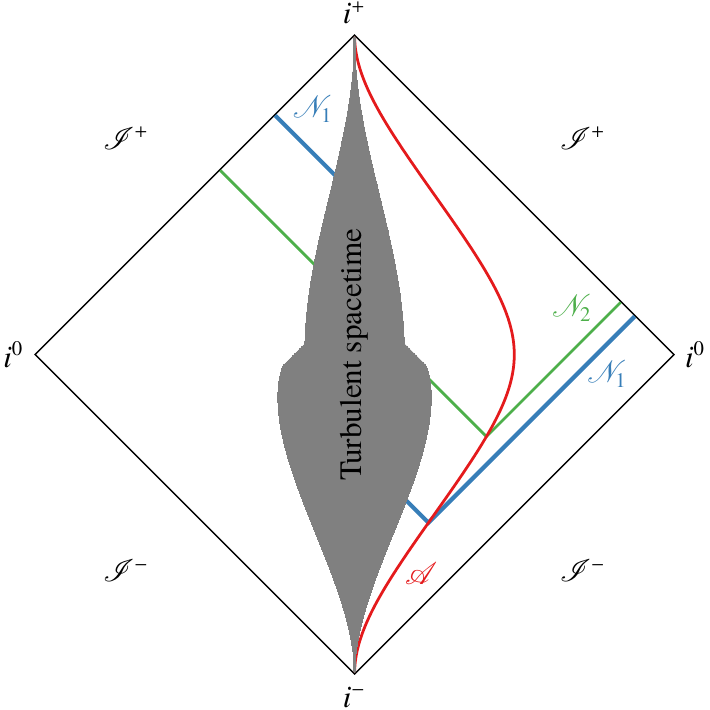}
  \caption{ \label{fig:SupertranslationAmbiguity} %
    \CapName{Null rays in complicated spacetimes} When the interior of
    the spacetime is not Minkowski, we cannot expect to construct
    retarded-time coordinates globally based on null cones.  The
    interior of this diagram is a rough cartoon, in which the shaded
    region represents the space between two merging black holes.  Some
    of the null generators from any emitter in this spacetime must
    pass through this region, and may be affected in erratic ways.
    Given null rays near $\scriplus$, we cannot say whether or not
    they originated at the same point.  Clearly, then, it is too much
    to demand that general asymptotically flat spacetimes must have
    their coordinates given by a construction like the one given for
    Minkowski space.  Instead, we simply place requirements on the
    compactified spacetime in a neighborhood of $\scriplus$. %
    Conversely, it is too much to ask that a ``nice'' coordinate
    system on $\scriplus$ correspond to null cones that meet at one
    spacetime event in general.  This motivates our intuitive
    acceptance of supertranslations. %
  }
\end{figure}

As another, possibly more enlightening, consideration of this peculiar
nature of supertranslations, we can imagine light cones originating at
an emitter in an asymptotically flat spacetime containing some
nontrivial geometry.  In the example shown in
Fig.~\ref{fig:SupertranslationAmbiguity}, we see a simple cartoon of a
merging binary.  The emitter $\A$ gives off two null cones, $\N_{1}$
followed by $\N_{2}$.  The rays given off to the right intersect
$\scriplus$ as we would expect, $\N_{1}$ followed by $\N_{2}$.  The
rays given off to the left, however, behave more erratically.  Here,
the first null ray interacts strongly with the black holes and is
delayed, arriving at $\scriplus$ \emph{after} the null ray that was
emitted later.  Obviously, coordinates constructed from null cones of
$\A$ will be ``bad'' coordinates, with singularities resulting from
caustics of the null rays.

%%%%%%%%%%%%%%%%%%%%%%%%%%%%%%%%%%%%%%%%%%%%%%%%%%%%%%%%%%%%%%%%%%%%%%
%%%%%%%%%%%%%%%%%%%%%%%%%%%%%%%%%%%%%%%%%%%%%%%%%%%%%%%%%%%%%%%%%%%%%%

So for general asymptotically flat spacetimes, it is simply a bad idea
to expect that the retarded time coordinate should be constructible
from null rays emitted from a timelike worldline.  Instead, we should
only expect to have ``good'' or ``nice'' coordinates in a neighborhood
of $\scriplus$.  In fact, the motivation for the original paper by
Newman and Penrose that introduced the $\eth$
operator~\cite{Newman1966} was to impose a condition on $\scritime$ in
a neighborhood of $\scriplus$ to fix the $\ell>1$ supertranslation
freedom.  This is also (at least partially) the motivation for the
``good cut'' construction~\cite{Adamo2009}, the ``nice section''
construction~\cite{Dain2000}, and the ``regularized null cone cut''
construction~\cite{Kozameh2013A}.

\subsection{The complete BMS group}
\label{sec:complete-bms-group}

One final element is needed to complete the construction of the BMS
group.  In Sec.~\ref{sec:RotationsAndBoosts}, we assumed that the
origins of the two emitters coincided, but only looked at the effect
of a boost on the null cone emitted at that common origin.  Obviously,
at later times, the null cones by which those emitters extend their
local coordinates to $\scriplus$ will not originate at the same event;
there will also be some translation involved.  Thus, we expect that
the simple formula from the previous section
$\scritime' = \scritime - \st$ must be modified in some way by the
boost.

\begin{figure}
  % \tikzset{external/force remake}
  \includegraphics{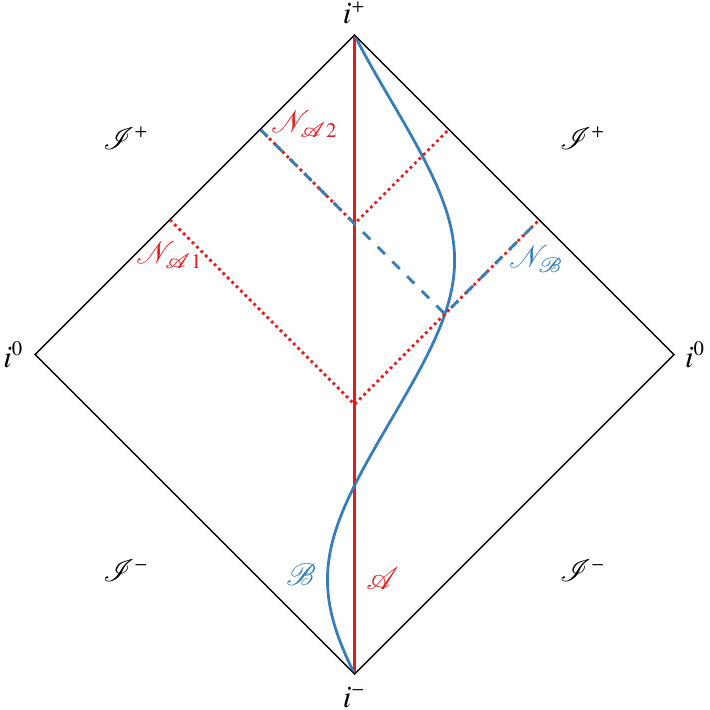}
  % \tikzset{external/force remake=false}
  \caption{ \label{fig:BoostShift} %
    \CapName{Effect of boost on coordinates of $\scriplus$ with
      $\scritime \neq 0$} Here, $\A$ and $\B$ represent emitters
    related by a simple boost.  Though their origins coincide, at some
    later time $\observertime_{\B}$ emitter $\B$ constructs the null
    cone $\N_{\B}$.  It is not hard to solve for the time
    $\observertime_{\A1} = \gamma(1-v) \observertime_{\B}$ at which
    emitter $\A$ must construct a null cone to overlap with the
    right-going null ray of $\N_{\B}$.  We can similarly solve for the
    time $\observertime_{\A2} = \gamma(1+v) \observertime_{\B}$ at
    which emitter $\A$ must construct a null cone to overlap with the
    left-going null ray of $\N_{\B}$.  %
  }
\end{figure}

The situation is easily described pictorially, as in
Fig.~\ref{fig:BoostShift}.
%%%%%%%%%%%%%%%%%%%%%%%%%%%%%%%%%%%%%%%%%%%%%%%%%%%%%%%%%%%%%%%%%%%%%%
%%%%%%%%%%%%%%%%%%%%%%%%%%%%%%%%%%%%%%%%%%%%%%%%%%%%%%%%%%%%%%%%%%%%%%
The origins of proper time for the observers coincide where their
paths cross.  At some later time $\observertime_{\B}$, emitter $\B$
constructs a null cone $\N_{\B}$.  A simple exercise in special
relativity shows that $\A$ must emit a null ray in direction
$\directionvec$ at time
$\observertime_{\A} = \gamma\, (1 - \fourvec{v} \cdot \directionvec)\,
\observertime_{\B}$
in order to reach the same point of $\scriplus$ as the null ray
emitted in that direction by $\B$.  Because we will encounter this
factor frequently, we define
\begin{equation}
  \label{eq:conformal-factor-dot-product}
  \asymptotic{k}
  \defined \frac{1} {\gamma\, (1 - \fourvec{v} \cdot \directionvec)}.
\end{equation}
As shown in Appendix~\ref{sec:conformal-factor}, this $\asymptotic{k}$
factor is also the conformal factor of a boost appropriate to the
spherical metric.  In this case, where the spatial origins coincide at
$\observertime_{\A} = \observertime_{\B} = 0$, we have the
transformation law $\scritime' = \asymptotic{k}\, \scritime$.

Finally, we can combine this with the supertranslation of
Eq.~\eqref{eq:supertranslation} and the angular effects of the Lorentz
transformation given by Eq.~\eqref{eq:stereographic_lorentz} to
find the general BMS transformation of coordinates on $\scriplus$,
representing an initial supertranslation, followed by a Lorentz
transformation:
\begin{subequations}
  \label{eq:BMS-transformation}
  \begin{align}
    \label{eq:BMS-transformation-time}
    \scritime'
    &= \asymptotic{k} \left( \scritime - \st \right) \\
    \label{eq:BMS-transformation-space}
    \asymptotic{\zeta}'
    &= \frac{a\, \asymptotic{\zeta} + b} {c\, \asymptotic{\zeta} + d}.
  \end{align}
\end{subequations}
Again, $(a, b, c, d)$ is a collection of complex coefficients
satisfying $a\,d - b\,c = 1$, representing the Lorentz transformation,
and $\st$ is an arbitrary real-valued square-integrable and
twice-differentiable function on the sphere.  The implementation of
these transformations to be described below will use
Eqs.~\eqref{eq:boost-angle-relationship} and~\eqref{eq:boost-rotor} to
represent Lorentz transformations rather than the stereographic
coordinates shown here.

It is also important to note that the transformation is constant; $a$,
$b$, $c$, $d$, $\asymptotic{k}$, \and $\st$ are all independent of
time.  This may seem to give us a static transformation---though we
know that a boost should, in some sense, result in a time-dependent
translation.  To simplify matters, we assume no rotation and $\st=0$,
leaving only a boost.  The transformation law for time in this case
might be rewritten as
$\gamma\, \scritime' = \scritime + \gamma\, \fourvec{v} \cdot
\directionvec\, \scritime'$.
We can interpret this as a rescaling of the time coordinate, in
agreement with the standard time dilation, along with a translation by
$\gamma\, \scritime'\, \fourvec{v}$, much as we might expect.
Interestingly, it is awkward to express this translation as being
proportional to $\scritime$ [\eg, by expanding the factor of
$\asymptotic{k}$ as spherical harmonics in
Eq.~\eqref{eq:BMS-transformation-time}], because this would imply that
a boost gives rise to a time-dependent \emph{super}translation.  This
suggests a minor subtlety of nomenclature when defining the
supertranslation, due to noncommutativity of the boost and
supertranslation.

In this section, we have built up the BMS group through heuristic
arguments in order to come to an intuitive and pedagogical
understanding of how coordinates change under a BMS
transformation---though of course, the same result is also obtained
through more rigorous methods~\cite{Bondi1962, Sachs1962, Sachs1962a}.
In particular, Sec. IV~C of Ref.~\cite{Sachs1962} describes the
associated Lie algebra $\bms$.  In short, the rotations and boosts
correspond to the standard generators of infinitesimal (Lorentz)
rotations in a plane of Minkowski space, while the generators of
supertranslations are given by the basis
$Y_{\ell,m}\, \partial / \partial \asymptotic{\scritime}$.  We will
not find this infinitesimal presentation directly useful, however,
because it is not easily applicable to finite transformations.
Moreover, we will see in Sec.~\ref{sec:transformations} that these
operators only account for the change in how coordinates label points,
but not changes in the waveforms themselves.

%%%%%%%%%%%%%%%%%%%%%%%%%%%%%%%%%%%%%%%%%%%%%%%%%%%%%%%%%%%%%%%%%%%%%%
%%%%%%%%%%%%%%%%%%%%%%%%%%%%%%%%%%%%%%%%%%%%%%%%%%%%%%%%%%%%%%%%%%%%%%
\section{Asymptotic flatness and transformations at a point}
\label{sec:AsymptoticFlatness}
Now, having seen the effects of the BMS transformation on
\emph{coordinates} on $\scriplus$, we need to understand the effects
on \emph{waveforms} measured at $\scriplus$.  We begin, in this
section, by examining the effect on the waveform at a single point,
where the transformation leaves that point fixed.  This will be
extended in Sec.~\ref{sec:impl-bms-transf} by allowing the point to
vary, which will involve the relatively simple task of evaluating the
known function at different points---in practice, requiring mostly
interpolation and other bookkeeping.

Though we will not yet vary the coordinates of our selected point, the
coordinates of \emph{neighboring} points will change.  For our
purposes, a waveform measures some piece of the differential structure
of spacetime.  But waveforms are not true scalars, in the sense that
they are not invariant under coordinate transformations---in fact,
they are inherently defined with respect to coordinates.  More
precisely, the tetrad with respect to which they are constructed is
\emph{defined} in terms of coordinates.  It is, of course, possible to
perform a coordinate transformation while leaving the tetrad fixed.
But this is not relevant; waveforms expressed in different coordinate
systems use different tetrads.  Therefore, a BMS transformation that
changes the coordinates of \emph{nearby} points should also change the
waveform \emph{at} the given point.

To make these ideas precise, we need to be more specific about our
representation of $\scriplus$, and the spacetime in a neighborhood of
$\scriplus$.  It will then be a relatively simple matter to calculate
the transformations of standard curvature quantities.  The reader who
is willing to take these results on faith may simply refer to
Eqs.~\eqref{eq:psi_n-prime}, \eqref{eq:sigma-prime},
\eqref{eq:h-prime}, and~\eqref{eq:news-prime}, and otherwise skip this
section.

\subsection{Asymptotically flat spacetime}
\label{sec:AsymptoticallyFlatSpacetime}

Numerous formulations describe the basic idea of asymptotic flatness,
most prominently developed by Penrose~\cite{Penrose1965a}.  For
definiteness, we will follow the development by
Moreschi~\cite{Moreschi1986, Moreschi1987}.  The essential idea is to
begin with a physical spacetime $(\finite{M}, \finite{\metric}_{ab})$,
and identify it with a portion of a model spacetime
$(\asymptotic{M}, \asymptotic{\metric}_{ab})$ representing the
asymptotic completion of the physical spacetime.  Here and in the
following, to simplify notation, quantities in the physical spacetime
will be represented by uppercase characters, while quantities in the
asymptotic spacetime will be represented by
lowercase.\footnote{Indices, of course, will not be included in this
  distinction; lowercase indices will denote tensor indices, while
  uppercase will denote spinor indices.}

We begin with the physical spacetime
$(\finite{M}, \finite{\metric}_{ab})$, which has Weyl spinor (the
spinor form of the standard Weyl tensor) $\finite{\psi}_{ABCD}$.  We
impose the assumption of (future) asymptotic flatness by requiring the
existence of another spacetime
$(\asymptotic{M}, \asymptotic{\metric}_{ab})$, with boundary
$\scriplus$ such that as topological spaces
$\scriplus = S^{2} \times \mathbb{R}$ and
$\finite{m} = \asymptotic{M} \setminus \scriplus$.  In particular, for
any point $\finite{p} \in \finite{M}$ we have a point identified as
$\finite{p} \in \asymptotic{m}$, so that we can interchangeably
describe any function at a point not on $\scriplus$ as being defined
either on $\finite{M}$ or $\asymptotic{m}$.  We further assume the
existence of a real-valued function $\asymptotic{\omega}$ that is
continuous on $\asymptotic{m}$ and smooth on $\finite{M}$, and
satisfies the following conditions:
\begin{enumerate}
\item $\left. \asymptotic{\omega} \right|_{\finite{M}} > 0$.
\item $\left. \asymptotic{\omega} \right|_{\scriplus} = 0$.
\item $\left. \d \asymptotic{\omega}\right|_{\scriplus} \neq 0$.
\end{enumerate}
Given this function, the spacetimes are also required to obey the
following conditions:
\begin{enumerate}[resume]
\item
  $\left. \asymptotic{\metric}_{ab} \right|_{\finite{M}} = \left.
    \finite{\metric}_{ab}\, \asymptotic{\omega}^{2}
  \right|_{\finite{M}}$.
  % \item
  %   $\left. \asymptotic{\metric}_{ab} \right|_{\finite{M}} =
  %   \left. \asymptotic{\omega}\, \finite{\metric}_{ab}
  %   \right|_{\finite{M}}$;
\item \label{it:geodesics} At every point of $\scriplus$, there ends a
  future-directed null geodesic of $\asymptotic{M}$.
\item In some neighborhood of $\scriplus$, there exist quantities
  $\hat{\asymptotic{R}}$ and $\tilde{\asymptotic{R}}$ on
  $\asymptotic{m}$ such that the Riemann tensor of
  $(\finite{M}, \finite{\gamma}_{ab})$ satisfies
  \begin{equation}
    \label{eq:Riemann_asymptotics}
    {\finite{R}_{abc}}^{d} = f(\asymptotic{\omega})\,
    {\hat{\asymptotic{R}}_{abc}}^{d} + {\tilde{\asymptotic{R}}_{abc}}^{d},
  \end{equation}
  where
  \begin{enumerate}
  \item $\d f / \d \asymptotic{\omega} > 0$,
  \item $\lim_{\asymptotic{\omega} \to 0} f = 0$,
  \item $\hat{\asymptotic{R}}$ is regular at $\scriplus$, and
  \item $\tilde{\asymptotic{r}}$ goes to zero faster than $f$ as
    $\asymptotic{\omega} \to 0$.
  \end{enumerate}
\end{enumerate}
The last condition is to be understood componentwise, with respect to
an orthogonal tetrad of $(\asymptotic{m}, \asymptotic{\gamma}_{ab})$
that is regular at $\scriplus$, like the one constructed below in
Eqs.~\eqref{eq:tetrad-as-differentials}.  We also \emph{define} a
spinor on $\finite{M}$ by
$\asymptotic{\psi}_{ABCD} \defined \asymptotic{\omega}^{-1}\,
\finite{\psi}_{ABCD}$,
which we can extend continuously to $\scriplus$.  Note, however, that
$\hat{\asymptotic{r}}$ and $\asymptotic{\psi}$ need not be the Riemann
tensor and Weyl spinor of
$(\asymptotic{m}, \asymptotic{\gamma}_{ab})$.

It is possible~\cite{Sachs1962, Sachs1962a, Moreschi1987} to choose
coordinates $(\scritime, \theta, \phi)$ on $\scriplus$, where
$\scritime$ labels a slice of $\scriplus$ with topology $S^{2}$ and
$(\theta, \phi)$ are the standard coordinates of the unit sphere.  The
latter are frequently expressed---at least for theoretical work---as
the usual stereographic coordinate $\asymptotic{\zeta}$ and its
complex conjugate $\bar{\asymptotic{\zeta}}$~\cite{Penrose1987}.
These coordinates can be extended into a neighborhood of $\scriplus$
by taking $\asymptotic{\omega}$ as an additional coordinate along
future-directed null geodesics, where $(\scritime, \theta, \phi)$
labels the geodesics.  Moreschi~\cite{Moreschi1987} showed that, up to
irrelevant gauge freedom, the $\asymptotic{\omega}$ function is
related to the luminosity distance $r_{\text{L}}$ by
\begin{equation}
  \label{eq:omega_as_r}
  \asymptotic{\omega} = \frac{1}{r_{\text{L}}} + \mathscr{O} \left(
    \frac{1} {r_{\text{L}}^{3}}  \right).
\end{equation}
These coordinates are essentially what are known as Bondi coordinates,
and allow the metric to be put in a particularly simple
form~\cite{Bondi1962, Sachs1962a}.  This form of the metric is
asymptotically invariant under BMS transformations.

An orthonormal spin dyad~\cite{Newman1962a, Newman1963, Geroch1973,
  Penrose1987, ODonnell2003}
$(\finite{\omicron}^{A}, \finite{\iota}^{A})$ and its asymptotic
counterpart $(\asymptotic{\omicron}^{A}, \asymptotic{\iota}^{A})$ can
also be defined related to these coordinates, such that we have
orthonormal tetrads
\begin{subequations}
  \label{eq:tetrad-as-spinors}
  \begin{align}
    \asymptotic{l}^{a}
    &\defined \sigma^{a}_{AA'}\, \asymptotic{\omicron}^{A}\,
      \asymptotic{\omicron}^{A'},
    &\finite{l}^{a}
    &\defined \sigma^{a}_{AA'}\, \finite{\omicron}^{A}\,
      \finite{\omicron}^{A'}, \\
    \asymptotic{m}^{a}
    &\defined \sigma^{a}_{AA'}\, \asymptotic{\omicron}^{A}\,
      \asymptotic{\iota}^{A'},
    &\finite{m}^{a}
    &\defined \sigma^{a}_{AA'}\, \finite{\omicron}^{A}\,
      \finite{\iota}^{A'}, \\
    \bar{\asymptotic{m}}^{a}
    &\defined \sigma^{a}_{AA'}\, \asymptotic{\iota}^{A}\,
      \asymptotic{\omicron}^{A'},
    &\bar{\finite{m}}^{a}
    &\defined \sigma^{a}_{AA'}\, \finite{\iota}^{A}\,
      \finite{\omicron}^{A'}, \\
    \asymptotic{n}^{a}
    &\defined \sigma^{a}_{AA'}\, \asymptotic{\iota}^{A}\,
      \asymptotic{\iota}^{A'},
    &\finite{n}^{a}
    &\defined \sigma^{a}_{AA'}\, \finite{\iota}^{A}\,
      \finite{\iota}^{A'},
  \end{align}
\end{subequations}
where $\sigma^{a}_{AA'}$ are the Infeld-van der Waerden symbols, and
at leading order in $\asymptotic{\omega}$ we have
\begin{subequations}
  \label{eq:tetrad-as-differentials}
  \begin{align}
    \asymptotic{l}_{a}
    &=  (\d u)_{a}
    &&\asymptoticallyequal \finite{L}_{a},
    \\
    \asymptotic{m}_{a}
    &= -\frac{\sqrt{2}} {1 + \asymptotic{\zeta}\,
      \bar{\asymptotic{\zeta}}} (\d\bar{\asymptotic{\zeta}})_{a}
    &&\asymptoticallyequal \asymptotic{\omega}\, \finite{M}_{a},
    \\
    \bar{\asymptotic{m}}_{a}
    &= -\frac{\sqrt{2}} {1 + \asymptotic{\zeta}\,
      \bar{\asymptotic{\zeta}}} (\d{\asymptotic{\zeta}})_{a}
    &&\asymptoticallyequal \asymptotic{\omega}\, \bar{\finite{M}}_{a},
    \\
    \asymptotic{n}_{a}
    &= -(\d \asymptotic{\omega})_{a}
    &&\asymptoticallyequal \asymptotic{\omega}^{2}\, \finite{N}_{a}.
  \end{align}
\end{subequations}
We also denote by
\begin{equation}
  \label{eq:asymptotic_eth}
  \finite{\eth} \asymptoticallyequal \asymptotic{\omega}\,
  \asymptotic{\eth}
\end{equation}
% In particular, considering the vector tetrad element $\finite{M}$ as
% a derivative operator on spin-weight $s=0$ fields, we have
% $\finite{\eth} = \finite{M}$, but we also have
% $\finite{M} = \asymptotic{\omega}\, \asymptotic{m}$ and
% $\asymptotic{m} = \asymptotic{\eth}$, which shows us
% Eq.~\eqref{eq:asymptotic_eth}.
the spin-raising differential operator introduced (at finite radius as
$\eth$) by Geroch, Held, and Penrose~\cite{Geroch1973}.\footnote{Note
  that the operator $\eth_{\text{NP}}$ originally introduced by Newman
  and Penrose~\cite{Newman1966} is generally different from the
  operator $\eth_{\text{GHP}} \identically \finite{\eth}$ introduced
  by Geroch, Held, and Penrose, in that only the latter has well
  defined transformation behavior under boosts.  There is also a
  discrepancy in the normalization such that
  $\eth_{\text{NP}} = \sqrt{2}\, \eth_{\text{GHP}}$ for scalar
  functions.}

This completes the basic framework we use to describe asymptotically
flat spacetimes, allowing us to understand the asymptotic behavior of
the physical fields.  Next, we will show how a BMS transformation
alters this framework, and use that result to find the changes in
curvature quantities expressed within this framework.

\subsection{Transformations}
\label{sec:transformations}
Equation~\eqref{eq:BMS-transformation} describes the general BMS
transformation.  However, this transformation is only defined
\emph{on} $\scriplus$.  Because the curvature quantities we are
interested in measure the differential structure of spacetime,
understanding those quantities requires understanding the
transformation \emph{in a neighborhood} of $\scriplus$.
Moreschi~\cite{Moreschi1986} found the general transformation to
first-order in $\asymptotic{\omega}$ that preserves the leading-order
Bondi form of the metric:
\begin{subequations}
  \label{eq:bms-transformation-extended}
  \begin{align}
    \breve{\scritime}
    &= \asymptotic{k} \left( \scritime - \st \right) -
      \asymptotic{\omega}\, \frac{\asymptotic{\eth} \scritime'\,
      \bar{\asymptotic{\eth}} \scritime'} {\asymptotic{k}}, \\
    \breve{\asymptotic{\zeta}}
    &= \frac{a\, \asymptotic{\zeta} + b} {c\, \asymptotic{\zeta} + d}
      - \asymptotic{\omega}\, \frac{\asymptotic{\eth} \scritime'\,
      \bar{\asymptotic{\eth}} \asymptotic{\zeta}' +
      \asymptotic{\eth} \asymptotic{\zeta}'\,
      \bar{\asymptotic{\eth}} \scritime'} {k}, \\
    \breve{\asymptotic{\omega}}
    &= k\, \asymptotic{\omega}.
  \end{align}
\end{subequations}
Here, $\scritime'$ and $\asymptotic{\zeta}'$ are the leading-order
terms in their respective equations---also given by the standard BMS
transformation of Eq.~\eqref{eq:BMS-transformation}.  This
transformation is defined in a neighborhood of $\scriplus$, so we can
evaluate the differentials in Eqs.~\eqref{eq:tetrad-as-differentials}
and take the limit as $\asymptotic{\omega} \to 0$, to find the
transformation laws for the tetrad and infer the effects on the spinor
basis:
\begin{subequations}
  \begin{align}
    \asymptotic{\omicron}'^{A}
    &= \frac{\e^{\i\, \SpinPhase / 2}} {\sqrt{\asymptotic{K}}}\, \left(
      \asymptotic{\omicron}^{A} - \frac{\asymptotic{\eth} u'}
      {\asymptotic{K}}\, \asymptotic{\iota}^{A} \right), \\
    \asymptotic{\iota}'^{A}
    &= \frac{\e^{-\i\, \SpinPhase / 2}} {\sqrt{\asymptotic{K}}}\,
      \asymptotic{\iota}^{A}.% \\
      % \asymptotic{\omicron}'_{A}
      % &= \sqrt{\asymptotic{K}}\, \e^{\i\, \SpinPhase / 2} \, \left(
      %   \asymptotic{\omicron}_{A} - \frac{\asymptotic{\eth} u'}
      %   {\asymptotic{K}}\, \asymptotic{\iota}_{A} \right), \\
      % \asymptotic{\iota}'_{A}
      % &= \sqrt{\asymptotic{K}}\, \e^{-\i\, \SpinPhase / 2} \,
      % \asymptotic{\iota}_{A}.
  \end{align}
\end{subequations}
Here, $\SpinPhase$ is the spin phase described in
Sec.~\ref{sec:RotationsAndBoosts} and Appendix~\ref{sec:rotor-boost}.
Because the curvature quantities are defined with respect to these
spinors and their spatial dependence, this is enough to calculate the
transformation laws of the curvature quantities.

The first and simplest set of curvature quantities we will need is the
collection of Newman-Penrose scalars.  The following are the
definitions of these scalars on $\scriplus$, along with their
leading-order relationship to the corresponding finite-radius scalars:
\begin{subequations}
  \label{eq:newman-penrose-scalars}
  \begin{align}
    \asymptotic{\psi}_{0}
    &\defined \asymptotic{\psi}_{ABCD}\, \asymptotic{\omicron}^{A}\,
      \asymptotic{\omicron}^{B}\,  \asymptotic{\omicron}^{C}
      \asymptotic{\omicron}^{D}
    &&\asymptoticallyequal \asymptotic{\omega}^{-5}\,
       \finite{\Psi}_{0}, \\
    \asymptotic{\psi}_{1}
    &\defined \asymptotic{\psi}_{ABCD}\, \asymptotic{\omicron}^{A}\,
      \asymptotic{\omicron}^{B}\,  \asymptotic{\omicron}^{C}
      \asymptotic{\iota}^{D}
    &&\asymptoticallyequal \asymptotic{\omega}^{-4}\,
       \finite{\Psi}_{1}, \\
    \asymptotic{\psi}_{2}
    &\defined \asymptotic{\psi}_{ABCD}\, \asymptotic{\omicron}^{A}\,
      \asymptotic{\omicron}^{B}\,  \asymptotic{\iota}^{C}
      \asymptotic{\iota}^{D}
    &&\asymptoticallyequal \asymptotic{\omega}^{-3}\,
       \finite{\Psi}_{2}, \\
    \asymptotic{\psi}_{3}
    &\defined \asymptotic{\psi}_{ABCD}\, \asymptotic{\omicron}^{A}\,
      \asymptotic{\iota}^{B}\,  \asymptotic{\iota}^{C}
      \asymptotic{\iota}^{D}
    &&\asymptoticallyequal \asymptotic{\omega}^{-2}\,
       \finite{\Psi}_{3}, \\
    \asymptotic{\psi}_{4}
    &\defined \asymptotic{\psi}_{ABCD}\, \asymptotic{\iota}^{A}\,
      \asymptotic{\iota}^{B}\,  \asymptotic{\iota}^{C}
      \asymptotic{\iota}^{D}
    &&\asymptoticallyequal \asymptotic{\omega}^{-1}\,
       \finite{\Psi}_{4}.
  \end{align}
\end{subequations}
Because $\asymptotic{\psi}_{ABCD}$ is a geometric object, it does not
transform under a change of coordinates, so the transformation law for
these scalars is given simply by replacing the spinors
$\asymptotic{\omicron}^{A}$ and $\asymptotic{\iota}^{A}$ with their
transformed values, which leads to a simple hierarchy with a basic
combinatorial pattern:
\begin{widetext}
  \begin{subequations}
    \label{eq:psi_n-prime}
    \begin{align}
      \label{eq:psi_0-prime}
      \asymptotic{\psi}'_{0}
      &= \frac{\e^{2\i\SpinPhase}} {\asymptotic{k}^{3}}\, \left[
        \asymptotic{\psi}_{0} %
        - 4\, \frac{\asymptotic{\eth} u'} {\asymptotic{K}}\,
        \asymptotic{\psi}_{1} %
        + 6\, \left( \frac{\asymptotic{\eth} u'} {\asymptotic{K}}
        \right)^{2}\, \asymptotic{\psi}_{2} %
        - 4\, \left( \frac{\asymptotic{\eth} u'} {\asymptotic{K}}
        \right)^{3}\, \asymptotic{\psi}_{3} %
        + \left( \frac{\asymptotic{\eth} u'} {\asymptotic{K}}
        \right)^{4}\, \asymptotic{\psi}_{4} \right], \\
      \label{eq:psi_1-prime}
      \asymptotic{\psi}'_{1}
      &= \frac{\e^{\i\SpinPhase}} {\asymptotic{k}^{3}}\, \left[
        \asymptotic{\psi}_{1} %
        - 3\, \frac{\asymptotic{\eth} u'} {\asymptotic{K}}\,
        \asymptotic{\psi}_{2} %
        + 3\, \left( \frac{\asymptotic{\eth} u'} {\asymptotic{K}}
        \right)^{2}\, \asymptotic{\psi}_{3} %
        - \left( \frac{\asymptotic{\eth} u'} {\asymptotic{K}}
        \right)^{3}\, \asymptotic{\psi}_{4} \right], \\
      \label{eq:psi_2-prime}
      \asymptotic{\psi}'_{2}
      &= \frac{1} {\asymptotic{k}^{3}}\, \left[
        \asymptotic{\psi}_{2} %
        - 2\, \frac{\asymptotic{\eth} u'} {\asymptotic{K}}\,
        \asymptotic{\psi}_{3} %
        + \left( \frac{\asymptotic{\eth} u'} {\asymptotic{K}}
        \right)^{2}\, \asymptotic{\psi}_{4} \right], \\
      \label{eq:psi_3-prime}
      \asymptotic{\psi}'_{3}
      &= \frac{\e^{-\i\SpinPhase}} {\asymptotic{k}^{3}}\, \left[
        \asymptotic{\psi}_{3} %
        - \frac{\asymptotic{\eth} u'} {\asymptotic{K}}\,
        \asymptotic{\psi}_{4} \right], \\
      \label{eq:psi_4-prime}
      \asymptotic{\psi}'_{4}
      &= \frac{\e^{-2\i\SpinPhase}} {\asymptotic{k}^{3}}\, \left[
        \asymptotic{\psi}_{4} \right].
    \end{align}
  \end{subequations}
\end{widetext}
The simplicity of this result is surprising because the Newman-Penrose
quantities $\finite{\Psi}_{n}$ represent the components of a tensor,
so at finite radius all of these components would mix with each other.
However, in the limit as $\scriplus$ is approached, the ``peeling-off
property''~\cite{Sachs1962, Sachs1962a, Penrose1965a} of
asymptotically flat spacetimes comes into play, as seen in the
right-hand column of Eqs.~\eqref{eq:psi_n-prime}, so that the pattern
emerges with lower-index quantities (\eg, $\asymptotic{\psi}_{0}$)
being irrelevant to the transformed values of higher-index quantities
(\eg, $\asymptotic{\psi}'_{4}$).  On the other hand, given this
peeling behavior, it may also seem surprising that effects from the
higher-index quantities do not overwhelm the lower-index scalars.  The
reason is that at finite radii $\asymptotic{\eth}$ is replaced by
$\asymptotic{\omega}^{-1}\, \finite{\eth}$, and since each
higher-index scalar appears in the expressions for lower-index scalars
accompanied by powers of $\asymptotic{\eth}$, the resulting factors of
$\asymptotic{\omega}^{-1}$ are exactly enough to cancel the dominance
of the higher-index scalars---to leading order in
$\asymptotic{\omega}$.

The most interesting remaining quantity is the spin coefficient
representing the shear:
\begin{equation}
  \label{eq:sigma}
  \asymptotic{\sigma} \defined \asymptotic{\omicron}^{A}\,
  \asymptotic{\omicron}^{B}\, \bar{\asymptotic{\iota}}^{B'}\,
  \finite{\nabla}_{BB'} \asymptotic{\omicron}_{A}.
\end{equation}
Because of the derivative, this is somewhat more difficult to evaluate
than the Newman-Penrose scalars.  However, after the suitable limit
has been taken, we arrive at the simple formula~\cite{Bondi1962,
  Sachs1962a, Moreschi1987}
\begin{equation}
  \label{eq:sigma-prime}
  \asymptotic{\sigma}' = \frac{\e^{2\i\SpinPhase}} {k} \left[
    \asymptotic{\sigma} - \asymptotic{\eth}^{2}\, \st  \right].
\end{equation}
This is consistent with Eq.~\eqref{eq:psi_4-prime} and the asymptotic
relation
\begin{equation}
  \label{eq:psi4_sigma}
  \asymptotic{\psi_{4}} = - \frac{\partial^{2}} {\partial u^{2}}
  \bar{\asymptotic{\sigma}}
\end{equation}
because
$\partial / \partial_{u'} = \frac{1}{k} \partial / \partial_{u}$, and
the BMS transformation is constant, so that the $\SpinPhase$, $k$, and
$\st$ functions are independent of $\scritime$.  The shear is also
related to the more commonly used~\cite{Ajith2007C} strain of the
transverse-traceless metric perturbation\footnote{It is worth pointing
  out that this relation is only true for the asymptotic fields.
  Trivially, we have
  $\asymptotic{h} \asymptoticallyequal r\, \finite{h}$, whereas
  $\asymptotic{\sigma} \asymptoticallyequal r^{2}\, \finite{\sigma}$.
  That is, the two finite-radius fields behave differently in the
  limit $r \to \infty$.  But terms at higher relative order in $1/r$
  may differ more substantially.}
$\asymptotic{h} = \bar{\asymptotic{\sigma}}$, which implies the
transformation law
\begin{equation}
  \label{eq:h-prime}
  \asymptotic{\h}' = \frac{\e^{-2\i\SpinPhase}} {k} \left[
    \asymptotic{\h} - \bar{\asymptotic{\eth}}^{2}\, \st  \right].
\end{equation}
Similarly, the Bondi news function~\cite{Bondi1962, Sachs1962a,
  Tamburino1966, Thorne1983, Helfer1993} satisfies
\begin{equation}
  \label{eq:N_sigma}
  \finite{N} \asymptoticallyequal \asymptotic{n} = - \frac{\partial}
  {\partial u} \bar{\asymptotic{\sigma}},
\end{equation}
which implies the transformation law
\begin{equation}
  \label{eq:news-prime}
  \asymptotic{n}' = \frac{\e^{-2\i\SpinPhase}} {k^{2}} \asymptotic{n}.
\end{equation}
We note, however, that these relationships between
$\asymptotic{\psi}_{4}$, $\asymptotic{\sigma}$, $\asymptotic{h}$, and
$\asymptotic{n}$ are only valid asymptotically, and only in Bondi
coordinates; more generally, the relationships would be more
complicated.

The expressions given here for the transformations of the waveform
quantities are fairly simple, and can all be constructed given the
waveforms and a choice of transformation---as described by the
functions $k$, $\SpinPhase$, and $\st$.  However, these expressions
hide a complication: all of the quantities involved are functions of
position.  To actually implement a BMS transformation, we need to know
how to express these functions in terms of the coordinates, both old
and new.  This requires combining the ideas of the present section
with those of the previous section.

%%%%%%%%%%%%%%%%%%%%%%%%%%%%%%%%%%%%%%%%%%%%%%%%%%%%%%%%%%%%%%%%%%%%%%
%%%%%%%%%%%%%%%%%%%%%%%%%%%%%%%%%%%%%%%%%%%%%%%%%%%%%%%%%%%%%%%%%%%%%%
\section{Implementation of BMS transformations of waveforms}
\label{sec:impl-bms-transf}

We assume that the field is known in some observer's frame
$\observer$, as a function of that observer's Bondi coordinates
throughout some portion of $\scriplus$.  A second observer
$\observer'$ is related to the first by some known BMS transformation
as in Eqs.~\eqref{eq:BMS-transformation}.  In particular the frame of
$\observer'$ can be obtained from $\observer$ by an initial
supertranslation $\st$, followed by a frame rotation $\FrameRotor$,
followed by a boost of velocity $\fourvec{v}$.  Our objective will be
to find the field as decomposed into modes of a spin-weighted
spherical-harmonic expansion, at a series of discrete retarded times
$\{\scritime_{i'}'\}$.

The first step is to find the mode weights of all quantities we will
need in frame $\observer$.  At the most basic level, we need the modes
of the waveform in question.  We denote this waveform by $\waveform$,
which may represent $\asymptotic{h}$, $\asymptotic{\psi}_{4}$, or any
of the other quantities discussed in Sec.~\ref{sec:transformations}.
A waveform $\waveform$ of spin weight $s$ will be decomposed into
modes of the spin-weighted spherical harmonic expansion as
\begin{equation}
  \label{eq:SWSHModes}
  \waveform(\scritime, \theta, \phi) = \sum_{\ell,m}
  \waveform^{\ell,m}(\scritime)\, \sYlm{\ell,m} (\theta, \phi),
\end{equation}
where the relevant data are the modes $\waveform^{\ell,m}(\scritime)$.
In practice, the sum over $\ell$ extends up to some maximum integer
$\ellmax$.

If $\waveform$ represents a Newman-Penrose quantity
$\asymptotic{\psi}_{n}$ with $n<4$, we also need all the higher-index
Newman-Penrose quantities as shown in Eqs.~\eqref{eq:psi_n-prime}, as
well as the quantity $\asymptotic{\eth} \scritime' / \asymptotic{k}$.
For $\asymptotic{\sigma}$ and $\asymptotic{h}$, we will need
$\asymptotic{\eth}^{2} \st$ (or its complex conjugate).  Given an
arbitrary function $\sfunc$ of spin weight $s$, we can calculate the
modes of the differentiated quantity $\asymptotic{\eth} \sfunc$ in
terms of the modes of the original function as\footnote{These
  equations differ from the similar Eq.~(3.22) of Newman and
  Penrose~\cite{Newman1966} and Eqs.~(2.7) of Goldberg
  \etal~\cite{Goldberg1967} by the factor $1/\sqrt{2}$ here.  As noted
  previously, this is because the operator here is---up to the factor
  given in Eq.~\eqref{eq:asymptotic_eth}---identical to the one given
  by Geroch, Held, and Penrose~\cite{Geroch1973}, which intentionally
  introduced the $1/\sqrt{2}$ factor.}
\begin{equation}
  \label{eq:eth_modes}
  \Big( \asymptotic{\eth} \sfunc \Big)^{\ell,m} =
  \begin{cases}
    0 & \ell < \abs{s+1}, \\
    \sqrt{ \frac{(\ell-s) (\ell+s + 1)} {2}}\, \sfunc^{\ell, m} &
    \text{otherwise}.
  \end{cases}
\end{equation}
Similarly, we have
\begin{equation}
  \label{eq:ethbar_modes}
  \Big( \bar{\asymptotic{\eth}} \sfunc \Big)^{\ell,m} =
  \begin{cases}
    0 & \ell < \abs{s-1}, \\
    -\sqrt{ \frac{(\ell+s) (\ell-s + 1)} {2}}\, \sfunc^{\ell, m} &
    \text{otherwise}.
  \end{cases}
\end{equation}
We will assume that the supertranslation $\st$ is given directly in
terms of its modes.  This makes it trivial to compute either
$\asymptotic{\eth}^{2} \st$ or $\bar{\asymptotic{\eth}}^{2} \st$,
noting that $\st$ has spin weight $s=0$, while $\asymptotic{\eth} \st$
has spin weight $s=1$ and $\bar{\asymptotic{\eth}} \st$ has spin
weight $s=-1$.  On the other hand, to compute
$\asymptotic{\eth} \scritime' / k = \scritime\, \asymptotic{\eth} k /
k - \asymptotic{\eth} \st$,
we need to know $k$ in terms of its spherical-harmonic modes.  This
could be done analytically with exact expressions involving Wigner's
$\D$ and $3$-$j$ functions.  For practical purposes, a more efficient
approach is to evaluate $k$ as given in
Eq.~\eqref{eq:conformal-factor-dot-product} on a series of grid
points, and feed the results into software that computes the modes, as
discussed below.

Now, given the modes of the various fields at some discrete set of
times $\{\scritime_{i}\}$, we need to be able to interpolate as a
function of time, because slices of constant $\scritime$ will not
typically correspond to slices of constant $\scritime'$.  Of course,
because of the direction dependence of these time slices,
interpolation of the mode weights themselves is not possible in
general.  Instead, we must transform the modes into a series of points
in physical space, interpolate the values of the field at each spatial
point to the appropriate time, and then transform back to modes.  The
current state-of-the-art numerical code for transforming between
physical space and modes is the \software{spinsfast}
package~\cite{Huffenberger2010, Huffenberger}.  The points in physical
space used by this package form an equiangular grid in
colatitude-longitude:
\begin{subequations}
  \begin{align}
    \theta_{j}' &= \frac{\pi\, j} {N_{\theta}-1} & \text{for}
    && j \in \{0, 1, \dots, N_{\theta}-1 \},
    \intertext{and}
    \phi_{k}' &= \frac{2\, \pi\, k} {N_{\phi}} & \text{for}
    && k \in \{0, 1, \dots, N_{\phi}-1 \}.
  \end{align}
\end{subequations}
Note that the poles, $\theta_{j}'=0$ and $\pi$ are each covered by
$N_{\phi}$ pairs of $(\theta, \phi)$ values, but each such pair
represents a different alignment of the tangent basis at that point.
For the sake of accuracy, it is best to choose $N_{\theta} > 2\ellmax$
and $N_{\phi} > 2\ellmax$~\cite{Huffenberger}.  In practice, it seems
to be sufficient to simply choose
$N_{\theta} = N_{\phi} = 2\ellmax+1$.  Of course, this grid is given
in the frame of $\observer'$; since the waveform is given in the frame
of $\observer$, we need to know the points in that frame corresponding
to the points $\{(\theta_{j}', \phi_{k}')\}$.  Moreover, a
spin-weighted field in $\observer'$ is defined with respect to the
tangent vectors to the sphere, canonically defined in terms of the
$(\theta', \phi')$ coordinates.  Thus, we also need to know what these
tangent vectors correspond to in the basis of $\observer$.

Adapting the discussion of Sec.~\ref{sec:RotationsAndBoosts}, we begin
by defining the rotor (in quaternion notation)
\begin{equation}
  \label{eq:R_j_k}
  \rotor{R}_{j, k}' \defined \e^{\phi_{k}'\, \fourvec{z}/2}\,
  \e^{\theta_{j}'\, \fourvec{y}/2},
\end{equation}
where $\fourvec{x}$, $\fourvec{y}$, and $\fourvec{z}$ are the
orthonormal basis vectors of $\observer$.  Note the mixing of
coordinates from $\observer'$ with basis elements of $\observer$.  We
then define the unit vector
\begin{equation}
  \label{eq:directionvec_j_k_prime}
  \directionvec_{j,k}' = \rotor{R}_{j, k}'\, \fourvec{z}\,
  \rotor{R}_{j, k}'^{-1},
\end{equation}
which points in the direction $(\theta_{j}', \phi_{k}')$ as measured
by $\observer$, and define the angle
\begin{equation}
  \label{eq:Theta_prime}
  \Theta_{j,k}' \defined \arccos \frac{\fourvec{v} \cdot
    \directionvec_{j,k}'} {\lvert \fourvec{v} \rvert\,
    \lvert \directionvec_{j,k}' \rvert}.
\end{equation}
The equivalent angle in the unprimed frame is
\begin{equation}
  \label{eq:Theta}
  \Theta_{j,k} = 2 \arctan \left[ \e^{-\rapidity} \tan
    \frac{\Theta_{j,k}'}{2} \right],
\end{equation}
where $\rapidity = \artanh \abs{\fourvec{v}}$ is the rapidity.  Then, we
can define
\begin{equation}
  \label{eq:boost-rotor_j_k}
  \BoostRotor_{j, k} \defined \exp \left[ \frac{\Theta_{j,k}' -
      \Theta_{j,k}} {2} \frac{ \directionvec_{j,k}' \times \fourvec{v} }
    {\abs{ \directionvec_{j,k}' \times \fourvec{v} }} \right],
\end{equation}
unless $\Theta' = \Theta = 0$ or $\pi$, in which case we simply have
$\BoostRotor_{j,k} = 1$.  Finally, we arrive at the required rotor
\begin{equation}
  \label{eq:net_rotor}
  R_{j,k} \defined \BoostRotor_{j, k}\, \FrameRotor\, R_{j,k}'.
\end{equation}
The physical point labeled by $(\theta_{j}', \phi_{k}')$ in
$\observer'$ is given by
\begin{equation}
  \label{eq:directionvec_j_k}
  \directionvec_{j,k} = \rotor{R}_{j, k}\, \fourvec{z}\,
  \rotor{R}_{j, k}^{-1}
\end{equation}
in $\observer$, while the complex tangent vector $\fourvec{m}_{j,k}'$
at that point in $\observer'$ corresponds to the vector
\begin{equation}
  \label{eq:m_j_k}
  \fourvec{m}_{j,k} = \rotor{R}_{j, k}\, \frac{\fourvec{x} + \i\,
    \fourvec{y}} {\sqrt{2}}\, \rotor{R}_{j, k}^{-1}
\end{equation}
in $\observer$.  The spin phase is determined by the relative rotation
between $\fourvec{m}_{j,k}$ as given here and the natural canonical
$\fourvec{m}$ vector given at the same point by $\observer$.  This was
depicted in Fig.~\ref{fig:BoostedGrids}, and is explained in more
detail in Appendix~\ref{sec:rotor-boost}.

More directly, we can evaluate any field, along with its appropriate
spin-phase factor, by evaluating the mode-weighted spin-weighted
spherical harmonics directly as functions of $\rotor{R}_{j, k}$.  As
detailed in Appendix~B of Ref.~\cite{Boyle2014}, this is made possible
by redefining the spin-weighted spherical harmonics to be functions of
a single unit-quaternion argument in terms of Wigner's $\D$ matrices
as\footnote{This relationship was originally noted by Goldberg
  \etal~\cite{Goldberg1967}, though they essentially restricted the
  possible rotations to rotors of the form
  $\rotor{R} = \e^{\phi\, \fourvec{z}/2}\, \e^{\theta\,
    \fourvec{y}/2}$.
  The problem with such a limited interpretation is that the
  spin-weighted spherical harmonics so defined \emph{do not} transform
  among themselves under rotations, and are incapable of expressing
  the correct spin-phase behavior.  By expanding the meaning of the
  spherical harmonics in this way we eliminate those problems, while
  maintaining agreement with the original definition and standard
  usage.}
\begin{equation}
  \label{eq:SWSH}
  \sYlm{\ell,m} (\rotor{R}) \defined \sqrt{ \frac{2\ell+1} {4\pi} }
  \D^{(\ell)}_{-s, m} ( \rotor{R} ).
\end{equation}
Thus, for example, when transforming the strain $\asymptotic{h}$, part
of the right-hand side of Eq.~\eqref{eq:h-prime} can be calculated
very simply as
\begin{equation}
  \label{eq:mode_sum}
  \left. \e^{-2\i\SpinPhase} \left[ \asymptotic{\h} -
      \bar{\asymptotic{\eth}}^{2}\, \st \right]
  \right\rvert_{\theta_{j}', \phi_{k}'} = \sum_{\ell,m}
  \left[\asymptotic{h}^{\ell,m} - \left( \bar{\asymptotic{\eth}}^{2}
      \st \right)^{\ell,m} \right] \mTwoYlm{\ell,m}(\rotor{R}_{j,k}).
\end{equation}
Note that no additional manipulation is required to find the
spin-phase factor $\e^{-2\i\SpinPhase}$; it is implicitly calculated by
$\mTwoYlm{\ell,m}(\rotor{R}_{j,k})$.  There is, however, the remaining
factor of $1/\asymptotic{k}$ to calculate.  Including this factor is
best done by evaluating this factor as [compare
Eq.~\eqref{eq:conformal-factor-dot-product}]
\begin{equation}
  \label{eq:conformal_factor_factor}
  \frac{1} {\asymptotic{k}} = \gamma \left( 1 - \fourvec{v} \cdot
    \directionvec_{j,k} \right),
\end{equation}
then multiplying this result by the result of
Eq.~\eqref{eq:mode_sum}.  In a similar way, other waveforms can be
computed as necessary by pointwise combination of the relevant
quantities given in the transformation laws of
Sec.~\ref{sec:transformations}.

Proceeding in this way for all values of the discrete indices, we
obtain the waveform values
$\waveform'(\scritime_{i,j,k}', \theta_{j}', \phi_{k}')$, where 
\begin{equation}
  \label{eq:indexed_time}
  \scritime_{i,j,k}' = \asymptotic{k}(\theta_{j}', \phi_{k}')\, \left[
    \scritime_{i} - \st(\theta_{j}', \phi_{k}') \right].
\end{equation}
Next, we simply need to interpolate these values in each direction to
a corresponding set of times $\{\scritime_{i'}'\}$ representing some
target time slices of observer $\observer'$.  There is a minor
ambiguity here, in that this set of times is somewhat arbitrary.  In
practice, the input data may be sampled unevenly in time---for
example, to provide better resolution of the merger-ringdown portion
of a waveform, while reducing the amount of data representing the slow
inspiral.  It would presumably be best to retain this sampling in the
transformed data set.  To a reasonable approximation, this can be done
by assigning
\begin{equation}
  \label{eq:u_prime_indexed}
  \scritime_{i}' = \frac{1}{\gamma} \left( \scritime_{i} -
    \st^{0,0} / \sqrt{4\pi} \right),
\end{equation}
which is the value of $\scritime'$ for which the average value of
$\scritime$ over the sphere (on the slice of constant $\scritime'$) is
precisely $\scritime_{i}$.  To clarify the notation,
$\{\scritime_{i,j,k}'\}$ is the set of time coordinates already
present in the data, whereas $\{\scritime_{i}'\}$ is the set of times
to which we might wish to interpolate.

However, we must deal with a subtlety first.  In some directions,
interpolation to some of the values of $\scritime_{i}'$ given by
Eq.~\eqref{eq:u_prime_indexed} would require data at times earlier
than $\scritime_{0}$ or later than $\scritime_{N_{\scritime} - 1}$.
This is because we have simply used the average value to derive
Eq.~\eqref{eq:u_prime_indexed}, while neglecting the direction
dependence.  To avoid extrapolation, then, we must restrict the set
$\{\scritime_{i}'\}$ to the range of times
$\scritime_{\text{min}}' \leq \scritime' \leq
\scritime_{\text{max}}'$, where
\begin{subequations}
  \label{eq:min_max_times}
  \begin{align}
    \label{eq:min_time}
    \scritime_{\text{min}}'
    &= \max_{j,k}\, \scritime_{0,j,k}',
    \\
    \scritime_{\text{max}}'
    &= \min_{j,k}\, \scritime_{N_{\scritime}-1,j,k}'.
  \end{align}
\end{subequations}
We denote the resulting subset by $\{ \scritime_{i'}' \}$, which is
the final set of times to which we will interpolate the data.  The
index $i'$ is used to indicate that it comes from a slightly different
indexing set than the index $i$ used for the input data.

Though the construction of $\{ \scritime_{i'}' \}$ suggested here is
by no means unique, we will always be limited to using a \emph{proper}
subset of the input data, whenever the boost and supertranslation
components with $\ell>0$ are nontrivial, because some of the input
time steps will correspond to slices of $\scritime'$ for which the
input data represent an incomplete sphere, and thus insufficient data
for computing spin-weighted spherical-harmonic modes.  Nonetheless,
this choice of $\{ \scritime_{i'}' \}$ is well defined and easy to
implement, it roughly preserves the sampling of the input data, and it
uses the input data to nearly the fullest possible extent.

Finally, for each value of $(i', j, k)$, we interpolate the waveform
values $\waveform'(\scritime_{i,j,k}', \theta_{j}', \phi_{k}')$ in
time to $\waveform'(\scritime_{i'}', \theta_{j}', \phi_{k}')$.  For
each $i'$, we then feed these values into a software package like
\software{spinsfast} to obtain the modes as measured by observer
$\observer'$, thus arriving at our goal: the set of modes
$\waveform'^{\ell,m}$ for each time $\scritime_{i'}'$.  The entire
transformation is implemented in the python module \software{scri},
which is included in the supplemental materials provided with this
paper~\cite{boyle2016supplement}.

%%%%%%%%%%%%%%%%%%%%%%%%%%%%%%%%%%%%%%%%%%%%%%%%%%%%%%%%%%%%%%%%%%%%%%
%%%%%%%%%%%%%%%%%%%%%%%%%%%%%%%%%%%%%%%%%%%%%%%%%%%%%%%%%%%%%%%%%%%%%%
\section{Effects of transformations on waveform modes}
\label{sec:underst-effects-bms}

It will be instructive to observe the effect of typical
transformations on waveform modes.  Because of the peculiar nature of
$\scriplus$ and the highly nonlinear behavior of waveforms under these
transformations, we will not be able to rely on any intuition for
transformations of multipole moments that we may have gained in
studying electromagnetism, for example.  However, we can take
advantage of the fact that mode decompositions are linear, so that it
is sufficient to observe the transformation of a single mode at a
time.  In particular, we will define input waveforms having a single
nonzero mode.  For further simplicity, that mode will behave as a pure
phase rotation at constant angular velocity.  We can then transform
this model waveform, and see how the power in the chosen mode leaks
out into other modes.  Because rotations are already well
understood---in fact, they behave identically to the more familiar
rotations of spin-zero spherical harmonics---we will focus here only
on translations and boosts.

%%%%%%%%%%%%%%%%%%%%%%%%%%%%%%%%%%%%%%%%%%%%%%%%%%%%%%%%%%%%%%%%%%%%%%
\begin{figure*}
  % \tikzset{external/force remake}
  \subfloat{%
    \label{fig:ModeTranslation-2_2_x}%
    \includegraphics{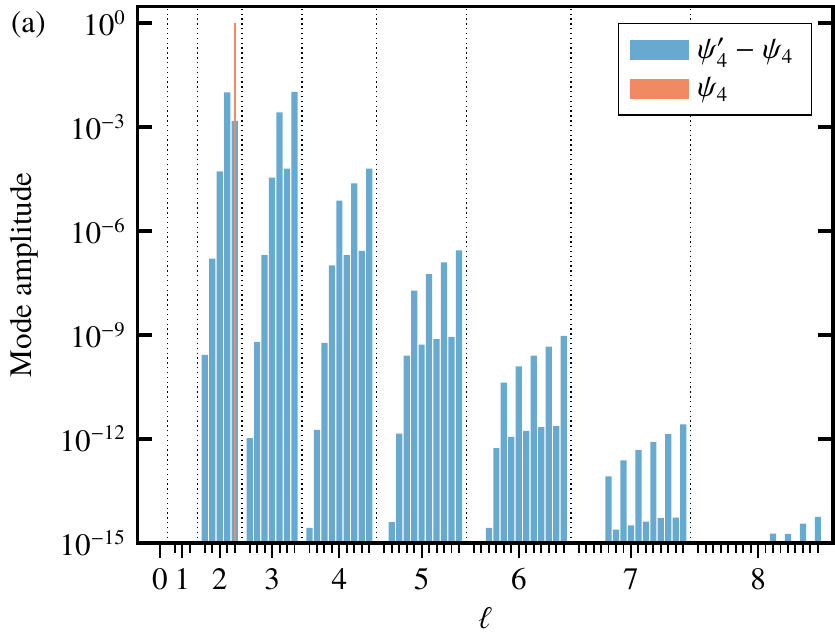}
  }%
  \hfill%
  \subfloat{%
    \label{fig:ModeTranslation-2_2_z}%
    \includegraphics{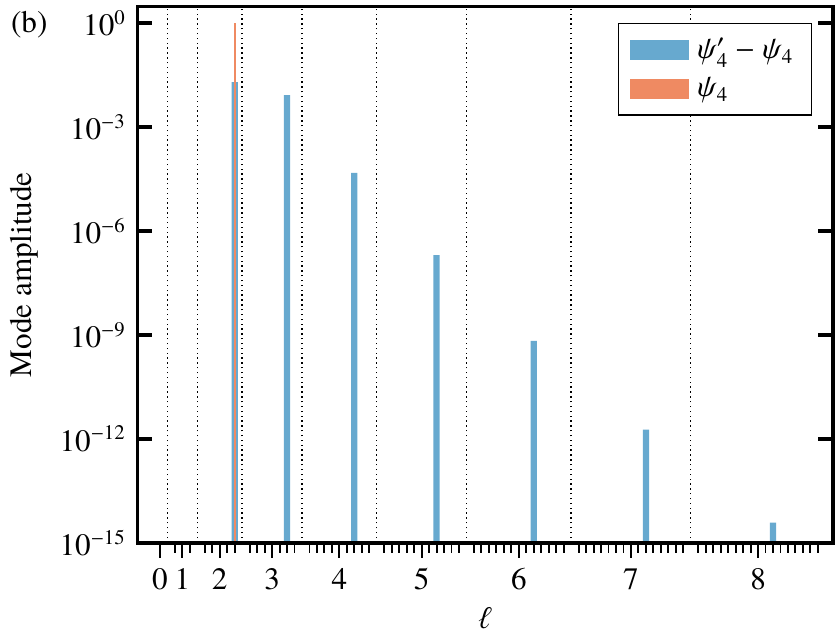}
  }%
  \\[10pt] %
  \subfloat{%
    \label{fig:ModeTranslation-4_2_x}%
    \includegraphics{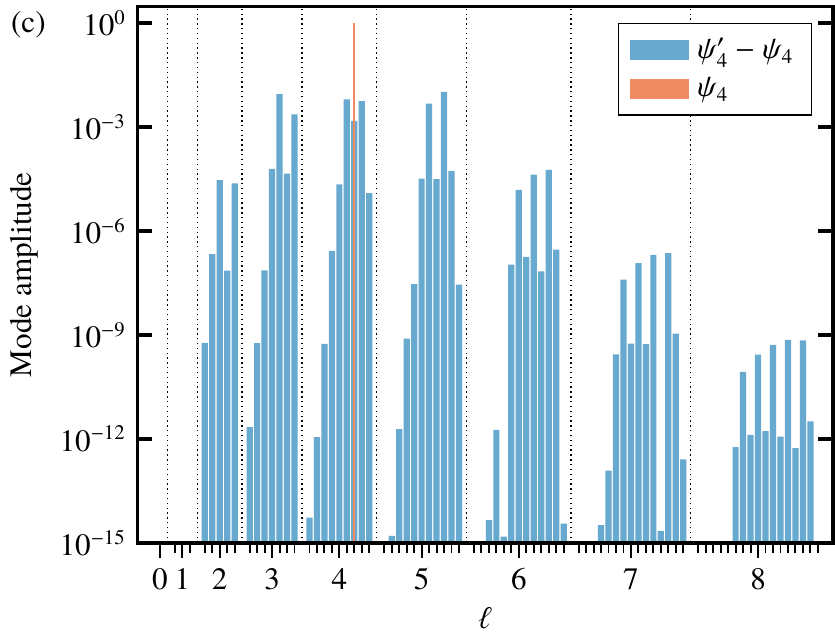}
  }%
  \hfill%
  \subfloat{%
    \label{fig:ModeTranslation-4_2_z}%
    \includegraphics{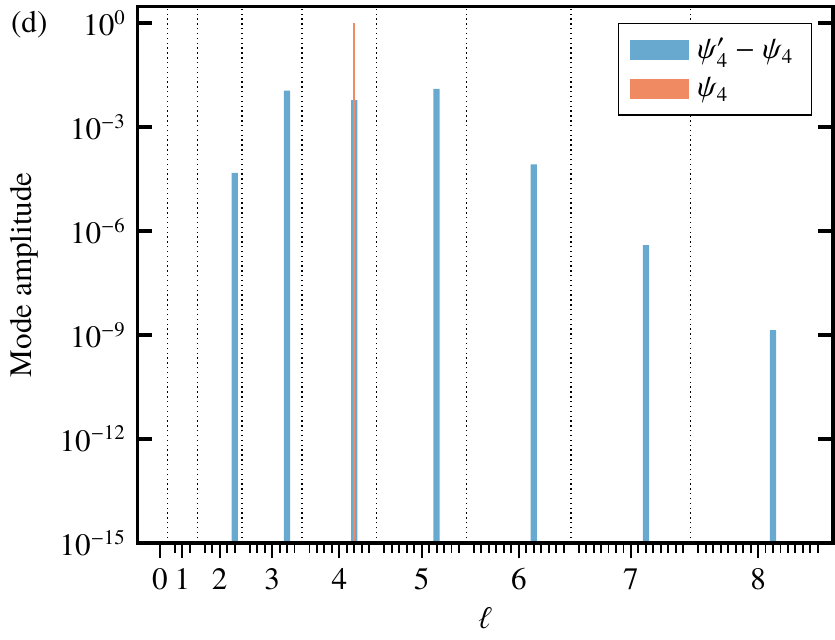}
  }%
  % \tikzset{external/force remake=false}
  \caption{ \label{fig:ModeTranslation} %
    \CapName{Mode transformations under translation} These plots show
    the changes to the amplitudes of the waveform modes when the
    system is translated.  The modes are grouped by $\ell$ value, with
    individual $m$ values increasing from $-\ell$ on the left to
    $\ell$ on the right in each group.  In each case, the initial
    waveform $\asymptotic{\psi}_{4}$ is made up of a single mode, as
    in Eq.~\eqref{eq:test_waveform}.  In the two upper panels,
    \protect\subref{fig:ModeTranslation-2_2_x} and
    \protect\subref{fig:ModeTranslation-2_2_z}, the nonzero mode is
    $(\ellnonzero, \mnonzero) = (2, 2)$; in the two lower panels,
    \protect\subref{fig:ModeTranslation-4_2_x} and
    \protect\subref{fig:ModeTranslation-4_2_z}, the nonzero mode is
    $(\ellnonzero, \mnonzero) = (4, 2)$.  These waveforms are then
    transformed to $\asymptotic{\psi}_{4}'$ by a translation of
    magnitude $0.1M$.  The panels on the left,
    \protect\subref{fig:ModeTranslation-2_2_x} and
    \protect\subref{fig:ModeTranslation-4_2_x}, depict a translation
    in the $x$ direction (the same translation in the $y$ direction
    would look almost identical here); the panels on the right,
    \protect\subref{fig:ModeTranslation-2_2_z} and
    \protect\subref{fig:ModeTranslation-4_2_z}, depict a translation
    in the $z$ direction.  We see that a translation in the $x$
    direction tends to move power into modes with a wide variety of
    $(\ell, m)$ values, whereas a translation in the $z$ direction
    only moves power into modes with the same $m$ values as the
    original waveform.  This is a result of the fact that the
    simulated waveform is effectively rotating about the $z$ axis, so
    a $z$ translation preserves a certain amount of symmetry, whereas
    the $x$ translation violates that symmetry.  As explained in the
    text, the power leakage is roughly given by powers of the product
    of displacement and frequency, which is roughly
    $0.03$ % Variable parameter value!
    in this case.  The frequency was chosen to be typical of
    frequencies seen just prior to the merger stage of comparable-mass
    binaries.  Earlier during the inspiral portion, the frequencies
    will be an order of magnitude smaller, and the size of these
    effects correspondingly smaller.  %
  }
\end{figure*}
%%%%%%%%%%%%%%%%%%%%%%%%%%%%%%%%%%%%%%%%%%%%%%%%%%%%%%%%%%%%%%%%%%%%%%

To be precise, let us choose the nonzero mode
$(\ellnonzero, \mnonzero)$ and define our model waveform by its modes
as
\begin{equation}
  \label{eq:test_waveform}
  \asymptotic{\psi}_{4}^{\ell, m}(\scritime) =
  \delta^{\ell,m}_{\ellnonzero, \mnonzero} \e^{\i\, \omega\, \scritime}.
\end{equation}
Here, $\omega$ represents an angular velocity.  For purposes of
illustration, let us choose $\omega=0.3\frac{1}{M}$, which is a
% Variable parameter value!
typical value for the $(\ell,m) = (2,2)$ mode of comparable-mass
binaries just before merger, where $M$ is the total mass of the
system.

As a first, example, we see the effect of translations in
Fig.~\ref{fig:ModeTranslation}.  The four cases shown here correspond
to $(\ellnonzero, \mnonzero) = (2, 2)$ or
$(\ellnonzero, \mnonzero) = (4, 2)$, and translations of
$\st = 0.1M\, \sin\theta\, \cos\phi$ or $\st = 0.1M\, \cos\theta$.
These are displacements of $0.1M$ in the $x$ and $z$ directions,
respectively, corresponding to typical displacements found in the
publicly available catalog of waveforms from the SXS
collaboration~\cite{Mroue2013}, as will be discussed further in
Sec.~\ref{sec:RemovingDriftFromNumericalWaveforms}.  The $z$
displacement evidently has a very simple effect on the modes; power is
transferred to all other $\ell$ modes with $m = \mnonzero$, where the
transferred power goes roughly as $\epsilon^{\abs{\ell-\ellnonzero}}$
for some parameter $\epsilon \approx 0.01$.  This simplicity is a
% Variable parameter value!
result of the fact that the waveform of Eq.~\eqref{eq:test_waveform}
is effectively rotating about the $z$ axis, so a translation along
that direction preserves a great deal of the symmetry of the system.
A similar but far more complicated pattern can be seen in the $x$
translations, where now the power is transferred into essentially all
modes.  And while there is a similar dependence in $\ell$---where the
coupling seems to get smaller exponentially with
$\abs{\ell - \ellnonzero}$---there is a more complicated dependence on
$m$.

These patterns can be understood by looking at the effect of a
translation on the time coordinate.  Kelly and Baker~\cite{Kelly2013}
pointed out that the effect on $\asymptotic{\psi}_{4}$ of a
supertranslation $\st$ (without any accompanying boost or rotation)
can typically be approximated by the first few terms of the
Taylor-series expansion\footnote{The term
  $-\st\, \partial / \partial u$ in
  Eq.~\eqref{eq:approx_effect_of_supertranslation_general} is the
  generator of the supertranslation $\st$, as described at the end of
  Sec.~\ref{sec:complete-bms-group} and in Ref.~\cite{Sachs1962}.
  Thus, this equation is simply the exponentiation of that element of
  the Lie algebra $\bms$, which gives us the corresponding element of
  the Lie group BMS.  It must be noted, however, that such
  exponentiation is not typically a sufficient method for transforming
  waveform data.  For example, Eq.~\eqref{eq:psi_4-prime} shows that
  we generally also have a factor
  $\e^{-2\i\SpinPhase} / \asymptotic{k}^{3}$ in the transformation
  law.  In our particular case, this factor happens to be $1$, which
  is why exponentiation works.  More generally, the generator of a
  boost would not supply the correct factor of $\asymptotic{k}$.  But
  even for supertranslations, exponentiation would fail to correctly
  transform other quantities.  For example, in transforming
  $\asymptotic{h}$ [Eq.~\eqref{eq:h-prime}], the term
  $-\bar{\asymptotic{\eth}}^{2} \st$ would not appear.  A more extreme
  example is provided by $\asymptotic{\psi}_{0}$
  [Eq.~\eqref{eq:psi_0-prime}]; action of the $\bms$ generators would
  fail to supply the terms $\asymptotic{\psi}_{1}$ through
  $\asymptotic{\psi}_{4}$.  Moreover, because of the infinite nature
  of this expansion, it may be useful for gaining qualitative insight
  into the approximate coupling between modes, but it is not useful
  for accurate implementation of these transformations.} %
\begin{subequations}
  \label{eq:approx_effect_of_supertranslation}
  \begin{equation}
    \label{eq:approx_effect_of_supertranslation_general}
    \asymptotic{\psi}_{4}'(\scritime', \theta', \phi') =
    \sum_{j=0}^{\infty} \frac{1}{j!} \left( - \st(\theta, \phi)\,
      \frac{\partial}{\partial \scritime} \right)^{j}
    \asymptotic{\psi}_{4}(\scritime, \theta, \phi), 
  \end{equation}
  where $(\theta, \phi) = (\theta', \phi')$.  With
  Eq.~\eqref{eq:test_waveform}, this specializes to
  \begin{equation}
    \label{eq:approx_effect_of_supertranslation_specific}
    \asymptotic{\psi}_{4}'(\scritime', \theta', \phi') =
    \sum_{j=0}^{\infty} \frac{1}{j!} \Big( - \i\, \omega\, \st(\theta,
    \phi) \Big)^{j} \mTwoYlm{\ellnonzero, \mnonzero}(\theta, \phi)\,
    \e^{\i\, \omega\, \scritime}.
  \end{equation}
\end{subequations}
In each case shown in Fig.~\ref{fig:ModeTranslation}, $\st$ is an
$\ell=1$ function, and thus couples with
$\mTwoYlm{\ellnonzero, \mnonzero}$ to progressively higher
orders---and hence at larger ``distances'' from
$(\ellnonzero, \mnonzero)$, in some sense---with increasing values of
the summation index $j$.  On the other hand, these couplings also
include progressively higher powers of $\omega$ times the amplitude of
$\st$, roughly $0.03$, and thus progressively smaller amplitudes.
% Variable parameter value!
Besides the factor of $1/j!$, there are further geometric factors
involved in the normalization of the spin-weighted spherical
harmonics, which means that the ratios of power in the various modes
do not follow a particularly simple pattern, but it is clear that
these considerations lead to the correct qualitative behavior
and---when accounting for the factorial and geometric factors---the
correct quantitative behavior.

%%%%%%%%%%%%%%%%%%%%%%%%%%%%%%%%%%%%%%%%%%%%%%%%%%%%%%%%%%%%%%%%%%%%%%
\begin{figure*}%[!thb]
  % \tikzset{external/force remake}
  \subfloat{%
    \label{fig:ModeBoost-2_2_x}%
    \includegraphics{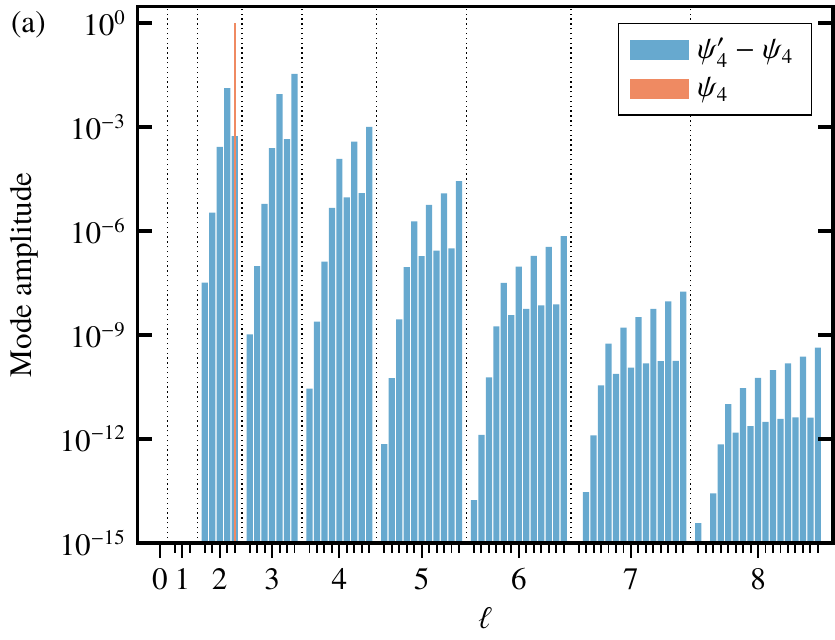}
  }%
  \hfill%
  \subfloat{%
    \label{fig:ModeBoost-2_2_z}%
    \includegraphics{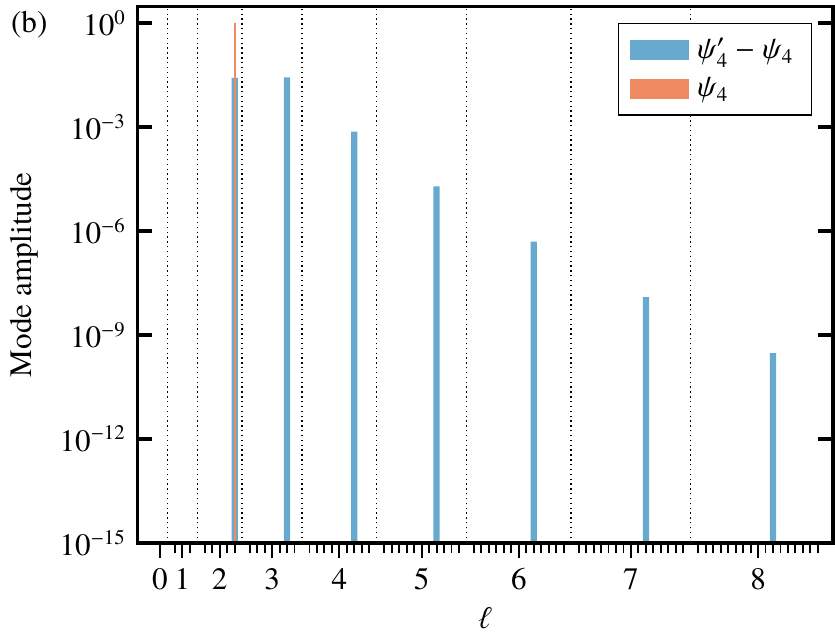}
  }%
  \\[10pt] %
  \subfloat{%
    \label{fig:ModeBoost-4_2_x}%
    \includegraphics{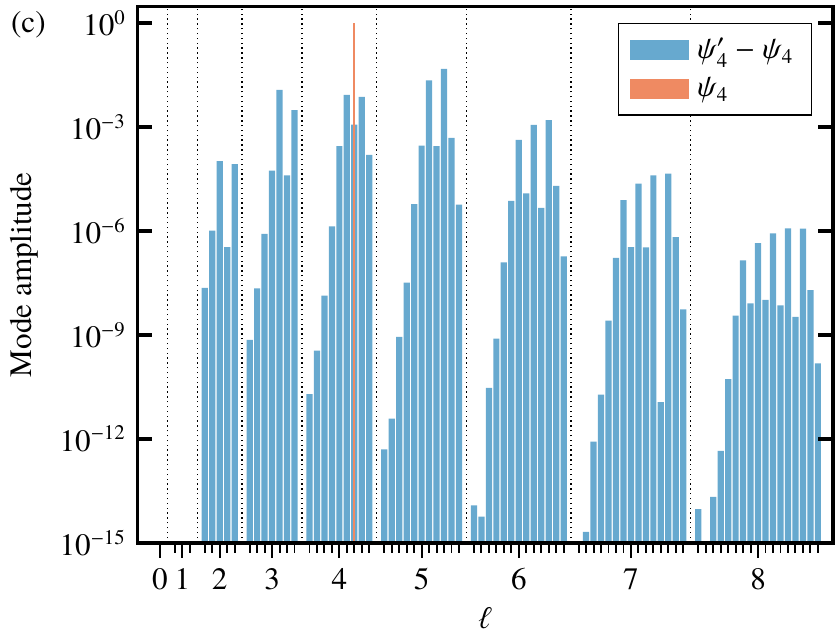}
  }%
  \hfill%
  \subfloat{%
    \label{fig:ModeBoost-4_2_z}%
    \includegraphics{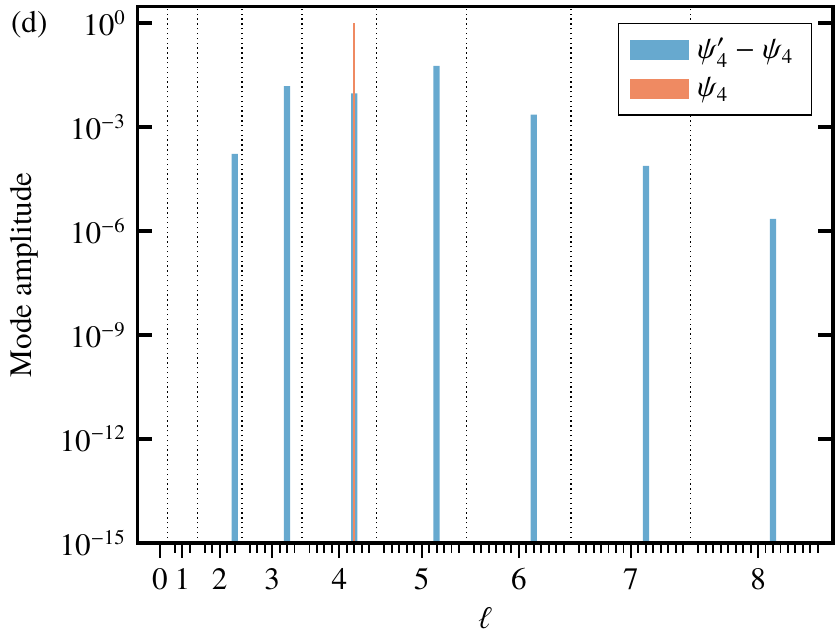}
  }%
  % \tikzset{external/force remake=false}
  \caption{ \label{fig:ModeBoost} %
    \CapName{Mode transformations under boost} These plots show the
    changes to the amplitudes of the waveform modes when the system is
    boosted.  These are similar to the plots in
    Fig.~\ref{fig:ModeTranslation}, except that there is no
    translation, only a boost.  Again, the initial waveform in the
    upper panels has nonzero mode $(\ellnonzero, \mnonzero) = (2, 2)$;
    in the two lower panels, the nonzero mode is
    $(\ellnonzero, \mnonzero) = (4, 2)$.  In this case, the two panels
    on the left depict a boost of speed $0.01c$ in the $x$ direction,
    % Variable parameter value!
    while the two panels on the right depict a boost of speed $0.01c$
    % Variable parameter value!
    in the $z$ direction.  The mode amplitudes change as a function of
    time in this case, because the boost is essentially a
    time-dependent supertranslation, as described at the end of
    Sec.~\ref{sec:complete-bms-group}.  The quantities shown here
    correspond to $\scritime=\scritime'=0$, so that the translation
    induced by the boost is actually zero.  Despite the difference in
    their construction, these results are qualitatively very similar
    to those of Fig.~\ref{fig:ModeTranslation}.  However, the boost
    used to produce these figures is orders of magnitude greater than
    any found in the numerical waveforms considered below.  The basic
    conclusion is that the boost matters in those cases only to the
    extent that even a small boost can induce a significant
    translation over the course of a long simulation. %
  }
\end{figure*}
%%%%%%%%%%%%%%%%%%%%%%%%%%%%%%%%%%%%%%%%%%%%%%%%%%%%%%%%%%%%%%%%%%%%%%

The nonzero input waveform mode in each of these cases has amplitude
$1$.  Of course, the effect of mode mixing is linear, so the plots in
Fig.~\ref{fig:ModeTranslation} should essentially be read as
fractional coupling between the modes.  For example, in
Fig.~\ref{fig:ModeTranslation-2_2_x}, we see that a little more than
\SI{1}{\percent} of the power in
% Variable parameter value!
the $(2,2)$ mode is mixed into the $(2,1)$ and $(3,3)$ modes.  But in
many cases, the physical $(2,2)$ mode is strongly dominant over either
of these modes, so that the expected ratio of amplitudes would be less
than \SI{1}{\percent}.  In such cases, the measured $(2,1)$ and
$(3,3)$ modes would
% Variable parameter value!
actually be primarily made up of power leaking from the $(2,2)$ mode.
And while we might typically expect the frequency of the $(2,1)$ mode
in real binary systems to be roughly $1/2$ that of the $(2,2)$ mode,
and the frequency of the $(3,3)$ mode to be roughly $3/2$ that of the
$(2,2)$ mode, the frequency of the mixed component would be nearly the
same as that of the $(2,2)$ mode.  Taken together, these features
can provide a signature of mixing due to transformations.

As we have seen, translations comprise a special case of
supertranslations having $\ell=1$.  Similar behavior results from
supertranslations with $\ell>1$, except that the coupling between
modes is more extensive.  For example, if the original waveform has
nonzero mode $(\ellnonzero, \mnonzero) = (2, 2)$, a supertranslation
with nonzero $(\ell,m)=(2,0)$ component couples power at a roughly
equal level into both the $(3,2)$ and $(4,2)$ modes of the transformed
waveform.  It should also be noted that supertranslations with
$\ell>1$ can directly alter the value of, for example, the strain
waveform $\asymptotic{h}$ through the term
$-\bar{\asymptotic{\eth}}^{2} \st$ in Eq.~\eqref{eq:h-prime}.  The
operator $\bar{\asymptotic{\eth}}^{2}$ eliminates the modes of $\st$
with $\ell \leq 1$, as it must for a field of spin weight $s=-2$.

We can make a similar comparison of waveform modes before and after a
boost.  Figure~\ref{fig:ModeBoost} shows essentially the same thing as
Fig.~\ref{fig:ModeTranslation}, except that instead of translations,
the waveforms have been subjected to boosts.  The speed of the boost
is $\beta = 0.01c$ in each case, directed in either the $x$ or $z$
% Variable parameter value!
direction.  The most obvious feature here is the remarkable similarity
between Figs.~\ref{fig:ModeTranslation} and~\ref{fig:ModeBoost}.  The
coupling due to translation falls off more quickly with increasing
distance from the dominant mode, but the general patterns are very
similar.  The time has been chosen as $\scritime = \scritime' = 0$, so
that only the boost itself factors into this transformation.  This
means that the translation induced by the boost, as discussed near the
end of Sec.~\ref{sec:complete-bms-group}, is zero.  In this case, only
the movement of the points around the sphere---as depicted in
Fig.~\ref{fig:BoostedGrids}---comes into the transformation.  In
particular, the transformation law given by Eq.~\eqref{eq:psi_4-prime}
is
\begin{equation}
  \label{eq:psi_4-prime-explicitly}
  \asymptotic{\psi}_{4}'(0, \theta', \phi') = \e^{-2\i \SpinPhase}
  \gamma^{3}\, (1 - \fourvec{v} \cdot \directionvec)^{3}\,
  \asymptotic{\psi}_{4}(0, \theta', \phi').
\end{equation}
The spin factor $\e^{-2\i\SpinPhase}$ is primarily just transforming the
tangent bases to avoid singularities in the basis vector fields, and
has no strong effect on the modes in this case.  The value of
$\gamma^{3}$ is approximately $1.00015$, which accounts for the change
% Variable parameter value!
to the $(2,2)$ mode, to reasonable accuracy.  The third factor
multiplies the waveform by roughly
$1-3\, \beta\, \sin\theta\, \cos\phi$ for the boost in the $x$
direction, and roughly $1 - 3\, \beta\, \cos\theta$ for the boost in
the $z$ direction.  But this factor can only explain coupling between
modes with $\Delta \ell = \pm 3$, whereas we clearly see more
extensive coupling in Fig.~\ref{fig:ModeBoost}.  In fact, if we expand
the boost rotor of Eq.~\eqref{eq:boost-rotor} in powers of $\beta$,
and use that to expand the argument of
$\asymptotic{\psi}_{4}(0, \theta', \phi')$ in a Taylor series, we find
another factor multiplying $\asymptotic{\psi}_{4}(0, \theta, \phi)$:
\begin{multline}
  \label{eq:geometric_factor}
  1 - \abs{\fourvec{v} \times \directionvec} + \frac{1}{2}
  \abs{\fourvec{v} \times \directionvec}\, \fourvec{v}
  \cdot \directionvec + \dots \\
  =
  \begin{cases}
    1 - \beta \sin\phi + \frac{1}{4}\, \beta^{2} \sin 2\phi\,
    \sin\theta + \dots
    & \text{$x$ boost,} \\
    1 - \beta \sin\theta + \frac{1}{4}\, \beta^{2} \sin 2\theta +
    \dots
    & \text{$z$ boost.}
  \end{cases}
\end{multline}
Because of the geometry, the largest couplings from this factor are
typically several times smaller than the couplings from the
$(1 - \fourvec{v} \cdot \directionvec)^{3}$ factor.  That is, the
largest peaks in Fig.~\ref{fig:ModeBoost} will be dominated by the
$3\, \beta$ term, but smaller peaks with $\abs{\Delta \ell} > 3$ will
be dominated (and in fact made possible) by the more complicated
factor of Eq.~\eqref{eq:geometric_factor}.

In the plots of Fig.~\ref{fig:ModeBoost}, $3\, \beta = 0.03$, which is
% Variable parameter value!
also the approximate scale of the effects of the translation for the
plots of Fig.~\ref{fig:ModeTranslation}.  This explains why the
magnitude of the coupling is so similar in the two cases, at least for
the dominant coupling terms.  Of course, there is no $1/j!$ term for
the boost couplings, as in
Eqs.~\eqref{eq:approx_effect_of_supertranslation}.  This explains why
the couplings in Fig.~\ref{fig:ModeTranslation} should fall off so
much faster than those in Fig.~\ref{fig:ModeBoost}.

In fact, this numerical equality between the couplings for
translations and boosts is the reason $\beta = 0.01c$ was chosen for
% Variable parameter value!
these examples, to ease comparison between the effects of a
translation and of a boost.  But we must note that this value was
chosen entirely for the purpose of illustration; it is an order of
magnitude larger than the largest speed found in the SXS catalog
discussed below, and several orders of magnitude larger than typical
speeds.  This might appear to suggest that the effect of the boost
itself is entirely negligible for those simulations.  However, we have
thus far only described the transformation due to a boost on the
$\scritime = \scritime' = 0$ slice.  At any later time $\scritime'$,
an additional coupling is present, which is essentially identical to a
translation by $\gamma\, \scritime'\, \fourvec{v}$, as we can see from
the arguments toward the end of Sec.~\ref{sec:complete-bms-group}.
Even a very small boost can build up to a significant translation over
the course of a long simulation.  In fact, we will find that near
merger, boosts play a significantly more important role than
translations in the SXS catalog.

%%%%%%%%%%%%%%%%%%%%%%%%%%%%%%%%%%%%%%%%%%%%%%%%%%%%%%%%%%%%%%%%%%%%%%
%%%%%%%%%%%%%%%%%%%%%%%%%%%%%%%%%%%%%%%%%%%%%%%%%%%%%%%%%%%%%%%%%%%%%%
\section{Removing drift from numerical waveforms}
\label{sec:RemovingDriftFromNumericalWaveforms}

%%%%%%%%%%%%%%%%%%%%%%%%%%%%%%%%%%%%%%%%%%%%%%%%%%%%%%%%%%%%%%%%%%%%%%
\begin{figure}
  % \tikzset{external/force remake}
  \includegraphics{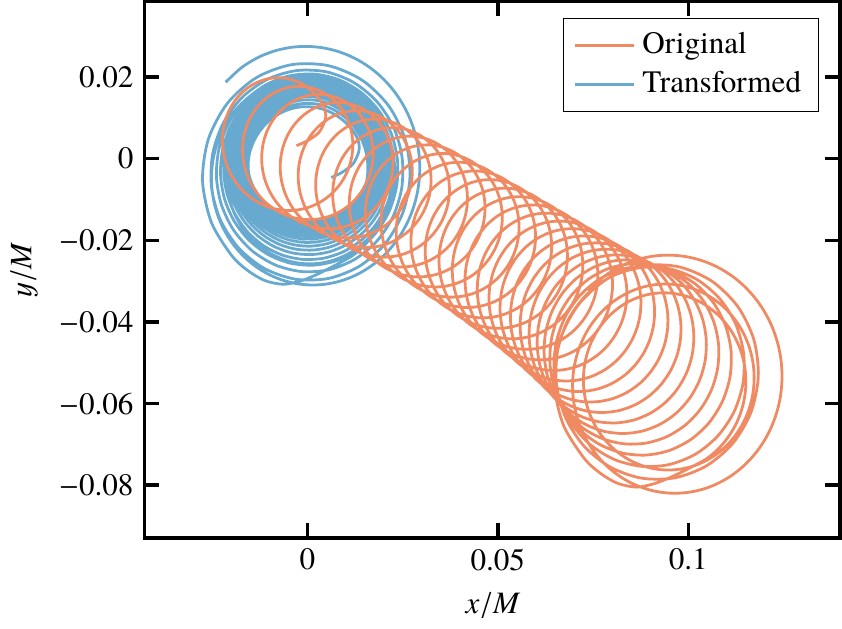}
  % \tikzset{external/force remake=false}
  \caption{ \label{fig:CoM} %
    \CapName{Center of mass motion} This plot shows the coordinates in
    the $x$-$y$ plane of the center of mass for system
    \texttt{SXS:BBH:0004}, throughout the inspiral of the system.
    Motion in the $z$ direction is far smaller.  The coordinates are
    given in units of $M = M_{1} + M_{2}$, the total Christodoulou
    mass~\cite{Christodoulou1970} of the system.  The raw data from
    the simulation results in the curve labeled ``Original''.  This is
    the motion of the center of mass, as seen in the same frame in
    which the waveforms are measured.  There is a small initial
    offset, as well as a strong drift velocity.  We can also apply a
    spatial translation and boost to the system, in which case the
    center of mass appears to rotate more simply around the origin, as
    seen in the curve labeled ``Transformed''. %
  }
\end{figure}
%%%%%%%%%%%%%%%%%%%%%%%%%%%%%%%%%%%%%%%%%%%%%%%%%%%%%%%%%%%%%%%%%%%%%%

%%%%%%%%%%%%%%%%%%%%%%%%%%%%%%%%%%%%%%%%%%%%%%%%%%%%%%%%%%%%%%%%%%%%%%
\begin{figure*}
  % \tikzset{external/force remake}
  \subfloat{%
    \label{fig:waveform-raw}%
    \includegraphics{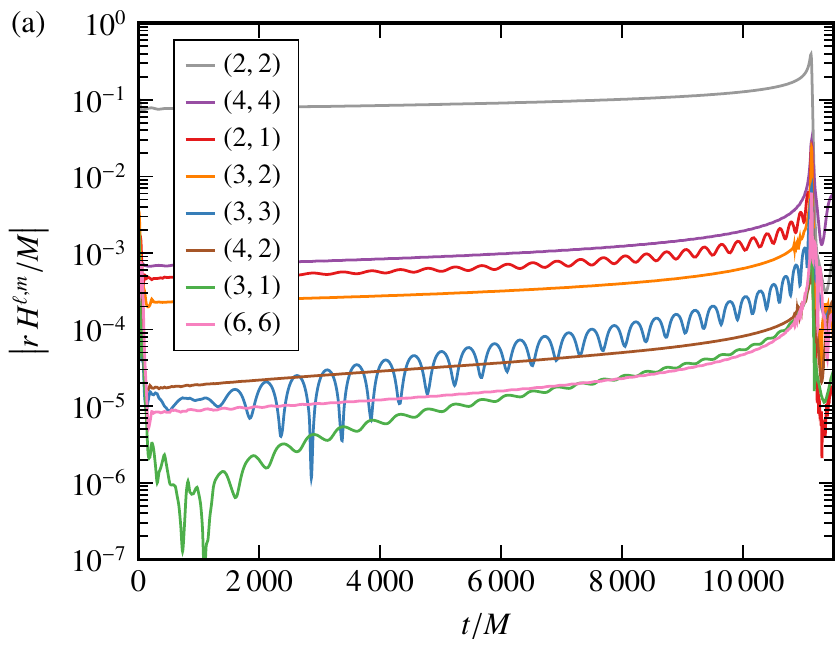}%
  }%
  \hfill%
  \subfloat{%
    \label{fig:waveform-corrected}%
    \includegraphics{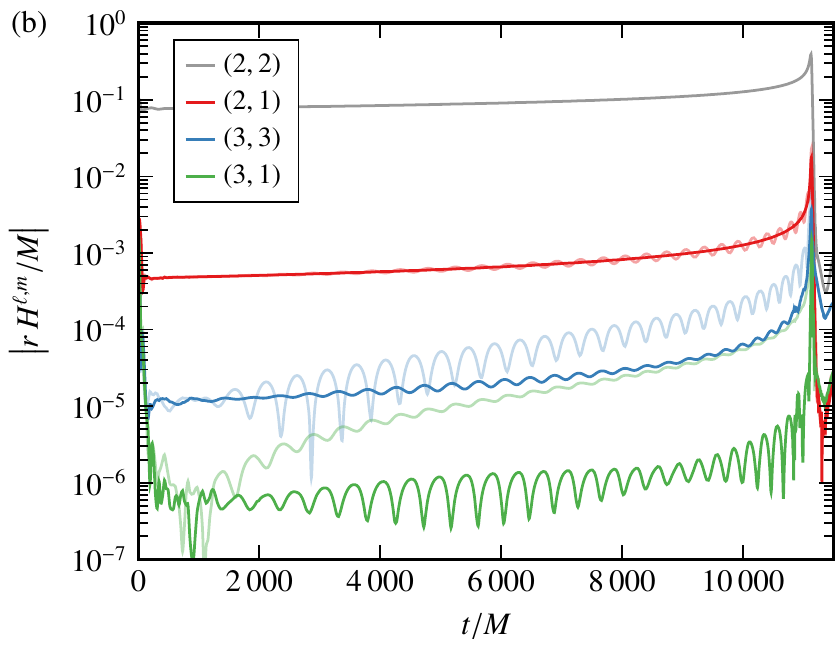}%
  }%
  % \tikzset{external/force remake=false}
  \caption{ \label{fig:waveform} %
    \CapName{Original and corrected waveform data} These plots show
    the waveform data for \texttt{SXS:BBH:0004}, and the effects of
    the transformation.  The plot on the left,
    \protect\subref{fig:waveform-raw}, shows the raw waveform obtained
    from the SXS catalog, using extrapolation with polynomials of
    order $N=3$~\cite{Boyle2009, Taylor2013}.  The eight most
    significant modes are shown.  Of these, the $(2,1)$, $(3,3)$, and
    $(3,1)$ modes exhibit substantial oscillations.  Oscillations are
    somewhat unexpected because analytical models of this system
    contain no such features.  However, these are also the modes that
    couple most strongly to the dominant $(2,2)$ mode under a
    translation in the $x$-$y$ plane, as shown in
    Fig.~\ref{fig:ModeTranslation-2_2_x}.  The plot on the right,
    \protect\subref{fig:waveform-corrected}, shows the $(2,2)$ mode
    and the three oscillating modes (removing the other modes for
    clarity) after the waveform has been transformed by the
    translation and boost given in
    Eqs.~\eqref{eq:0004-transformation}.  For comparison, the original
    modes are also shown in the corresponding colors, with lower
    opacity.  The effect of the transformation is largest near merger,
    when the translation induced by the boost is largest and the
    frequency is highest.  Despite the crude way in which the
    transformation parameters were determined, the transformation
    itself eliminates the oscillations of the $(2,1)$ mode, while
    reducing the overall amplitudes of the $(3,3)$ and $(3,1)$ modes
    by as much as a factor of $20$.  %
  }
\end{figure*}
%%%%%%%%%%%%%%%%%%%%%%%%%%%%%%%%%%%%%%%%%%%%%%%%%%%%%%%%%%%%%%%%%%%%%%

To demonstrate one way in which BMS transformations are important at a
practical level, we examine the publicly available catalog of
simulations from the SXS collaboration~\cite{Mroue2013, sxs_catalog}.
First, we will illustrate a particular system to see unexpected
effects in its waveform modes, and see how these effects can be
reduced by applying a spatial translation and a boost derived from the
simulation data.  Then, we will briefly examine the size of the
translation and boost for other simulations in the catalog.

The first simulation we consider is labeled in the waveform catalog as
\href{\SXSwaveforms/data/DisplayMetadataFile.php/?id=SXS:BBH:0004}%
{\texttt{SXS:BBH:0004}}, and represents an (approximately) equal-mass
system in which one black hole has dimensionless spin
$S_{1}/M_{1}^{2} = 0.5$ along the $-z$ axis, while the other black
hole has no spin.  This system is interesting because it is not
precessing, and so retains enough symmetry to allow us to
unambiguously identify some curious features.  But it is nonetheless
not perfectly symmetric, and thus exhibits those nontrivial features.

We can see the first example of nontrivial features in this system by
simply plotting the center of mass.  Using the Christodoulou masses
and coordinate positions of the black holes, we form the usual center
of mass.\footnote{These quantities are all stored in the waveform
  catalog in files named \texttt{Horizons.h5}.}  The result is plotted
in Fig.~\ref{fig:CoM}.  Because the system is not symmetric, we expect
to see some asymmetry in the emission of gravitational waves in the
orbital ($x$-$y$) plane~\cite{Blanchet2006a, Boyle2014}, and thus some
force in this plane.  But that force should have roughly constant
magnitude on the orbital timescale, and should simply rotate with the
system.  So we expect the center of mass to be pushed around in a
circle.  This is essentially what we find in the data.  The center of
mass starts nearly at the origin, so this circle is initially not
centered on the origin.  But there is another strong effect: an
overall drift.  Evidently, this drift is due to residual linear
momentum in the initial data.  For future evolutions,
Ref.~\cite{Ossokine2015} introduced a method to eliminate such
residual momentum from the initial data.  However, for the present
waveform catalog and any future simulations in which a large initial
translation is present, or a significant recoil develops during the
inspiral, we must transform the data to eliminate the offset and
drift.

The approach taken here is crude, but will serve the purpose of
illustration.  By minimizing the average distance between the center
of mass and the origin, we can find the optimal translation and boost,
as described in Appendix~\ref{sec:estim-transl-boost}.  For this
system, the results are
\begin{subequations}
  \label{eq:0004-transformation}
  \begin{gather}
    \delta \fourvec{x}
    = \left(-9.1 \times 10^{-3}, 7.8 \times 10^{-3}, -4.0 \times
      10^{-9} \right), \\
    % [-0.00914851263322, 0.007805991613215365,
    % -4.035633351434781e-09]
    \fourvec{v}
    = \left( 9.4 \times 10^{-6}, -5.3 \times 10^{-6}, 2.6 \times
      10^{-12} \right).
    % [9.3837263335e-06, -5.30182283777258e-06, 2.645692680506991e-12]
  \end{gather}
\end{subequations}
Over the course of this $\roughly \num{11000}\,M$ simulation, the
small boost grows into a larger translation than the initial offset
$\delta \fourvec{x}$.  Applying this transformation to the center of
mass we see a much cleaner-looking curve, essentially orbiting the
origin, in Fig.~\ref{fig:CoM}.  Although the center of mass measured
in this way is based on coordinates, and thus susceptible to all the
vagaries of gauge in the most extreme regions of the simulated
spacetime, we will nonetheless find that the same transformation
applied to the waveform removes features that we would not expect
based on naive analytical models.

%%%%%%%%%%%%%%%%%%%%%%%%%%%%%%%%%%%%%%%%%%%%%%%%%%%%%%%%%%%%%%%%%%%%%%
\begin{figure*}%[!thb]
  % \tikzset{external/force remake}
  \subfloat{%
    \label{fig:CatalogCoM_i}%
    \includegraphics{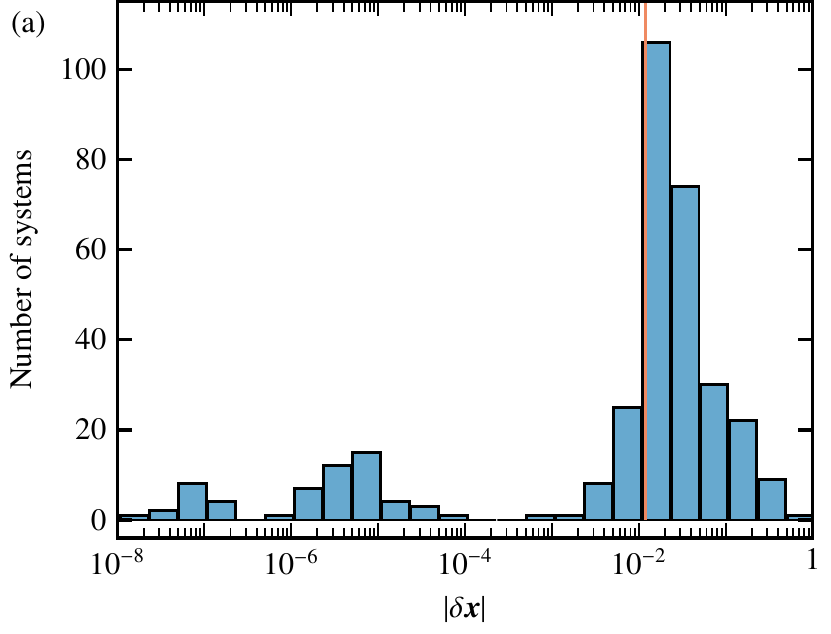}
  }%
  \hfill%
  \subfloat{%
    \label{fig:CatalogCoM_f}%
    \includegraphics{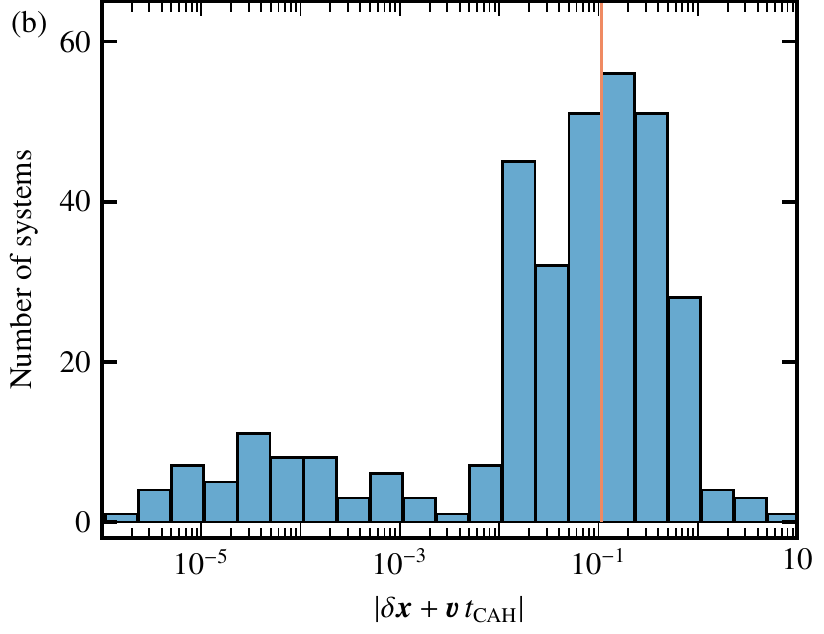}
  }%
  % \tikzset{external/force remake=false}
  \caption{ \label{fig:CatalogCoM} %
    \CapName{Survey of translations and boosts in the SXS catalog}
    These plots show the number of systems with a given initial
    displacement, \protect\subref{fig:CatalogCoM_i}, and a given
    displacement at merger, \protect\subref{fig:CatalogCoM_f}.  In the
    latter plot, $t_{\text{CAH}}$ is the time at which a common
    apparent horizon is found, which is a typical definition of the
    merger time.  The data include every system in the SXS catalog,
    but only the highest-resolution instance of each system.  The
    vertical red line in each plot shows the value for the system
    \texttt{SXS:BBH:0004}, which is the one described in
    Sec.~\ref{sec:RemovingDriftFromNumericalWaveforms} and
    Figs.~\ref{fig:CoM} and~\ref{fig:waveform}.  Systems with small
    values in each plot are typically simple systems with little or no
    spin, or high symmetry; larger values generally indicate
    asymmetries in the masses, unequal spins, and especially strongly
    precessing systems.  The displacement at merger is dominated in
    most cases by the translation due to boost, rather than initial
    displacement---though the boost actually \emph{reduces} the
    displacement from its initial value in roughly one fifth of the
    systems in the catalog.  %
  }
\end{figure*}
%%%%%%%%%%%%%%%%%%%%%%%%%%%%%%%%%%%%%%%%%%%%%%%%%%%%%%%%%%%%%%%%%%%%%%

Figure~\ref{fig:waveform-raw} shows the largest modes\footnote{This is
  as measured shortly before merger.  We ignore modes with negative
  $m$ values because, for this system, they have essentially the same
  magnitudes as their counterparts with positive $m$ values.  Also,
  the $(2,0)$ mode comes out above the $(4,2)$ mode when extrapolating
  with $N=3$ polynomials, but is evidently not
  trustworthy~\cite{Taylor2013}, so we ignore it.} %
in the waveform.  This is the original data taken from the SXS
catalog.  The $(2,2)$ mode is entirely dominant, as expected.  The
post-Newtonian model~\cite{Blanchet2006a} of this system predicts
smooth, monotonic waveform amplitudes during the inspiral.  Yet the
$(2,1)$, $(3,3)$, and $(3,1)$ modes exhibit distinctive oscillations
that are not visible in the other modes.  These modes also have the
largest couplings with the $(2,2)$ mode under translation in the
$x$-$y$ plane, as seen in Fig.~\ref{fig:ModeTranslation-2_2_x}, and
the oscillations are at the same frequency as the $(2,2)$ mode.  These
facts strongly suggest that the oscillations are caused by mode
coupling due to the motion of the center of mass.  In fact, we can
even predict the size of these couplings.  The original system ends up
translated from the origin by about $0.1M$ at merger, and the
frequency of the $(2,2)$ mode just prior to merger exceeds $0.3/M$.
These were the parameters used to construct
Fig.~\ref{fig:ModeTranslation-2_2_x}, which means that the couplings
shown in that plot should be roughly the couplings found in this
waveform near merger.  Specifically, we expect to find mode couplings
in this waveform starting at just over \SI{1}{\percent} of the
magnitude of the $(2,2)$ mode near merger.

Figure~\ref{fig:waveform-corrected} shows the dominant mode---which is
not visibly changed at this scale---and the oscillating modes after
the transformation of Eq.~\eqref{eq:0004-transformation} has been
applied to the waveform.  For comparison, the original modes are shown
in the same colors with lower opacity.  In each case, the effect of
the transformation is smallest at the beginning of the simulation,
when the offset and frequency are smallest; conversely, it is largest
at merger, when the offset and frequency are largest.  This is just as
we would expect, given the arguments of
Sec.~\ref{sec:underst-effects-bms}.  Moreover, we can look at the
changes to these modes as a fraction of the $(2,2)$ amplitude, and
find that they do agree nicely with
Fig.~\ref{fig:ModeTranslation-2_2_x}: the $(2,1)$ and $(3,3)$ modes
change by just over \SI{1.3}{\percent} at merger, and the $(3,1)$ mode
by around \SI{0.35}{\percent}.  Finally, we can subtract the
transformed waveform from the original data and measure the frequency
of the difference; for each of these three modes, we find that it
matches the frequency of the $(2,2)$ mode, rather than the frequency
of the transformed mode.  These facts all suggest that the changes to
these three modes are primarily undoing leakage of the $(2,2)$ mode.

The oscillations have been essentially removed from the $(2,1)$ mode.
This mode is the third-largest overall, after the $(2,2)$ and $(4,4)$
modes.  Yet its amplitude is altered in this transformation-induced
coupling by several percent throughout the inspiral, growing to
\SI{30}{\percent} at merger (relative to the transformed values).  The
oscillations of the much smaller $(3,3)$ mode are reduced
substantially, though not entirely eliminated.  This is not
surprising, given that the change in the waveform is so large, and the
method of choosing the transformation of
Eq.~\eqref{eq:0004-transformation} so crude.  In fact small changes in
the parameters used to choose the transformation ($t_{i}$ and $t_{f}$
in Appendix~\ref{sec:estim-transl-boost}) lead to significant changes
in the smoothness of this transformed mode, suggesting that there may
be better choices.  However, the remarkable feature of this
transformation is the size of the change, which ranges up to
\SI{350}{\percent} late in the inspiral (again, relative to the
transformed values).  Even more extreme is the change in the $(3,1)$
mode, which reaches typical values over \SI{2000}{\percent} toward the
end of the inspiral.

We can conclude from this that throughout almost the entire
simulation, the $(3,3)$ and $(3,1)$ modes as given in the original
data from the SXS catalog are entirely dominated by coupling from the
$(2,2)$ mode, while the $(2,1)$ mode is strongly affected---though not
completely dominated.  This means that any attempt to use these modes
without accounting for the effect of the residual velocity in the
initial data will be prone to errors.

The mode couplings we have seen here are all caused by a very small
residual velocity, which leads to an anomalous translation of just
$0.1M$ around merger.  It may be surprising that such large effects
can follow from such a seemingly small cause.  But it is more
surprising that this anomalous translation is typical of the
simulations in the SXS catalog.  Figure~\ref{fig:CatalogCoM} shows the
initial and final displacements of the center of mass for every system
in the SXS catalog.  We can see that the size of the transformation in
\texttt{SXS:BBH:0004} is fairly typical of systems in the catalog.  In
fact, the translation near merger [Fig.~\ref{fig:CatalogCoM_f}] for
this system (\num{0.108}) is slightly above the median (\num{0.070}),
and just half the mean (\num{0.216}).

Closer inspection of the data show that all the systems with very
small translations (less than about $10^{-2}$ in
Fig.~\ref{fig:CatalogCoM_f}) are symmetric, with equal masses and
spins, and the spins are all aligned with the orbital axis.  If the
masses or spin magnitudes are not equal, there is generally a larger
translation.  Still larger translations are typically found in systems
for which one or both black holes have spin components in the orbital
plane.  On the other hand, simulations that run for longer have more
opportunity to develop a large translation; the very largest values
result from very long simulations, rather than extraordinarily
asymmetric physics.

In this section, we have found that applying translations and boosts
determined from the orbital trajectories of a simulation in the SXS
catalog can have a very large effect on the distribution of power in
the modes, and can diminish unmodeled features in the waveform.
Moreover, we have seen that this particular system is fairly typical
of systems in the SXS catalog, with numerous systems expected to
exhibit significantly larger mode couplings.  The mode transformations
described in this paper can be expected to substantially improve the
agreement between analytical waveform models and numerical waveforms
in these cases.

%%%%%%%%%%%%%%%%%%%%%%%%%%%%%%%%%%%%%%%%%%%%%%%%%%%%%%%%%%%%%%%%%%%%%%
%%%%%%%%%%%%%%%%%%%%%%%%%%%%%%%%%%%%%%%%%%%%%%%%%%%%%%%%%%%%%%%%%%%%%%

\section{Effects on data analysis for gravitational-wave detectors}
\label{sec:effects-data-analys}

The previous section showed that seemingly small transformations can
have pronounced effects on waveform modes.  Having understood the
nature and origin of these effects, we can now address the issue of
what must be done about them.  This section briefly discusses the
impact these transformations have on two aspects of detections of
gravitational waves: the production of waveform models, and the
construction of template banks for searches in detector data.

The first step in detecting a gravitational wave is to devise a model
of a waveform we might expect to find in the data.  From astrophysical
arguments, the most reliable candidates for detection are mergers of
compact binary systems.  Because of the nonlinear nature of mergers,
they can only be modeled accurately by computers.  On the other hand,
using a computer to generate the entire signal is simply unrealistic
for most of the expected systems~\cite{Kumar2014}.  Thus, at some
level, waveform modeling must combine numerical and analytical
results.  But because the waveforms come from different approaches, we
should expect to find differences in their gauges---as amply
demonstrated by the boosts and translations discovered in the SXS
data, which will not naturally appear in any analytical model.

These gauge differences will have real impacts on any model that uses
numerical waveforms.  For example, when ``calibrating''
effective-one-body models~\cite{Taracchini2014, Damour2014C,
  nagar2015}, the analytical waveform must be aligned to the numerical
waveform.  If the numerical waveform has spurious features, the
waveforms will appear to align poorly, so the calibration will be less
than optimal and result in inaccurate waveforms.  Other
phenomenological waveform models~\cite{Santamaria2010A, Damour2014D}
and surrogate models~\cite{Field2014, Blackman2015a} would experience
the same biases, trying to fit simple formulas to waveforms with
effectively random gauge effects.  Similarly, when constructing hybrid
waveform models~\cite{Kumar2014}, the hybrids will be imperfect or
even discontinuous in the region where one switches from analytical to
numerical data.

As mentioned in Sec.~\ref{sec:Introduction}, some of these gauge
freedoms---time translations and rotations---are entirely familiar,
and routinely dealt with simply by applying a gauge transformation to
one waveform to minimize some measure of the difference between the
waveforms.  In principle at least, this approach could also be
extended to the full BMS group, though the supertranslations would
obviously be represented only up to some finite spherical-harmonic
order, and the numerical implementation may be delicate.  It may also
be feasible to resolve the gauge ambiguities using any of various
methods presented in the literature~\cite{Moreschi2004, Helfer2007,
  Adamo2011, Kozameh2013A}, though it is not clear that such an
approach would be numerically feasible.  In any case, a simplistic
approach like the one found in
Sec.~\ref{sec:RemovingDriftFromNumericalWaveforms} is presumably a
helpful first step.

Now, assuming that we have a waveform model for a particular
astrophysical system produced with the appropriate care for gauge
ambiguities, searches of detector data require a family of template
waveforms---specific instances of waveforms from the broader class
making up the model.  As noted in Sec.~\ref{sec:Introduction}, the
signal measured along any simple curve on $\scriplus$ of constant
spatial coordinates is a good approximation for the signal measured by
some inertial observer, because the metric in Bondi gauge is
\emph{manifestly} asymptotically flat.  Here, we consider the signal
to be measured as a function of the retarded-time coordinate
$\asymptotic{\scritime}$ along some direction
$(\asymptotic{\theta}, \asymptotic{\phi})$, in which case the limiting
process as $\asymptotic{r} \to \infty$ is well defined and the signals
at finite radius and on $\scriplus$ can be compared meaningfully.  Of
course, since BMS transformations preserve Bondi gauge, we can apply
any BMS transformation to generate another curve on $\scriplus$, and
another corresponding waveform.  We might worry that we would need a
separate template for each member of the BMS group---or at least for
some discrete sampling of the BMS group.  But given its infinite
dimensionality, this could still be a very large or even impossible
task.

Fortunately, the situation is not quite so dire.  We only need to
generate templates for elements of the BMS group that produce
\emph{detectably distinct} waveforms.  But there are degeneracies
among the templates created in this way, particularly among the
supertranslations.  Given that our detector will lie along a single
direction from the source, the supertranslation $\st$ will be
evaluated along a particular direction.  Ignoring the Lorentz
transformation for the moment, the retarded time transforms as
$\scritime' = \scritime - \st(\theta, \phi)$.  As far as its effect on
the time variable is concerned, all those infinitely many degrees of
freedom in $\st$ reduce to a single number.  In principle, the angular
dependence of $\st$ does lead to a transformation of the quantity
$\asymptotic{\h}$ measured by a gravitational-wave detector, as shown
in Eq.~\eqref{eq:h-prime}.  However, the term
$\bar{\asymptotic{\eth}}^{2}\, \st$ is constant in time, and so is not
detectable.  Thus, for detections along a single line of sight from
the source, the entire supertranslation sector of the BMS group is
reduced to a single time offset.

Factoring the supertranslations out of the BMS group leaves us with
the familiar Lorentz group of rotations and boosts.  The rotations
determine the sky position of the detector relative to the source---or
equivalently the orientation of the source relative to the detector.
\footnote{There is still another rotation that must be accounted for
  in data analysis, related to the orientation of the detector
  relative to the source---or equivalently the sky position of the
  source relative to the detector.  This second rotation is partially
  degenerate with the first, in that it will also affect the spin
  phase.  The other two degrees of freedom in this rotation determine
  the detector's sensitivity to the signal via the ``antenna
  pattern''.  However, since this rotation relates the detector's
  orientation to the coordinates we have already been dealing with, it
  does not fit comfortably within the scope of this discussion.} %
We can further separate rotations into a rotation along the line of
sight between detector and source, and two other degrees of freedom
that we might call latitude and longitude.  This first rotation is
directly degenerate with the spin phase $\SpinPhase$ described in
Sec.~\ref{sec:RotationsAndBoosts}.  But both this and the time offset
are ``extrinsic'' parameters, already dealt with in searches by simply
finding the element of a discrete Fourier transform with the largest
magnitude~\cite{Finn1992, Brown2007, Boyle2011c}.  The remaining
rotational degrees of freedom are described in more detail
elsewhere~\cite{Capano2014, OShaughnessy2014, kyutoku2014}.  In brief,
it appears that accounting for them could provide benefits for
localization and parameter estimation, but could actually be
counterproductive for detection.  The impact of the boost degrees of
freedom is likely to be much smaller, and indistinguishable from an
error in the total mass of the system.

To summarize this section, let us reiterate how it is that
supertranslations are so important for waveform modeling, but not
important for detection.  Supertranslations are important to waveform
models for two reasons: (1) the models must be able to describe the
waveform in any direction from the source; and (2) at some point we
generally need to compare or combine two different models, so the
gauge freedom must be accounted for.  On the other hand, a detector
lies along a single direction from the source, which means that all
the degrees of freedom in the supertranslation are degenerate.  If we
had a network of detectors located in significantly different
directions from a source, and we wished to combine their information,
we would need some control over their relative time offsets which
could be considered equivalent to supertranslation degrees of freedom.
This is not expected to be a pressing concern in the near future.

%%%%%%%%%%%%%%%%%%%%%%%%%%%%%%%%%%%%%%%%%%%%%%%%%%%%%%%%%%%%%%%%%%%%%%
%%%%%%%%%%%%%%%%%%%%%%%%%%%%%%%%%%%%%%%%%%%%%%%%%%%%%%%%%%%%%%%%%%%%%%
\section{Conclusions}
\label{sec:Conclusions}

There is no such thing as a gauge-invariant gravitational waveform.
It is possible to find a gauge-\emph{fixed} waveform---for example,
one measured in the standard Bondi gauge.  However, even this is not a
particular coordinate system, but an infinite-dimensional class of
equally acceptable systems.  We can transform between members of the
class using any element of the infinite-dimensional
Bondi-Metzner-Sachs group, which shows that the class is very large
indeed.  Moreover, we have seen that such transformations can affect
the waveform dramatically, even when the transformation seems to be
small.  This means that any comparison between waveforms---whether
numerical, analytical, or even experimentally measured
waveforms---will be affected by the gauge in which the waveform is
expressed (or equivalently, the frame in which the waveform is
measured).  There is no obvious preferred frame.  Instead, all we can
(and should) do is to insist that the waveforms are at least in the
\emph{same} frame.  Doing so requires understanding the BMS group, and
how its elements transform waveforms.

This paper has explored the BMS group, and illustrated some of its
impact on gravitational-wave analysis.  We began with a thorough and
pedagogical introduction to the group itself, to provide a common
starting point to be used in the remainder of the paper.  We then
examined asymptotically flat spacetime, and found how the BMS group
transforms various types of waveforms.  We then used these insights to
see how such transformations can be implemented in practice.  This is
applied in the python package \software{scri} accompanying this paper
on its arXiv page.  The following section then showed how these
transformations should affect the spin-weighted spherical-harmonic
modes of a waveform with simplified numerical models, and we found
good agreement with analytical approximations for the leading-order
couplings.  Anomalous translations and boosts were found in the
publicly available SXS catalog.  A particular example was used to show
that the original data contains large effects from these anomalies,
including modes that are several to dozens of times larger than they
would be expected to be.  These modes can be transformed to simplify
their structure, and bring them more closely in line with what is
expected from analytical models.  However, more complicated systems
will have even larger mode couplings.  The size of the coupling is
expected to scale roughly linearly with the size of the translation
involved---since the direct contribution of the boost is relatively
small compared to the influence of the translation it gives rise
to---and some simulations in the SXS catalog have translations almost
100 times greater than the example system.  Finally, we discussed the
effect of the BMS gauge freedom on data analysis for
gravitational-wave detectors, showing that it must be accounted for
when creating model waveforms, but the supertranslations do not
complicate searches.

The waveforms found in the SXS catalog are not wrong, \perse; but they
contain effects that may not be expected.  For example, they will not
be consistent with the usual post-Newtonian waveforms; using the raw
waveforms to construct hybrids with \pN waveforms would result in
mismatches between the modes.
Using raw waveforms to calibrate effective-one-body
waveforms~\cite{Taracchini2014, Damour2014C, nagar2015}, surrogate
models~\cite{Field2014, Blackman2015a}, or other phenomenological
waveform models~\cite{Santamaria2010A, Damour2014D} would degrade the
quality of the numerous fits inherent to the calibration process, by
subjecting them to effectively random noise in the input.  A broader
and deeper survey of the effects of these transformations on waveforms
in the SXS catalog will be the subject of an upcoming
paper~\cite{Schmidt2016}.

Essentially, we have a more general form of the familiar alignment
problem in which arbitrary time and phase offsets need to be removed.
Those simple alignments are just special cases of the one described
here, restricted to the subgroup of BMS transformations consisting of
time translations and rotations about the $z$ axis.  This more general
alignment problem will necessitate using more general elements of the
BMS group.  With the algorithm presented in this paper, we can begin
to investigate ways to achieve such alignment.  Previous
investigations have suggested ways of using asymptotic data to
determine the center of mass, and more generally resolve the
supertranslation ambiguity~\cite{Moreschi2004, Helfer2007, Adamo2011,
  Kozameh2013A}.  While these are promising theoretical developments,
additional work will be needed to make these methods
practicable---towards which the present work is a crucial first step.

%%%%%%%%%%%%%%%%%%%%%%%%%%%%%%%%%%%%%%%%%%%%%%%%%%%%%%%%%%%%%%%%%%%%%%
\begin{acknowledgments}
  It is my pleasure to thank David Nichols and Leo Stein for helpful
  comments on the paper draft; and Scott Field, {\'{E}}anna Flanagan,
  Dan Hemberger, Larry Kidder, Richard O'Shaughnessy, Sergei Ossokine,
  Harald Pfeiffer, Mark Scheel, Patricia Schmidt, and Saul Teukolsky
  for useful conversations.  This project was supported in part by the
  Sherman Fairchild Foundation and by NSF Grants PHY-1306125 and
  AST-1333129.

  The computations discussed in this paper were performed on the
  \texttt{Zwicky} cluster hosted at Caltech by the Center for Advanced
  Computing Research, which was funded by the Sherman Fairchild
  Foundation and the NSF MRI-R\textsuperscript{2} program; and on the
  \texttt{GPC} and \texttt{Gravity} clusters at the SciNet HPC
  Consortium, funded by: the Canada Foundation for Innovation under
  the auspices of Compute Canada; the Government of Ontario; Ontario
  Research Fund--Research Excellence; and the University of Toronto.
\end{acknowledgments}

%%%%%%%%%%%%%%%%%%%%%%%%%%%%%%%%%%%%%%%%%%%%%%%%%%%%%%%%%%%%%%%%%%%%%%
%%%%%%%%%%%%%%%%%%%%%%%%%%%%%%%%%%%%%%%%%%%%%%%%%%%%%%%%%%%%%%%%%%%%%%
% \appendix* % Use \appendix* if there is just one appendix
\appendix % Use \appendix if there are multiple appendices

\section{Conventions}
\label{sec:conventions}

We start with some fiducial frame
$(\fourvec{t}, \fourvec{x}, \fourvec{y}, \fourvec{z})$, and some
corresponding observer $\observer$.  A spacetime event is a point
$\point$, and is represented by some vector corresponding to the
displacement from the origin of $\observer$ to that point.  The point
$\point$ can be given coordinates
$(\point_{t}, \point_{x}, \point_{y}, \point_{z})$ such that its
corresponding vector is
$\point_{t}\,\fourvec{t} + \point_{x}\,\fourvec{x} +
\point_{y}\,\fourvec{y} + \point_{z}\,\fourvec{z}$.

Another observer $\Rotated{\observer}$ moves at velocity
$\threevec{v}$ relative to $\observer$, which means that the location
of the spatial origin of $\Rotated{\observer}$ relative to the
(absolute) origin of $\observer$ is of the form
$\eta\, (\fourvec{t} + \threevec{v})$ for some $\eta$, where we assume
the speed of light is $c=1$.  We also define the following shorthand
notations:
\begin{subequations}
  \label{eq:velocity_parameters}
  \begin{gather}
    \beta \defined \lvert\, \threevec{v}\, \rvert, \\
    \rapidity \defined \artanh \beta, \\
    \gamma \defined \frac{1} {\sqrt{1 - \beta^{2}}}, \\
    \threevec{\rapidity} \defined \rapidity\, \frac{\fourvec{v}}
    {\beta}.
  \end{gather}
  It is worth noting the convenient identities
  \begin{gather}
    \gamma \equiv \cosh \rapidity, \\
    \beta\, \gamma \equiv \sinh \rapidity, \\
    \gamma(1+\beta) \equiv \cosh\rapidity + \sinh\rapidity \equiv
    \e^{\rapidity}, \\
    \gamma(1-\beta) \equiv \cosh\rapidity - \sinh\rapidity \equiv
    \e^{-\rapidity}, \\
    \frac{1}{2} \ln \frac{1+\beta} {1-\beta} \equiv \rapidity.
  \end{gather}
\end{subequations}
The frame
$(\Rotated{\fourvec{t}}, \Rotated{\fourvec{x}}, \Rotated{\fourvec{y}},
\Rotated{\fourvec{z}})$
of observer $\Rotated{\observer}$ is \emph{defined} by the relations
\begin{subequations}
  \label{eq:frame_of_moving_observer}
  \begin{align}
    \Rotated{\fourvec{t}} &\defined \rotor{B}\, \fourvec{t}\,
                            \rotor{B}^{-1}, \\
    \Rotated{\fourvec{x}} &\defined \rotor{B}\, \fourvec{x}\,
                            \rotor{B}^{-1}, \\
    \Rotated{\fourvec{y}} &\defined \rotor{B}\, \fourvec{y}\,
                            \rotor{B}^{-1}, \\
    \Rotated{\fourvec{z}} &\defined \rotor{B}\, \fourvec{z}\,
                            \rotor{B}^{-1},
  \end{align}
\end{subequations}
where $\rotor{B}$ is a Lorentz rotor:
\begin{equation}
  \label{eq:lorentz_rotor}
  \rotor{B} \defined \e^{- \threevec{\rapidity}\, \fourvec{t} / 2} = \cosh
  \frac{\rapidity}{2} - \frac{\fourvec{\rapidity}\,
    \fourvec{t}}{\rapidity}\, \sinh \frac{\rapidity}{2}.
\end{equation}
Here, we use the formalism of Geometric Algebra~\cite{Hestenes2002,
  Doran2010, Hestenes2015} to describe the boost.  In particular, the
term $\fourvec{\rapidity}\, \fourvec{t}$ represents the geometric (or
Clifford) product between these two vectors.  Because
$\fourvec{\rapidity}$ is a spatial vector, this product
$\fourvec{\rapidity}\, \fourvec{t}$ is a pure bivector
$\fourvec{\rapidity} \wedge \fourvec{t}$ representing the hyperplane
spanned by $\fourvec{\rapidity}$ and $\fourvec{t}$.  The quantity
$\rotor{B}$ is a mild generalization of a unit quaternion (also called
a rotor), except that now ``rotations'' need not be confined to
spatial planes; the vectors spanning the plane of rotation can now
include time components---as in this case.  For simplicity, we have
also specialized to the case where there is no additional rotation.
If there is some additional rotation, it can easily be absorbed by
redefining the frame of $\observer$, or simply replacing every
occurrence of $\rotor{B}$ with $\rotor{B}\, \FrameRotor$, where
$\FrameRotor$ represents the required (purely spatial) rotation of the
$\observer$ frame.

Using the form of $\rotor{B}$ given above, we have
$\Rotated{\fourvec{t}} = \gamma\, (\fourvec{t} + \threevec{v})$, which
agrees with our earlier statement, because any point at the spatial
origin of $\Rotated{\observer}$ will be of the form
$\eta'\, \Rotated{\fourvec{t}} = \eta'\, \gamma\, (\fourvec{t} +
\threevec{v}) = \eta\, (\fourvec{t} + \threevec{v})$
for some $\eta = \eta'\, \gamma$.  We should also note that
$\Rotated{\observer}$ observes $\observer$ moving with velocity
$-\Rotated{\fourvec{v}} = -\rotor{B}\, \fourvec{v}\, \rotor{B}^{-1} =
-\gamma\, (\fourvec{v} + \beta^{2}\, \fourvec{t})$,
which is a purely spatial vector in $\Rotated{\observer}$, with
magnitude $\beta$.

\section{Spherical coordinates}
\label{sec:spherical-coordinates}
Spherical coordinates are defined as usual, so that a point on the
sphere at position $\directionvec$ has coordinates $(\theta, \phi)$
when the angle between $\directionvec$ and $\fourvec{z}$ is $\theta$,
and the angle between $\fourvec{x}$ and the projection of
$\directionvec$ onto the $x$-$y$ plane is $\phi$.  Then, any point
$\directionvec$ may be represented by a rotor
$\rotor{R}_{\theta,\phi}$ as
\begin{equation}
  \label{eq:n_hat}
  \directionvec = \rotor{R}_{\theta,\phi}\, \fourvec{z}\,
  \rotor{R}_{\theta,\phi}^{-1},
\end{equation}
where
\begin{equation}
  \label{eq:R_theta_phi}
  \rotor{R}_{\theta,\phi} \defined \e^{\phi\, \fourvec{y}\, \fourvec{x}/2}\,
  \e^{\theta\, \fourvec{x}\, \fourvec{z}/2}.
\end{equation}
Similarly, the observer $\Rotated{\observer}$ can represent a point as
\begin{equation}
  \label{eq:n_hat_rotated}
  \Rotated{\directionvec} = \Rotated{\rotor{R}}_{\Rotated{\theta},
    \Rotated{\phi}}\, \Rotated{\fourvec{z}}\,
  \Rotated{\rotor{R}}_{\Rotated{\theta}, \Rotated{\phi}}^{-1},
\end{equation}
where
\begin{equation}
  \label{eq:R_theta_phi_rotated}
  \Rotated{\rotor{R}}_{\Rotated{\theta}, \Rotated{\phi}} \defined
  \e^{\Rotated{\phi}\, \Rotated{\fourvec{y}}\,
    \Rotated{\fourvec{x}}/2}\, \e^{\Rotated{\theta}\,
    \Rotated{\fourvec{x}}\, \Rotated{\fourvec{z}}/2} = \rotor{B}\,
  \rotor{R}_{\Rotated{\theta}, \Rotated{\phi}}\, \rotor{B}^{-1}.
\end{equation}
Note that the final form above is written using basis vectors from the
frame of $\observer$, but uses the coordinates measured by
$\Rotated{\observer}$.

This presentation of spherical coordinates in terms of the
corresponding rotor is useful, not only because the point itself may
be expressed as in Eqs.~\eqref{eq:n_hat} and~\eqref{eq:n_hat_rotated},
but also because the corresponding tangent vectors are easily
expressed.  For example, if $\fourvec{\theta}$ and $\fourvec{\phi}$
are the standard tangent vectors, we have
\begin{subequations}
  \label{eq:tangent_vectors}
  \begin{align}
    \fourvec{t} &= \rotor{R}_{\theta,\phi}\, \fourvec{t}\,
                  \rotor{R}_{\theta,\phi}^{-1}, \\
    \fourvec{\theta} &= \rotor{R}_{\theta,\phi}\, \fourvec{x}\,
                       \rotor{R}_{\theta,\phi}^{-1}, \\
    \fourvec{\phi} &= \rotor{R}_{\theta,\phi}\, \fourvec{y}\,
                     \rotor{R}_{\theta,\phi}^{-1}, \\
    \directionvec &= \rotor{R}_{\theta,\phi}\, \fourvec{z}\,
                  \rotor{R}_{\theta,\phi}^{-1}.
  \end{align}
\end{subequations}
This suggests the use of rotors more generally as a better way to keep
track of a basis frame than retaining all four vectors separately.  We
also take this opportunity to define another frame
\begin{subequations}
  \label{eq:prime_frame}
  \begin{align}
    \fourvec{t}'
    &\defined \rotor{R}_{\Rotated{\theta},
      \Rotated{\phi}}\, \fourvec{t}\, \rotor{R}_{\Rotated{\theta},
      \Rotated{\phi}}^{-1} \equiv \fourvec{t}, \\
    \fourvec{\theta}'
    &\defined \rotor{R}_{\Rotated{\theta}, \Rotated{\phi}}\,
      \fourvec{x}\, \rotor{R}_{\Rotated{\theta}, \Rotated{\phi}}^{-1}, \\
    \fourvec{\phi}'
    &\defined \rotor{R}_{\Rotated{\theta}, \Rotated{\phi}}\,
      \fourvec{y}\, \rotor{R}_{\Rotated{\theta}, \Rotated{\phi}}^{-1}, \\
    \directionvec'
    &\defined \rotor{R}_{\Rotated{\theta},
      \Rotated{\phi}}\, \fourvec{z}\, \rotor{R}_{\Rotated{\theta},
      \Rotated{\phi}}^{-1}.
  \end{align}
\end{subequations}
In the frame of $\observer$, the last three are pure spatial vectors.
Once they are boosted they will be pure spatial vectors in the frame
of $\Rotated{\observer}$ and, along with $\Rotated{\fourvec{t}}$, will
comprise the correct frame for a point on the sphere at coordinates
$(\Rotated{\theta}, \Rotated{\phi})$, as measured by
$\Rotated{\observer}$:
\begin{subequations}
  \begin{align}
    \Rotated{\fourvec{t}} &= \rotor{B}\, \fourvec{t}'\,
                            \rotor{B}^{-1}, \\
    \Rotated{\fourvec{\theta}} &= \rotor{B}\, \fourvec{\theta}'\,
                                 \rotor{B}^{-1}, \\
    \Rotated{\fourvec{\phi}} &= \rotor{B}\, \fourvec{\phi}'\,
                               \rotor{B}^{-1}, \\
    \Rotated{\directionvec} &= \rotor{B}\, \directionvec'\,
                            \rotor{B}^{-1}.
  \end{align}
\end{subequations}
This shows that we can extend Eqs.~\eqref{eq:tangent_vectors} to use
Lorentz rotors (generalizing from pure spatial rotors) to keep track
of all possible basis frames related by a Lorentz transformation,
rather than retaining all four vectors separately.

\section{Rotor of a boost}
\label{sec:rotor-boost}
Any future-directed null vector may be represented by $\observer$ up
to a positive scaling as
\begin{equation}
  \label{eq:null_vector}
  \fourvec{l} \defined \directionvec + \fourvec{t} \equiv
  \rotor{R}_{\theta,\phi}\, (\fourvec{z} + \fourvec{t})\,
  \rotor{R}_{\theta,\phi}^{-1}.
\end{equation}
Note that the rotor has no effect on $\fourvec{t}$, as it is an
entirely spatial rotor.  Similarly, observer $\Rotated{\observer}$ may
express any future-directed null vector via
\begin{subequations}
  \label{eq:rotated_null_vector}
  \begin{align}
    \Rotated{\fourvec{l}}
    &\defined
      \Rotated{\directionvec} + \Rotated{\fourvec{t}} \equiv
      \Rotated{\rotor{R}}_{\Rotated{\theta}, \Rotated{\phi}}\,
      (\Rotated{\fourvec{z}} + \Rotated{\fourvec{t}})\,
      \Rotated{\rotor{R}}_{\Rotated{\theta}, \Rotated{\phi}}^{-1} \\
    &=
      \rotor{B}\, \rotor{R}_{\Rotated{\theta}, \Rotated{\phi}}\,
      (\fourvec{z} + \fourvec{t})\, \rotor{R}_{\Rotated{\theta},
      \Rotated{\phi}}^{-1}\, \rotor{B}^{-1}.
  \end{align}
\end{subequations}
The final expression represents the conjugation by $\rotor{B}$ of a
vector expressed entirely in the basis of $\observer$, though using
coordinates as measured by $\Rotated{\observer}$.

We need to know the coordinates $(\theta, \phi)$ given
$(\Rotated{\theta}, \Rotated{\phi})$ such that $\fourvec{l}$ is a
positive scalar multiple of $\Rotated{\fourvec{l}}$, which is possible
if and only if $\fourvec{l}\, \Rotated{\fourvec{l}} = 0$.  (Again,
juxtaposition of the vectors $\fourvec{l}$ and $\Rotated{\fourvec{l}}$
denotes the geometric product.)  For now, let us assume that
$\rotor{B}$ is a boost along $\fourvec{z}$.  Then clearly
$\phi = \Rotated{\phi}$ since $\fourvec{y}$ and $\fourvec{x}$ are
unaffected.  We can calculate
\begin{subequations}
  \begin{align}
    \label{eq:parallel_null_condition}
    \fourvec{l}\, \Rotated{\fourvec{l}}
    &= \rotor{R}_{\theta,\phi}\, (\fourvec{z} +
      \fourvec{t})\, \rotor{R}_{\theta,\phi}^{-1}\, \rotor{B}\,
      \rotor{R}_{\Rotated{\theta}, \Rotated{\phi}}\, (\fourvec{z} +
      \fourvec{t})\, \rotor{R}_{\Rotated{\theta}, \Rotated{\phi}}^{-1}\,
      \rotor{B}^{-1} \\
    &= \e^{\theta\, \fourvec{x}\, \fourvec{z}/2}\, (\fourvec{z} +
      \fourvec{t})\, \e^{-\theta\, \fourvec{x}\, \fourvec{z}/2}\, \rotor{B}\,
      \e^{\Rotated{\theta}\, \fourvec{x}\, \fourvec{z}/2}\, (\fourvec{z} +
      \fourvec{t})\, \e^{-\Rotated{\theta}\, \fourvec{x}\, \fourvec{z}/2}\,
      \rotor{B}^{-1}.
  \end{align}
\end{subequations}
The key expression here is the Lorentz rotor
\begin{subequations}
  \begin{align}
    \label{eq:rotor_expansion}
    \rotor{L}
    &= \e^{-\theta\, \fourvec{x}\, \fourvec{z}/2}\, \e^{- \rapidity
      \fourvec{z}\, \fourvec{t}/2}\, \e^{\Rotated{\theta}\,
      \fourvec{x}\, \fourvec{z}/2} \\
    \begin{split}
      &= %
      \cosh \frac{\rapidity}{2}\, \cos\frac{\theta -
        \Rotated{\theta}}{2} - \fourvec{z}\, \fourvec{t} %
      \sinh \frac{\rapidity}{2}\, \cos\frac{\theta -
        \Rotated{\theta}}{2} \\ %
      &\quad + \fourvec{x}\, \fourvec{t} %
      \sinh \frac{\rapidity}{2}\, \sin\frac{\theta +
        \Rotated{\theta}}{2} - \fourvec{x}\, \fourvec{z} %
      \cosh \frac{\rapidity}{2}\, \sin\frac{\theta -
        \Rotated{\theta}}{2} .
    \end{split}
  \end{align}
\end{subequations}
In particular, if $\fourvec{l}\, \Rotated{\fourvec{l}}$ is a scalar,
we have
$\fourvec{l}\, \Rotated{\fourvec{l}} = (\fourvec{z} + \fourvec{t})\,
\rotor{L}\, (\fourvec{z} + \fourvec{t})\, \rotor{L}^{-1}$,
and if the latter expression is to be a scalar,
$\rotor{L}\, (\fourvec{z} + \fourvec{t})\, \rotor{L}^{-1}$ must have
no $\fourvec{x}$ component.  A simple argument from Geometric Algebra
shows that $\rotor{L}$ can only have terms involving $\fourvec{x}$ of
the form $\fourvec{x}\, (\fourvec{z} + \fourvec{t})$; terms of the
form $\fourvec{x}\, (\fourvec{z} - \fourvec{t})$ must vanish.  Using
the coefficients of $\fourvec{x}\, \fourvec{t}$ and
$\fourvec{x}\, \fourvec{z}$ above, some simple algebra shows us that
this implies that
\begin{subequations}
  \label{eq:boost_angle_effects}
  \begin{equation}
    \label{eq:boost_future_angle_effect}
    \tan \frac{\theta}{2} = \e^{-\rapidity}\, \tan
    \frac{\Rotated{\theta}}{2}.
  \end{equation}
  We can repeat this analysis for a past-directed null vector, and
  find the condition that
  $\rotor{L}\, (\fourvec{z} - \fourvec{t})\, \rotor{L}^{-1}$ must have
  no $\fourvec{x}$ component, which implies that
  \begin{equation}
    \label{eq:boost_past_angle_effect}
    \tan \frac{\theta}{2} = \e^{\rapidity}\, \tan
    \frac{\Rotated{\theta}}{2},
  \end{equation}
\end{subequations}
This is equivalent to flipping the sign or direction of the boost in
Eq.~\eqref{eq:boost_future_angle_effect}.  Note that
Eq.~\eqref{eq:boost_past_angle_effect} is the standard formula for
stellar aberration due to a boost,\footnote{See, \eg, Eq.~(1.3.5) of
  Ref.~\cite{Penrose1987}.} because an observer detects photons moving
into the future along past-directed null vectors.  Put another way, an
observer receiving null rays assigns a direction to a ray according to
where it \emph{came from}, rather than where it \emph{is going}; an
emitter assigns directions according to where the ray is going, rather
than where it would have come from---this is the reason for the sign
difference.

By looking at this more geometrically, we can eliminate the
requirement that the boost be in the $\fourvec{z}$ direction.  We
first dispense with the trivial case for which $\fourvec{v}$ and
$\directionvec$ are parallel or anti-parallel, in which case
$\directionvec = \directionvec'$.  Assuming henceforth the situation
is not so trivial, we note that $\fourvec{v}$, $\directionvec$, and
$\directionvec'$ all lie in the same plane, and angles between them
are governed by Eqs.~\eqref{eq:boost_angle_effects}.  To be specific,
define $\Rotated{\Theta}$ to be the angle measured by
$\Rotated{\observer}$ between $\Rotated{\fourvec{v}}$ and
$\Rotated{\directionvec}$, and similarly for $\Theta$.  We can
calculate $\Rotated{\Theta}$ in the $\observer$ frame as
\begin{equation}
  \label{eq:theta_rotated}
  \Rotated{\Theta} \defined \arccos \left[ \Rotated{\fourvec{v}}
    \cdot \Rotated{\directionvec} \right] \equiv \arccos \left[
    \fourvec{v} \cdot \left( \rotor{R}_{\Rotated{\theta},
        \Rotated{\phi}}\, \fourvec{z}\, \rotor{R}_{\Rotated{\theta},
        \Rotated{\phi}}^{-1} \right) \right].
\end{equation}
The corresponding value of $\Theta$ for future-directed (respectively
past-directed) null rays is simply
\begin{equation}
  \label{eq:boost_angle_effects_generally}
  \tan \frac{\Theta}{2} = \e^{\mp \rapidity}\, \tan
  \frac{\Rotated{\Theta}}{2}.
\end{equation}
Using this equation, we can find another useful relation between
$\fourvec{l}$ and $\Rotated{\fourvec{l}}$: the latter can be rotated
into the former with a rotation that is purely spatial in $\observer$.
Essentially, we simply rotate by $\Theta - \Rotated{\Theta}$ in the 
$\directionvec$-$\fourvec{v}$ plane.  The rotor that does this is
\begin{equation}
  \label{eq:rotor_of_boost}
  \BoostRotor \defined
  \exp \left[ \frac{\Theta - \Rotated{\Theta}}{2}\,
    \frac{\directionvec \wedge \fourvec{v}} {\lvert
      \directionvec \wedge \fourvec{v} \rvert} \right].
\end{equation}
With this rotor, we have
\begin{equation}
  \label{eq:parallel-null-rays-spatially}
  \rotor{R}_{\theta,\phi}\, (\fourvec{z} +
  \fourvec{t})\, \rotor{R}_{\theta,\phi}^{-1} = \BoostRotor\,
  \rotor{R}_{\Rotated{\theta}, \Rotated{\phi}}\, (\fourvec{z} +
  \fourvec{t})\, \rotor{R}_{\Rotated{\theta}, \Rotated{\phi}}^{-1}\,
  \BoostRotor^{-1}.
\end{equation}
Note that this equation does \emph{not} imply
$\rotor{R}_{\theta,\phi} = \BoostRotor\, \rotor{R}_{\Rotated{\theta},
  \Rotated{\phi}}$;
instead we have\footnote{We know that the extra factor in this
  equation is the most general possible such factor because: (1) it
  must be an even-grade element of unit norm, since all other elements
  of this equation have even grade and unit norm; (2) it must be
  purely spatial in $\observer$, since all other elements are purely
  spatial; (3) it must commute with $\fourvec{z} + \fourvec{t}$.
  Thus, it can only be a rotation in the $x$-$y$ plane.}
\begin{equation}
  \label{eq:rotor_boosted}
  \rotor{R}_{\theta, \phi}\, \e^{\SpinPhase\,
    \fourvec{x}\, \fourvec{y}/2} = \BoostRotor\,
  \rotor{R}_{\Rotated{\theta}, \Rotated{\phi}},
\end{equation}
for some angle $\SpinPhase$.  It turns out that this angle is the spin
phase described in Sec.~\ref{sec:RotationsAndBoosts}.  Though it will
never be necessary to compute this directly (except for the purposes
of visualizations like Fig.~\ref{fig:BoostedGrids}), we can rearrange
Eq.~\eqref{eq:spin-phase} and express $\SpinPhase$ as
\begin{equation}
  \label{eq:spin-phase}
  \SpinPhase= 2\log \left[ \rotor{R}_{\theta, \phi}^{-1}\, \BoostRotor\,
    \rotor{R}_{\Rotated{\theta}, \Rotated{\phi}} \right]\,
  \fourvec{y}\, \fourvec{x}.
\end{equation}
Of course, rather than computing this angle to evaluate spin-weighted
functions, we can just use the right-hand side of
Eq.~\eqref{eq:rotor_boosted} directly, and evaluate the spin-weighted
function on that rotor.

It may be helpful to see why this spin phase is a meaningful quantity
under a boost.  The fact that $\fourvec{v}$, $\directionvec$, and
$\directionvec'$ lie in the same plane and the fact that angles
between them are governed by
Eq.~\eqref{eq:boost_angle_effects_generally} are purely geometric
statements; they are independent of our basis frame.  We can use these
facts to express the value of a spin-weighted function in the boosted
frame in terms of the spin-weighted function in the original frame.
Assuming $\fourvec{v}$ and $\directionvec$ are not proportional to
each other, we know that the products
$\fourvec{t} \wedge \fourvec{v} \wedge \directionvec$ and
$\fourvec{t} \wedge \fourvec{v} \wedge \directionvec'$ are the same up
to some nonzero scalar multiple; they represent the same hyperplane.
Under the boost $\e^{-\fourvec{\rapidity}\, \fourvec{t}/2}$, the three
vectors $\fourvec{t}$, $\fourvec{v}$, and $\directionvec'$ transform
among themselves, which means that
$\Rotated{\fourvec{t}} \wedge \Rotated{\fourvec{v}} \wedge
\Rotated{\directionvec}$
also represents the same hyperplane.\footnote{That is, the latter
  product is the same as the former two products up to some other
  nonzero scalar multiple.  In fact, a straightforward calculation
  shows that
  $\csc \Rotated{\Theta}\, \Rotated{\fourvec{t}} \wedge
  \Rotated{\fourvec{v}} \wedge \Rotated{\directionvec} = \csc
  \Rotated{\Theta}\, \fourvec{t} \wedge \fourvec{v} \wedge
  \directionvec' = \csc\Theta\, \fourvec{t} \wedge \fourvec{v} \wedge
  \directionvec$.}
There is a unique axis orthogonal to this hyperplane.  In fact, we can
construct a unique unit vector $\fourvec{\Phi}$ along this axis by
defining
\begin{subequations}
  \label{eq:orthogonal_direction}
  \begin{align}
    \fourvec{\Phi} %
    &\defined \frac{\fourvec{t} \wedge \fourvec{v} \wedge
      \directionvec} {\beta\, \sin\Theta}\, \fourvec{t} \wedge
      \fourvec{\theta} \wedge \fourvec{\phi} \wedge \directionvec,
    \\ %
    &= \frac{\fourvec{t} \wedge \fourvec{v} \wedge
      \directionvec'} {\beta\, \sin\Rotated{\Theta}}\, \fourvec{t}
      \wedge \fourvec{\theta} \wedge \fourvec{\phi} \wedge
      \directionvec, \\ %
    &= \frac{\Rotated{\fourvec{t}} \wedge \Rotated{\fourvec{v}} \wedge
      \Rotated{\directionvec}} {\beta\, \sin\Rotated{\Theta}}\,
      \Rotated{\fourvec{t}} \wedge \Rotated{\fourvec{\theta}} \wedge
      \Rotated{\fourvec{\phi}} \wedge \Rotated{\directionvec}.
  \end{align}
\end{subequations}
% \begin{align}
%   \fourvec{\Phi} %
%   &= \fourvec{t} \wedge \fourvec{z} \wedge \fourvec{x}\, \fourvec{t}
%     \wedge \fourvec{x} \wedge \fourvec{y} \wedge \fourvec{z} \\ %
%   &= -\fourvec{z} \wedge \fourvec{x}\, \fourvec{x} \wedge
%   \fourvec{y} \wedge \fourvec{z} \\ %
%   &= -\fourvec{z} \wedge \fourvec{y} \wedge \fourvec{z} \\ %
%   &= \fourvec{y}% \\ %
% \end{align}
When $\fourvec{v} = \beta\, \fourvec{z}$, we have
$\fourvec{\Phi} = \fourvec{\phi}$---the usual basis vector.  But the
definition given in Eqs.~\eqref{eq:orthogonal_direction} is
geometrically invariant.  By construction $\fourvec{\Phi}$ is
orthogonal to $\fourvec{v} \wedge \fourvec{t}$, and so is invariant
under boosts.  More specifically, $\fourvec{\Phi}$ is a purely spatial
vector for both observers, orthogonal to the velocity, and lies in the
tangent space of the sphere at $\Rotated{\directionvec}$ for
$\Rotated{\observer}$ and at $\directionvec$ for $\observer$.  We can
therefore use it to compare directions in the tangent spaces for our
spin-weighted functions.

These invariance properties of the
$\fourvec{\Phi} = \Rotated{\fourvec{\Phi}}$ vector field allow us to
identify the alignment of the tangent space.  We choose the point on
the sphere designated by $\Rotated{\directionvec}$, with coordinates
$(\Rotated{\theta}, \Rotated{\phi})$.  This has a
standard~\cite{Ajith2007C} alignment of the tangent space given by
\begin{subequations}
  \label{eq:m_vectors}
  \begin{align}
    \label{eq:rotated-m_vector}
    \Rotated{\fourvec{m}}
    &\defined \frac{1}{\sqrt{2}}
      \left( \Rotated{\fourvec{\theta}} + \i\,
      \Rotated{\fourvec{\phi}} \right) \identically \rotor{B}\,
      \rotor{R}_{\Rotated{\theta}, \Rotated{\phi}}\, \frac{\fourvec{x} +
      \i\, \fourvec{y}} {\sqrt{2}}\, \rotor{R}_{\Rotated{\theta},
      \Rotated{\phi}}^{-1}\, \rotor{B}^{-1}. \\
    \intertext{Because this is a purely spatial vector in the frame of
    $\Rotated{\observer}$, but has a time component in the frame of
    $\observer$, direct comparison would be complicated.  However, we
    can define the similar vectors}
    \label{eq:m-prime_vector}
    \fourvec{m}'
    &\defined \frac{1}{\sqrt{2}}
    \left( \fourvec{\theta}' + \i\, \fourvec{\phi}' \right)
      \identically \BoostRotor\, \rotor{R}_{\Rotated{\theta},
      \Rotated{\phi}}\, \frac{\fourvec{x} + \i\, \fourvec{y}}
      {\sqrt{2}}\, \rotor{R}_{\Rotated{\theta}, \Rotated{\phi}}^{-1}\,
      \BoostRotor^{-1}, \\
    \label{eq:m_vector}
    \fourvec{m}
    &\defined \frac{1}{\sqrt{2}} \left( \fourvec{\theta} + \i\,
      \fourvec{\phi} \right) \identically \rotor{R}_{\theta, \phi}\,
      \frac{\fourvec{x} + \i\, \fourvec{y}} {\sqrt{2}}\,
      \rotor{R}_{\theta, \phi}^{-1},
  \end{align}
\end{subequations}
and the products
\begin{subequations}
  \label{eq:Phi_components}
  \begin{align}
    \label{eq:rotatedm-Phi}
    \Rotated{m}_{\Rotated{\Phi}}
    &\defined \Rotated{\fourvec{m}} \cdot \Rotated{\fourvec{\Phi}}, \\
    \label{eq:mprime-Phi}
    m'_{\Phi}
    &\defined \fourvec{m}' \cdot \fourvec{\Phi}, \\
    \label{eq:m-Phi}
    m_{\Phi}
    &\defined \fourvec{m} \cdot \fourvec{\Phi}.
  \end{align}
\end{subequations}
Almost by definition, we have
$\Rotated{m}_{\Rotated{\Phi}} \identically m'_{\Phi}$.  Thus, the
nontrivial comparison is between $m'_{\Phi}$ and $m_{\Phi}$.

To make this comparison, suppose that $\fourvec{l}$ and $\fourvec{l}'$
are as given in Eqs.~\eqref{eq:null_vector}
and~\eqref{eq:rotated_null_vector}.  We know that the rotors involved
in those expressions are related by Eq.~\eqref{eq:rotor_boosted}, so
we can calculate the relative alignment of the tangent spaces as
follows:
\begin{subequations}
  \begin{align}
    m'_{\Phi}
    &\defined \fourvec{m}' \cdot \fourvec{\Phi}, \\
    &= \left( \BoostRotor\, \rotor{R}_{\Rotated{\theta},
      \Rotated{\phi}} \frac{\fourvec{x} +
      \i\, \fourvec{y}} {\sqrt{2}}\,
      \bar{\rotor{R}}_{\Rotated{\theta}, \Rotated{\phi}} \,
      \bar{\BoostRotor} \right) \cdot \fourvec{\Phi}, \\
    &= \left( \rotor{R}_{\theta, \phi}\, \e^{\SpinPhase\,
      \fourvec{x}\, \fourvec{y}/2} \, \frac{\fourvec{x} +
      \i\, \fourvec{y}} {\sqrt{2}}\, \e^{-\SpinPhase\,
      \fourvec{x}\, \fourvec{y}/2} \, \bar{\rotor{R}}_{\theta, \phi}
      \right) \cdot \fourvec{\Phi}, \\
    &= \e^{\i\, \SpinPhase} \left( \rotor{R}_{\theta, \phi}\,
      \frac{\fourvec{x} + \i\, \fourvec{y}} {\sqrt{2}}\,
      \bar{\rotor{R}}_{\theta, \phi} \right) \cdot \fourvec{\Phi}, \\
    &= \e^{\i\, \SpinPhase}\, m_{\Phi}.
  \end{align}
\end{subequations}
This relation is exactly the one implied by Newman and Penrose's
original definition of spin at the beginning of Sec.~III in
Ref.~\cite{Newman1966}: they defined spin with respect to the
transformation $\fourvec{m}' = \e^{\i\, \SpinPhase}\, \fourvec{m}$.
This describes the relative alignment of the $\Rotated{\fourvec{m}}$
and $\fourvec{m}$ fields---except for any time component orthogonal to
$\fourvec{\Phi}$ produced by the relative boost.  Those additional
components cannot be accounted for simply by a rotation; they must be
accounted for by mixing between different components of the tensor in
question.  This is why we find various Newman-Penrose scalars on the
right-hand sides of Eqs.~\eqref{eq:psi_0-prime}
through~\eqref{eq:psi_3-prime}, for example.

This property of rotating the tangent space is very important, and is
the primary motivation for this more geometric approach.  That is
because we are dealing with spin-weighted functions, which means that
we need to know not only how points move around on the sphere, but
also how the tangent space to the sphere changes at each point under a
boost.  Using rotors allows us to automatically track both the change
of position and the change of the tangent basis.

\section{Conformal factor of a boost}
\label{sec:conformal-factor}
We can use spatial directions to label all the null directions from a
point, which has the topology of a sphere.  We define the metric on
this null sphere as the metric induced on the sphere of spatial
directions.  In that case, a boost induces a conformal transformation
of the null sphere, which means that we can find the conformal factor
of the boost.  In particular, for future-directed null rays, by simply
applying the transformation of $\theta$ given by
Eq.~\eqref{eq:boost_future_angle_effect} we can calculate
\begin{subequations}
  \label{eq:conformal_factors}
  \begin{equation}
    \label{eq:conformal_factor_future}
    \d \Rotated{\theta}^{2}  + \sin^{2} \Rotated{\theta}\,
    \d\Rotated{\phi}^{2} = \left[ \frac{1} {\gamma\, (1 - \fourvec{v}
        \cdot \directionvec)} \right]^{2}\, \left( \d \theta^{2}
      + \sin^{2} \theta\, \d\phi^{2} \right),
  \end{equation}
  and similarly for past-directed null rays with
  Eq.~\eqref{eq:boost_past_angle_effect}
  \begin{equation}
    \label{eq:conformal_factor_past}
    \d \Rotated{\theta}^{2}  + \sin^{2} \Rotated{\theta}\,
    \d\Rotated{\phi}^{2} = \left[ \frac{1} {\gamma\, (1 + \fourvec{v}
        \cdot \directionvec)} \right]^{2}\, \left( \d \theta^{2}
      + \sin^{2} \theta\, \d\phi^{2} \right).
  \end{equation}
\end{subequations}
So we define the conformal factors for future-directed ($k_{+}$) and
past-directed ($k_{-}$) null spheres as
\begin{equation}
  \label{eq:conformal_factor}
  k_{\pm} \defined \frac{1} {\gamma\, (1 \mp \fourvec{v} \cdot
    \directionvec)}.
\end{equation}
We know that the form of the metric is invariant under rotations, and
this the form of this conformal factor is clearly invariant under
rotations, so this is the correct conformal factor for boosts in any
direction.  In this paper, we are always dealing with future-directed
null rays, so we drop the subscript and just use
$\asymptotic{k} \defined k_{+}$.

\section{Estimating translation and boost in simulations}
\label{sec:estim-transl-boost}

The coordinate center of mass of a simulated compact binary presents
an imperfect representation of its motion.  Obviously, this can be
tainted by gauge effects, especially because the data are drawn from
the most dynamical and nonlinear portion of the simulated spacetime.
And while this may be a topic ripe for improvement, it is nonetheless
useful to have some way to illustrate the methods of this paper for
the waveform data of the SXS catalog.  In that spirit, this appendix
presents a simple method for estimating the translation and boost,
given the coordinate tracks and Christodoulou masses of the black
holes.  As noted in
Sec.~\ref{sec:RemovingDriftFromNumericalWaveforms}, the data can be
obtained from the \texttt{Horizons.h5} file accompanying each waveform
in the SXS catalog.

Denoting by $\xCoM(t)$ the coordinate location of the center of mass,
as a function of the coordinate time, and in units where the total
mass of the system is $1$, we can define the quantity
\begin{equation}
  \label{eq:objective-function}
  \Xi(\delta \fourvec{x}, \fourvec{v}) = \int_{t_{i}}^{t_{f}}
  \abs{\xCoM - (\delta \fourvec{x} + \fourvec{v}\, t)}^{2}\, \d t.
\end{equation}
This measures the distance between the origin and the center of mass
of a system transformed by $(\delta \fourvec{x}, \fourvec{v})$,
integrated over some range of times.  We can minimize this quantity
over the transformation to find the optimal transformation.  This
minimum can be found analytically by defining two moments of the
center of mass:
\begin{equation}
  \label{eq:CoM-moments}
    \fourvec{x}_0 = \int_{t_{i}}^{t_{f}} \xCoM(t)\, \d t
    \quad \text{and} \quad
    \fourvec{x}_1 = \int_{t_{i}}^{t_{f}} t\, \xCoM(t)\, \d t.
\end{equation}
Then, the minimum is given by
\begin{subequations}
  \label{eq:optimal-transformation}
  \begin{gather}
    \delta \fourvec{x} = \frac{4\, (t_{f}^{2} + t_{f}\, t_{i} +
      t_{i}^{2})\, \fourvec{x}_{0} - 6\, (t_{f}+t_{i})\,
      \fourvec{x}_{1}}
    {(t_{f} - t_{i})^{3}}, \\
    \fourvec{v} = \frac{12\, \fourvec{x}_{1} - 6\, (t_{f}+t_{i})\,
      \fourvec{x}_{0}} {(t_{f} - t_{i})^{3}}.
  \end{gather}
\end{subequations}
The moments can be computed by numerical integration of the data, and
simply plugged into these formulas to find the desired transformation.

The only free parameters in this case are the limits of integration,
$t_{i}$ and $t_{f}$.  In principle, these could span the entire time
for which there are two separate apparent horizons in the data.  In
some cases, as when a simulation needs to be aligned with another
simulation or an analytical waveform, for example, it would likely be
better to restrict this time span to the same times over which the
waveforms are being aligned.  In this case, however, where we are
simply interested in finding estimates for the motion of the systems,
we can be somewhat more liberal.  The initial time should be delayed
slightly, to allow junk radiation to settle down, so that the black
holes can be measured accurately.  For simplicity and definiteness, we
will set $t_{i}$ to be \SI{1}{\percent} of the entire time for which
data is available.  On the other hand, bizarre features are sometimes
present in the SXS catalog close to merger.  To avoid these, and to
lessen the impact of true physical recoils that develop close to
merger, we similarly set $t_{f}$ to be \SI{10}{\percent} before the
end of the data.

With this simple recipe in hand, we can apply it to the entire SXS
catalog very easily.  The results are shown for \texttt{SXS:BBH:0004}
in Eq.~\eqref{eq:0004-transformation}, and are aggregated for the
entire catalog in Fig.~\ref{fig:CatalogCoM}.  Again, this is a very
crude and gauge-sensitive measure of the motion of the system.  It
should no doubt be improved in future work.  But for the purposes of
illustration in this paper, it seems to be sufficient.

%%%%%%%%%%%%%%%%%%%%%%%%%%%%%%%%%%%%%%%%%%%%%%%%%%%%%%%%%%%%%%%%%%%%%%
%%%%%%%%%%%%%%%%%%%%%%%%%%%%%%%%%%%%%%%%%%%%%%%%%%%%%%%%%%%%%%%%%%%%%%
%% References

%% Try this if the last two columns before the bib don't break nicely
\vspace{0.1in}

\vfil

%% Try this if missing footnote on page with references:
% \clearpage

%% Bibtex sometimes uses these automatically, so we have to make sure
%% they have their original meanings when it does
\let\c\Originalcdefinition %
\let\d\Originalddefinition %
\let\i\Originalidefinition

%% Include any .bib files that need to be referenced, with `.bib`
%% removed, separated by commas without spaces.
\bibliography{References,References2}
%% My standard way of working with references is to use the `zotelo`
%% emacs mode, where I tell zotelo which collection in my Zotero
%% database I want to use with `zotelo-set-collection` (which may be
%% updated with `zotelo-update-database`).  The collection is then
%% exported as bibtex into the first file named above.

%%%%%%%%%%%%%%

\end{document}